\documentclass[%
 reprint,
 amsmath,amssymb,
 aps,
prl,
]{revtex4-2}

\usepackage{amsthm}
\usepackage{graphicx}
\usepackage{dcolumn}
\usepackage{bm}
\usepackage{quantikz}
\usepackage{tikz}
\tikzset{
  quantikz/classical wire/.append style={
    preaction={draw, line width=0.6pt, white}
  }
}
\newtheorem*{definition}{\textbf{Definition}}
\newtheorem*{proposition}{\textbf{Proposition}}
\newtheorem*{theorem}{\textbf{Theorem}}
\newtheorem*{lemma}{Lemma}
\newenvironment{proofS}{%
  \par\noindent\textbf{Proof.}---\normalfont
}{\hfill$\square$\par}
\makeatletter
\let\orig@addcontentsline\addcontentsline
\newcommand{\TOCdisable}{\renewcommand{\addcontentsline}[3]{}}
\newcommand{\TOCenable}{\let\addcontentsline\orig@addcontentsline}
\makeatother
\begin{document}
\TOCdisable

\preprint{APS/123-QED}

\title{Verification of general multi-qudit pure states}

\author{Xiao-Dong Zhang\textsuperscript{1}}
\author{Bin-Bin Cai\textsuperscript{1,2}}
 \email{cbb@finu.edu.cn}
\author{Song Lin\textsuperscript{1}}
 \email{lins95@fjnu.edu.cn}
\affiliation{\textsuperscript{1}College of Computer and Cyber Security, Fujian Normal University, Fuzhou 350117, China\\
\textsuperscript{2}Digital Fujian Internet-of-Things Laboratory of Environmental Monitoring, Fujian Normal University, Fuzhou 350117, China}
\date{\today}

\begin{abstract}

Verifying prepared quantum states is crucial for hybrid systems whose subsystems may have different local dimensions. In this paper, we present a generalized stabilizer framework and associated test that apply to general multi-qudit states, including composite-dimensional and hybrid architectures. Using only adaptive local measurements, our method verifies qutrit–qubit states, arbitrary two-qubit pure states, Bell/Bell-like, GHZ/GHZ-like, graph, hypergraph, multigraph, and multihypergraph states, with efficiencies matching or surpassing the best known schemes.
\end{abstract}

\maketitle
Entangled quantum states are key resources for quantum communication, computation, sensing, and networked quantum information processing \cite{gisin2007quantum,divincenzo1995quantum,degen2017quantum}. A central practical task is quantum-state verification, namely verifying that an untrusted device prepares the intended state. This is crucial for verifiable blind quantum computation and quantum networks\cite{hayashi2015verifiable,morimae2017verification,takeuchi2019resource,drmota2024verifiable,quan2023verifiable,pappa2012multipartite,mccutcheon2016experimental,han2021optimal,unnikrishnan2022verification}. Quantum-state verification has developed from early GHZ- and graph-state verification \cite{pappa2012multipartite,hayashi2015verifiable,mccutcheon2016experimental} into a framework with rigorous and near-optimal efficiency guarantees, including adversarial and device-independent settings \cite{pallister2018optimal,zhu2019efficient1,zhu2019general,takeuchi2018verification,govcanin2022sample,xu2024measurement}. High-performance protocols now exist for many important families of states—Bell states, arbitrary two-qubit states, GHZ and GHZ-like states, Dicke and phased Dicke states, stabilizer, graph, and hypergraph states, and have been optimized for efficiency, robustness, and scalability \cite{wang2019optimal,li2019efficient,yu2019optimal,li2021minimum,zhu2019optimal,kalev2019validating,dangniam2020optimal,li2020optimal,han2021optimal,hayashi2019verifying,unnikrishnan2022verification,xu2024efficient,li2023robust,zhu2019efficient,tao2022verification,zhu2024efficient,chen2023efficient,zhang2025verification,chen2025quantum}. Quantum memory can even push the efficiency close to theoretical limits \cite{chen2025quantum}. At the same time, experiments now stabilize entanglement across subsystems of different local dimension, such as a superconducting qutrit–qubit pair \cite{brown2022trade}. This creates a pressing need to verify hybrid-dimensional states, not only uniform multi-qubit registers.

Beyond the above classes of states, there has been substantial progress toward more general multi-qubit verification. Liu \textit{et al.} gave a systematic construction for generic multi-qubit states \cite{liu2023efficient}, and Huang \textit{et al.} experimentally certified almost all multi-qubit states using only a few single-qubit measurements \cite{huang2025certifying}. Beyond qubits, Li and Zhu proposed a universal qudit scheme based on the Schmidt decomposition and mutually unbiased bases (MUBs) \cite{li2025universal}. That scheme relies on the existence of a complete set of MUBs, which is only known in prime and prime-power dimensions and is not generally available in composite dimensions \cite{li2025universal}. Therefore, it  cannot work on arbitrary qudit systems or heterogeneous architectures with different local dimensions, such as the qutrit–qubit platform mentioned above \cite{brown2022trade}. Despite advances in efficiency, robustness, adversarial security, and device independence \cite{zhu2019general,takeuchi2018verification,govcanin2022sample,xu2024measurement,zhang2025verification}, there is still no dimension-independent, efficient method that covers general hybrid systems.

In this paper, we propose a dimension-independent and efficient verification method for quantum states in a general multi-qudit system, i.e., a composite system of several subsystems whose local Hilbert-space dimensions may differ \cite{daboul2003quantum}. We first introduce generalized stabilizers for general multi-qudit quantum states, inspired by generalized stabilizers in qubit systems \cite{zhang2021efficient}, and then develop generalized stabilizer testing based on adaptive local measurements. Based on the generalized stabilizer testing, the presented quantum-state verification protocol applies uniformly to qubit, qudit, and general multi-qudit systems with nonuniform local dimensions, without assuming complete MUB sets. In this process, we first analyze prime dimensions and then extend to composite dimensions using suitable hybrid operators. The presented scheme enables practical, high-efficiency verification for qutrit–qubit states, arbitrary two-qubit pure states, qubit and qudit Bell/Bell-like states, GHZ/GHZ-like states, graph and hypergraph states, stabilizer states, and recently proposed multigraph and multihypergraph states \cite{fields2022optimal,li2020optimal,rossi2013quantum,qu2013encoding,steinhoff2017qudit,xiong2018qudit,gottesman1996class,gottesman1998fault,hostens2005stabilizer,zhang2024quantum}, with required numbers of measurements that are lower than or comparable to the best existing schemes.

\textit{Framework for verification protocols}.---Following Refs.~\cite{pallister2018optimal,zhu2019efficient1,zhu2019general}, we adopt the standard framework for quantum-state verification in the nonadversarial setting. Suppose a device produces a sequence of states $\sigma_i=\left| \psi \right\rangle \left\langle \psi \right|$ or $\left\langle \psi_{\epsilon} \right|\sigma_i\left| \psi_{\epsilon}  \right\rangle \leq 1-\epsilon$, $i=1,2,3,\cdots$, whose infidelity with $\left| \psi \right\rangle \left\langle \psi \right|$ is at least $\epsilon$. Operationally, these $\sigma_i$ are claimed to be the target state $\left|\psi \right\rangle$, but they may in fact be arbitrary. We test $n_{opt}$ copies $\sigma_1,\sigma_2,\ldots,\sigma_{n_{opt}}$ and draw a conclusion such as: ``The prepared state has fidelity $99\%$ with the target, with confidence level $90\%$.''. Each test is a binary projective measurement $\{P_j, I-P_j\}$, where $P_j$ denotes ``pass'', $I-P_j$ denotes ``fail'', and $I$ is identity. We choose $P_j$ with probability $\mu_j$, from a fixed set $\mathcal{S}$ such that $P_j\left| \psi \right\rangle=\left| \psi \right\rangle$ for all $P_j\in\mathcal{S}$. The overall verification operator is
$\Omega=\sum_{j}\mu_jP_j$. If all $n_{opt}$ tests pass, we accept the source as preparing $\left|\psi \right\rangle$. Otherwise, we reject it. For any state with $\left\langle \psi \right|\sigma_i\left| \psi \right\rangle \leq 1-\epsilon$, i.e., any state that is at least $\epsilon$-far in infidelity from the target, the maximum probability of passing all tests is
\begin{equation}
\hspace{-7mm}
\max_{\left\langle \psi \right|\sigma_i\left| \psi \right\rangle \leq 1-\epsilon}{\text{tr}(\Omega\sigma_i)=1-[1-\beta(\Omega)]\epsilon=1-\nu(\Omega)\epsilon},
\label{eq:01}
\end{equation}
where $\beta(\Omega)$ is the second-largest eigenvalue of $\Omega$, and $\nu(\Omega)=1-\beta(\Omega)$ is the spectral gap. To verify the target state with infidelity at least $\epsilon$ and significance level $\delta$, the number of tests $n_{opt}$ required satisfies \cite{pallister2018optimal,zhu2019efficient1,zhu2019general}
\begin{equation}
\normalsize
\begin{aligned}
 n_{opt}=\left\lceil \frac{\ln\delta}{\ln[1-\nu(\Omega)\epsilon]}\right\rceil\leq\left\lceil \frac{\ln\delta^{-1}}{\nu(\Omega)\epsilon}\right\rceil.
\end{aligned}
\label{eq:02}
\end{equation}

\textit{Generalized stabilizer states}.—Zhang \textit{et al.}~\cite{zhang2021efficient} extended the notion of qubit stabilizer states to qubit generalized stabilizer states.

\begin{definition}[Qubit generalized stabilizer state~\cite{zhang2021efficient}]\label{def1}
For an $n$-qubit system, a generalized stabilizer state $\left| \psi \right\rangle$ is the unique eigenstate with eigenvalue $+1$ of $n$ mutually commuting, independent generalized stabilizer operators $g_0,\cdots,g_{n-1}$, where each $g_i$ is Hermitian and unitary. The set $S=\{g_0,\cdots,g_{n-1}\}$ is called the set of generalized stabilizer generators.
\end{definition}
\begin{proposition}[Ref.~\cite{zhang2021efficient}]\label{prop1}
For any pure state $\left| \psi \right\rangle \in \mathcal{H}^{\otimes n}$, there exists a generalized stabilizer set $S_\psi$ that uniquely determines $\left| \psi \right\rangle$.
\end{proposition}

Before defining generalized stabilizers for general multi-qudit states, we recall the structure of a general multi-qudit system. Following Ref.~\cite{daboul2003quantum}, the basis of such a system can be written as
\begin{equation}
\normalsize
\hspace{-2mm}
\begin{aligned}
\left| k_0 \right\rangle\otimes\cdots\otimes\left| k_{n-1} \right\rangle
\in
\mathcal{H}_{d_0}\otimes\cdots\otimes\mathcal{H}_{d_{n-1}},
k_i\in\mathbb{Z}_{d_i}.
\end{aligned}
\label{eq:03}
\end{equation}
A general multi-qudit state $\left| \psi \right\rangle$ takes the form
\begin{equation}
\normalsize
\begin{aligned}
\left| \psi \right\rangle
=U_{\psi}
\left| 0_{d_0} \right\rangle\otimes\cdots\otimes\left| 0_{d_{n-1}} \right\rangle,
\end{aligned}
\label{eq:04}
\end{equation}
where the subscript indicates the corresponding local dimension. If two or more of the $d_i$ differ, the multi-qudit system in Eq.~\eqref{eq:03} is called a \textit{hybrid system}. If all $d_0,\cdots,d_{n-1}$ are prime numbers, it is called a \textit{prime-dimensional multi-qudit system}. We use the following notation for single-particle gates in a hybrid system. The superscript denotes the local dimension of the particle on which the gate acts, and the subscript denotes its position. For example, $Z_k^{(d_k)}$ is the $d_k$-dimensional generalized Pauli-$Z$ acting on the particle at position $k\in\mathbb{Z}_n$. In a non-hybrid system, where $d_0=\cdots=d_{n-1}$, no explicit distinction is required, the superscript is omitted and the subscript is dropped when the position is clear. A representative hybrid quantum operation $U_\psi$ is the \textit{hybrid controlled-phase gate} from Ref.~\cite{daboul2003quantum},
$\mathcal{CZ}_{(j\leftarrow k)}^{(d_j\otimes d_k)}:=\sum_{i_k=0}^{d_k-1} \scriptstyle\left(Z_j^{(d_j)}\right)^{i_k}\otimes\left|i_k\right\rangle\left\langle i_k\right|$. For example, when $d_0=3$ and $d_1=2$, let $U_{\psi_1}=\mathcal{CZ}_{(0\leftarrow1)}^{(3\otimes2)}({QFT}^{(3)}\otimes H)$. The corresponding quantum state is
$|\psi_1\rangle
=U_{\psi_1}|0_3\rangle|0_2\rangle
=\frac{1}{\sqrt{6}}\big[
(|0_3\rangle+|1_3\rangle+|2_3\rangle)\otimes|0_2\rangle
+(|0_3\rangle+\omega_3|1_3\rangle+\omega_3^2|2_3\rangle)\otimes|1_2\rangle
\big]$, where $\omega_3=e^{2\pi \mathbf{i}/3}$ and $\mathbf{i}=\sqrt{-1}$. Such hybrid quantum states also arise in practice, e.g., in the \textit{qutrit--qubit system} of Ref.~\cite{brown2022trade}. Now, we give the following definition and proposition.

\begin{definition}[\textit{Generalized stabilizer states for prime-dimensional multi-qudit systems}]\label{def2}
For a multi-qudit system with prime local dimensions $d_0,\cdots,d_{n-1}$, a generalized stabilizer state $\left| \psi \right\rangle \in \mathcal{H}_{d_0}\otimes\cdots\otimes\mathcal{H}_{d_{n-1}}$ is the unique eigenstate with eigenvalue $+1$ of $n$ mutually commuting, independent generalized stabilizer operators $g_0,\cdots,g_{n-1}$, each of which is unitary. The set $S=\{g_0,\cdots,g_{n-1}\}$ is called the set of generalized stabilizer generators.
\end{definition}

\begin{proposition}[Proof provided in the Supplemental Material]\label{prop2}
For a multi-qudit system with prime local dimensions $d_0,\cdots,d_{n-1}$, and for any pure state $\left| \psi \right\rangle \in \mathcal{H}_{d_0}\otimes\cdots\otimes\mathcal{H}_{d_{n-1}}$, there exists a generalized stabilizer set $S_\psi$ that uniquely determines $\left| \psi \right\rangle$.
\end{proposition}

The relation between prime-dimensional quantum states and their generalized stabilizers is (proof provided in the Supplemental Material)
\begin{equation}
    \hspace{-6mm}
    \normalsize
    \begin{aligned}
        \left| \psi \right\rangle \left\langle \psi \right|
    &=\frac{1}{\prod_{i=0}^{n-1}d_i}
    \sum_{h_0=0}^{d_0-1}\sum_{h_1=0}^{d_1-1}\cdots\sum_{h_{n-1}=0}^{d_{n-1}-1}
    \left(\prod_{k=0}^{n-1}g_k^{h_k}\right)
    \\&=\frac{1}{\prod_{i=0}^{n-1}d_i}
    \prod_{i=0}^{n-1}\left(\sum_{j=0}^{d_i-1}g_i^j\right).
    \end{aligned}
    \label{eq6}
\end{equation}
Unlike in the qubit case, the generators of qudit generalized stabilizers are not Hermitian in general. This difference arises because the qubit Pauli $Z$ operator is Hermitian, whereas the qudit Pauli $Z$ operator is not. If ${Z_i}^\dagger = Z_i$, then $(U_\psi Z_i U_\psi^\dagger)^\dagger = U_\psi Z_i U_\psi^\dagger$. Otherwise this equality fails. For example, qubit hypergraph states are qubit generalized stabilizer states with Hermitian stabilizer generators \cite{rossi2013quantum,qu2013encoding},  while qudit hypergraph states are qudit generalized stabilizer states whose stabilizer generators are non-Hermitian \cite{steinhoff2017qudit,xiong2018qudit}. 

We now consider composite local dimensions. For a single particle with dimension $d=4$ or $d=6$, the states $\left|0_4\right\rangle$ and $\left|0_6\right\rangle$ can be written as $\left|0_2\right\rangle\left|0_2\right\rangle$ and $\left|0_3\right\rangle\left|0_2\right\rangle$, respectively. Thus, a single-particle state with composite dimension can be represented as a multi-particle state with prime dimensions, which may be hybrid or non-hybrid. Under this identification,
$U_\psi\left|0_4\right\rangle = U_\psi\left|0_2\right\rangle\left|0_2\right\rangle$ and
$U_\psi\left|0_6\right\rangle = U_\psi\left|0_3\right\rangle\left|0_2\right\rangle$.
In general, if a single-particle Hilbert space has dimension $d={p_1}^{\alpha_1}\cdots{p_m}^{\alpha_m}$, then
\begin{equation}
    U_\psi\left| 0_d \right\rangle
    =U_\psi\left| 0_{p_1} \right\rangle^{\otimes\alpha_1}\otimes\cdots\otimes\left| 0_{p_m} \right\rangle^{\otimes\alpha_m}.
\end{equation}
Now consider an arbitrary $n$-particle state $\left| \psi \right\rangle$, where the $i$th particle has local dimension $d_i={p_{i,1}}^{\alpha_{i,1}}{p_{i,2}}^{\alpha_{i,2}}\cdots{p_{i,m_i}}^{\alpha_{i,m_i}}$. The state $\left| \psi \right\rangle$ can be written as
\begin{equation}
\hspace{0mm}
\normalsize
\begin{aligned}
\left| \psi \right\rangle&=U_\psi\left| 0_{d_0} \right\rangle\otimes\cdots\otimes\left| 0_{d_{n-1}} \right\rangle\\[-1mm]
    &=U_\psi\bigotimes_{i=0}^{n-1}\left| 0_{p_{i,1}} \right\rangle^{\otimes\alpha_{i,1}}\otimes\cdots\otimes\left| 0_{p_{i,m_i}} \right\rangle^{\otimes\alpha_{i,m_i}}.
\label{eq5}
\end{aligned}
\end{equation}
In this representation, $\left| \psi \right\rangle$ becomes a prime-dimensional state with $\sum_{i=1}^n \left(\sum_{j=1}^{m_i}\alpha_{i,j}\right)$ particles. This is only a relabeling of subsystems, and physically, the state remains in the original system. Our purpose is to construct the associated generalized stabilizers. By the generalized stabilizer definition and existence result established above, $\left| \psi \right\rangle$ is uniquely specified by a set of generalized stabilizer generators. Hence we can obtain generalized stabilizers for general multi-qudit states, including both prime- and composite-dimensional systems.

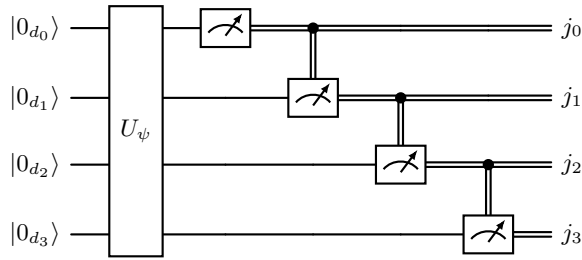
\begin{figure}[t]
  \centering
  \resizebox{0.92\columnwidth}{!}{
  \begin{quantikz}[row sep=0.38cm, column sep=0.50cm]
  \lstick{$|0_{d_0}\rangle$}
    & \gate[wires=4]{U_{\psi}}
    & \meter{}  
    & \ctrl[vertical wire=c]{1} \setwiretype{c}& \setwiretype{c} & \setwiretype{c}  &\setwiretype{c}\rstick{$j_0$}\\
  \lstick{$|0_{d_1}\rangle$}
    & 
    & 
    & \meter{}  
    & \ctrl[vertical wire=c]{1} \setwiretype{c} & \setwiretype{c}  & \setwiretype{c} \rstick{$j_1$}\\
  \lstick{$|0_{d_2}\rangle$}
    & 
    & 
    & 
    & \meter{} 
    & \ctrl[vertical wire=c]{1}\setwiretype{c} & \setwiretype{c}\rstick{$j_2$}\\
  \lstick{$|0_{d_3}\rangle$}
    & 
    & 
    &  
    & 
    & \meter{}
    & \setwiretype{c} \rstick{$j_{3}$}
  \end{quantikz}
  }%
  \caption{Adaptive multi-qudit measurement for $n=4$.}
  \label{fig:adaptive}
\end{figure}

\textit{Generalized stabilizer testing}.---For standard stabilizer states, the stabilizers are tensor products of Pauli operators, so stabilizer testing can be implemented by local Pauli measurements~\cite{kalev2019validating}. For generalized stabilizers, which are not tensor products of Pauli operators, the procedure is more involved.

Consider a general multi-qudit entangled state
$\left| \psi \right\rangle = U_\psi \left| 0_{d_0} \right\rangle \otimes \cdots \otimes \left| 0_{d_{n-1}} \right\rangle$.
We perform an adaptive local measurement as follows. First, measure the first particle in the basis
$\{U_0\left|0\right\rangle,\ldots,U_0\left|d_0-1\right\rangle\}$,
obtaining outcome $j_0$ and projecting the remaining $(n-1)$ particles onto an $(n-1)$-qudit pure state. Next, measure the second particle in the basis
$\{U_1\left|0\right\rangle,\ldots,U_1\left|d_1-1\right\rangle\}$,
obtaining $j_1$ and projecting the rest onto an $(n-2)$-qudit pure state. This process continues until only the last particle remains, which is a single-qudit pure state $U_{n-1}\left|0\right\rangle$. When that particle is finally measured in
$\{U_{n-1}\left|0\right\rangle,\ldots,U_{n-1}\left|d_{n-1}-1\right\rangle\}$,
the outcome is deterministically $j_{n-1}=0$. Figure~1 illustrates this measurement procedure for $n=4$. Such an adaptive procedure is feasible in practice. Since $\left| \psi \right\rangle$ is known, the measurement bases and the expected previous outcomes are known in advance. The $i$th measurement basis $\{U_i\left|i\right\rangle\}$ may depend on earlier bases and outcomes, i.e.,
$U_i \in \{ U_i(\mathbf{J}_i)\,|\,\mathbf{J}_i \in \mathcal{J} \}$,
where
$\mathcal{J}:=\mathbb{Z}_{d_0}\times\cdots\times\mathbb{Z}_{d_{i-1}}$
and
$\mathbf{J}_i=\{j_0,j_1,\cdots,j_{i-1}\}$
collects the outcomes of the first $i$ particles (with $\mathbf{J}_0=\varnothing$). The measurement order of particles may also be rearranged, as in measurement-based one-way quantum computation~\cite{raussendorf2001one}.

The measurement operator at step $i$ is $
P_i(j_i|\mathbf{J}_i):=U_i(\mathbf{J}_i)|j_i\rangle\langle j_i|U_i(\mathbf{J}_i)^\dagger$
. A full adaptive measurement path $\mathbf{J}$ defines the projector $\Pi(\mathbf{J}):=\prod_{i=0}^{n-1} P_i(j_i|\mathbf{J}_i)$, which satisfies
$\Pi(\mathbf{J})\Pi(\mathbf{J}')=\delta_{\mathbf{J},\mathbf{J}'}\Pi(\mathbf{J})$
and
$\sum_{\mathbf{J}\in\mathcal{J}}\Pi(\mathbf{J})=I$.
We assign to each path a complex weight
$\Lambda(\mathbf{J}):=\prod_{i=0}^{n-1}\lambda_{(i,j_i,\mathbf{J}_i)},
\qquad
\lambda_{(i,j_i,\mathbf{J}_i)} \in \mathbb{C}$, and define the overall measurement operator $M=\sum_{\mathbf{J}\in\mathcal{J}}\Lambda(\mathbf{J})\Pi(\mathbf{J})$, which collects all adaptive paths and their weights. Define the support of the state as $\text{Supp}(\psi):=\{\mathbf{J}\,|\,\mathbf{J}\in\mathcal{J},\ \Pi(\mathbf{J})|\psi\rangle\neq0\}$. If $\Lambda(\mathbf{J})=1$ for all $\mathbf{J}\in\text{Supp}(\psi)$, then
$M|\psi\rangle=|\psi\rangle$. In this case $M$ is a generalized stabilizer of $\left| \psi \right\rangle$, and the above adaptive measurement realizes its generalized stabilizer testing. If two operators $M$ and $M'$ differ only in the coefficients $\lambda_{(i,j_i,\mathbf{J}_i)}$ but use the same measurement bases, the same run of the procedure tests both stabilizers simultaneously. When $M$ happens to be a tensor product of Pauli operators, this construction reduces to standard stabilizer testing. 

As an example, consider the two-qubit state from Ref.~\cite{pallister2018optimal},
$\left| \psi_2\right\rangle=\sin{\theta}\left|00\right\rangle+\cos{\theta}\left|11\right\rangle$
(the Bell-like state of Ref.~\cite{fields2022optimal}).
The corresponding unitary is
$U_{\psi_2}=CNOT\cdot(R_y(\pi-2\theta)\otimes I)$,
and the generalized stabilizer generators are
$\{g_0^{(\psi_2)}=U_{\psi_2}Z_0U_{\psi_2}^\dagger,\,
g_1^{(\psi_2)}=U_{\psi_2}Z_1U_{\psi_2}^\dagger=Z\otimes Z\}$.
Explicitly,
\begin{equation}
    \normalsize
    \begin{aligned}
    g_0^{(\psi_2)}&=[\left| +\varphi_{+x} \right\rangle \left\langle +\varphi_{+x} \right|
    -\left| -\varphi_{+x} \right\rangle \left\langle -\varphi_{+x} \right|]
    \otimes\left| +x \right\rangle \left\langle +x \right|\\
    &-[\left| +\varphi_{-x} \right\rangle \left\langle +\varphi_{-x} \right|
    -\left| -\varphi_{-x} \right\rangle \left\langle -\varphi_{-x} \right|]
    \otimes\left| -x \right\rangle \left\langle -x \right|.
    \end{aligned}
\label{eq:07}
\end{equation}
Here,
$\left| \pm x \right\rangle=\frac{1}{\sqrt{2}}(\left|0\right\rangle\pm\left|1\right\rangle)$,
$\left| +\varphi_{+x} \right\rangle=\sin{\theta}\left|0\right\rangle+\cos{\theta}\left|1\right\rangle$,
$\left| -\varphi_{+x} \right\rangle=\cos{\theta}\left|0\right\rangle-\sin{\theta}\left|1\right\rangle$,
$\left| +\varphi_{-x} \right\rangle=\cos{\theta}\left|0\right\rangle+\sin{\theta}\left|1\right\rangle$,
and
$\left| -\varphi_{-x} \right\rangle=\sin{\theta}\left|0\right\rangle-\cos{\theta}\left|1\right\rangle$. The testing procedure for $g_1$ is as follows. Measure the second qubit (or equivalently the first, since $\left| \psi_2 \right\rangle$ is symmetric) in the basis $\{\left| \pm x \right\rangle\}$. If the outcome is $\left|+x\right\rangle$ (respectively $\left|-x\right\rangle$), measure the other qubit in the basis $\{\left| \pm\varphi_{+x} \right\rangle\}$ (respectively $\{\left| \pm\varphi_{-x} \right\rangle\}$). The test is accepted if the second measurement yields $\left|+\varphi_{+x}\right\rangle$ (or $\left|+\varphi_{-x}\right\rangle$). For $\theta=\frac{\pi}{4}$, $\left| \psi_2 \right\rangle$ becomes a Bell state, the generalized stabilizer reduces to the standard stabilizer $X\otimes X$, and the generalized stabilizer test reduces to standard stabilizer testing. Hence, stabilizer-state verification is a special case of the generalized stabilizer verification introduced below.

\textit{Verification based on generalized stabilizer testing}.---We now present a verification protocol for general multi-qudit pure states. The protocol is built entirely from generalized stabilizer testing and accepts the target state $\left|\psi\right\rangle$ with certainty.

Let $\mathcal{S}_\psi$ be the generalized stabilizer group of $\left|\psi\right\rangle$. We denote by $S_{Test}$ the subset of $\mathcal{S}_\psi$ whose elements can be implemented via generalized stabilizer testing. The $n$ generalized stabilizer generators are
$g_k = U_\psi Z_k U_\psi^\dagger$ for all $k\in\mathbb{Z}_n$.
Any element of the generalized stabilizer group can be written as
$g_{(h_0,\cdots,h_{n-1})}:=\prod_{k=0}^{n-1}g_k^{h_k}$,
with $h_k\in\mathbb{Z}_{d_k}$ for all $k\in\mathbb{Z}_n$.

\begin{theorem}[Proof provided in the Supplemental Material]
For arbitrary $n$ prime numbers $d_0,\cdots,d_{n-1}$, consider a generalized stabilizer state $\left| \psi \right\rangle \in \mathcal{H}_{d_0}\otimes\cdots\otimes\mathcal{H}_{d_{n-1}}$.
Let
$S_{Test}:=\big\{g_{(h_0,\cdots,h_{n-1})}=\prod_{k=0}^{n-1}g_{k}^{h_{k}}\big\}$ 
denote the set of generalized stabilizers that are testable. Assume that $S_{Test}$ contains at least all generalized stabilizer generators, and that it can be partitioned into $\tau$ subsets,
$S_{Test}=\bigcup_{i=0}^{\tau-1}S_{Test_i}$,
such that each subset $S_{Test_i}$ can be tested by a single measurement. Define
$C_{Test_i}=\{(h_0,\cdots,h_{n-1})\,|\,g_{(h_0,\cdots,h_{n-1})}\in S_{Test_i}\}$.
Let $\sum_{i=0}^{\tau-1}\mu_i=1$ with $\mu_i>0$ for all $i\in\mathbb{Z}_\tau$.
If
\begin{equation}
    \fontsize{8}{10}
    \lambda_{(j_0,\cdots,j_{n-1})}=\sum_{i=0}^{\tau-1}\frac{\mu_i}{|S_{Test_i}|}\bigg(\sum_{\substack{h_0,\cdots,h_{n-1}:\ \sum_{k=0}^{n-1}\frac{h_k\times j_k}{d_k}\in\mathbb{Z}\\
    (h_0,\cdots, h_{n-1})\in C_{Test_i}}}1\bigg),
\end{equation}
\begin{equation}
    \fontsize{8}{10}
    \begin{aligned}
        &\Omega_\psi=|\psi\rangle\langle\psi|+\sum_{\substack{(j_0,\cdots,j_{n-1})\\\neq(0,\cdots,0)}}\lambda_{(j_0,\cdots,j_{n-1})}U_\psi\big(\bigotimes_{k=0}^{n-1}\left| j_k \right\rangle \left\langle j_k\right|\big)U_\psi^\dagger
    \end{aligned}
    \label{eq:18}
\end{equation}
can be used to verify the state $\left|\psi\right\rangle$. The number of measurements required to achieve infidelity $\epsilon$ and significance level $\delta$ satisfies
\begin{equation}
    \fontsize{8}{10}
    \begin{aligned}
       n_{opt}^\psi\leq\bigg\lceil&\epsilon^{-1}\ln\delta^{-1}\bigg(1-\min_{\sum_{i=0}^{\tau-1}\mu_i=1}\bigg\{ \\[-3mm]&\max_{\substack{(j_0,\cdots,j_{n-1})\neq(0,\cdots,0),\\
j_k\in\mathbb{Z}_{d_k},\forall k\in\mathbb{Z}_{n}}}\big\{\lambda_{(j_0,\cdots,j_{n-1})}\}\bigg\}\bigg)^{-1}\bigg\rceil.
    \end{aligned}
    \label{eq:18}
\end{equation}
\end{theorem}
The above theorem applies to systems with prime local dimensions. For composite local dimensions, the definition of a generalized stabilizer state (see Eq.~(6)) implies that a state in a composite dimension can be represented as an equivalent prime-dimensional state. Thus any composite-dimensional quantum state can be rewritten as a prime-dimensional state, the corresponding verification measurement operator can be derived in that representation, and the verification then follows. Based on the theorem above, the verification measurement operator for $\left|\psi_1\right\rangle$ and $\left|\psi_2\right\rangle$ are
\begin{equation}
    \normalsize
    \begin{aligned}
    \Omega_{\psi_1}=&|\psi_1\rangle\langle\psi_1|+\frac{1}{2}U_{\psi_1}\big(|10\rangle\langle10|+|20\rangle\langle20| +|01\rangle\langle01|\big)U_{\psi_1}^\dagger,\\[-2mm] 
    \Omega_{\psi_2}=&\frac{1}{6}\big[3I\otimes I-\cos{2\theta}(Z\otimes I)+\sin{2\theta(X\otimes X)}\\[-2mm] 
    &+Z\otimes Z-\cos{2\theta}(I\otimes Z)-\sin{2\theta(Y\otimes Y)}\big].
    \end{aligned}
    \label{eq:12}
\end{equation}
Using the same theorem, we provide in the Supplemental Material verification procedures for several classes of quantum states, including $\left|\psi_1\right\rangle$, arbitrary two-qubit states (locally unitary equivalent to $\left|\psi_2\right\rangle$, and requiring the same number of measurements~\cite{pallister2018optimal}), qubit and qudit Bell/Bell-like states, GHZ/GHZ-like states, graph states, hypergraph states, and the recently proposed multigraph and multihypergraph states. Table~\ref{tab:measurements} summarizes the required measurement resources. Here, $n_{opt}^{\psi_2}$ is equal to the number of measurements required for Bell-state verification, and the Bell-state test operator in Ref.~\cite{pallister2018optimal} is exactly the special case of our $\Omega_{\psi_2}$ when $\theta=\frac{\pi}{4}$.  To the best of our knowledge, there are no existing verification schemes for direct comparison in the cases of multigraph and multihypergraph states or the qutrit--qubit state $\left|\psi_1\right\rangle$. In addition, we verify a generalized Bell-like state with $d=6$ to show how a composite-dimensional state can be mapped into a hybrid system of prime-dimensional subsystems. In summary, compared with the recent universal scheme of Ref.~\cite{li2025universal}, our method applies to hybrid states and reduces composite-dimensional systems to prime-dimensional hybrid systems. It is therefore more general, and the generalized-stabilizer framework is conceptually simpler and more direct than approaches based on mutually unbiased bases. 
\begin{table}[t]
  \caption{Verification cost for representative target states. $N(\varepsilon,\delta)$ is the number of state copies (rounds) required to certify infidelity $\varepsilon$ with significance level $\delta$.}
  \label{tab:compare-param}
  \begin{ruledtabular}
  \begin{tabular}{lccc}
    Target states & Ref. & $N_{\text{ref}}(\varepsilon,\delta)$ & $N_{\text{ours}}(\varepsilon,\delta)$ \\ 
    \hline
    $|\psi_1\rangle$ & -
      & -
      & $ \left\lceil2c_{\epsilon,\delta}\right\rceil$ \\[1pt]
    $|\psi_2\rangle$ & \cite{pallister2018optimal}
      & $\scriptstyle\left\lceil(2+\frac{1}{2}\sin{2\theta})c_{\epsilon,\delta}\right\rceil$
      & $\lceil\frac{3}{2}c_{\epsilon,\delta}\rceil$\\[2pt]
    $\scriptstyle\left|{GHZ}_n^d\right\rangle$ & \cite{li2020optimal}
      & $\left\lceil2c_{\epsilon,\delta}\right\rceil$
      & $\left\lceil3c_{\epsilon,\delta}\right\rceil$ \\[1pt]
      $\scriptstyle\left|\widetilde{GHZ}_n^d\right\rangle$ & \cite{li2020optimal}
      & $\left\lceil2c_{\epsilon,\delta}\right\rceil$
      & $\left\lceil3c_{\epsilon,\delta}\right\rceil$ \\[1pt]
      $\scriptstyle\left|G_1\right\rangle$,$\scriptstyle G_1\in\left\{G_n^d,\widetilde{G}_n^d\right\}$ & \cite{zhu2019efficient}
      & $\scriptstyle\left\lceil\chi_f(G_1)c_{\epsilon,\delta}\right\rceil$
      & $\scriptstyle\left\lceil \left(1+\frac{1}{d-1}\right)\chi(G_1)c_{\epsilon,\delta}\right\rceil$ \\[1pt]
      $\scriptstyle\left|G_2\right\rangle$, $\scriptstyle G_2\in\scriptstyle\left\{\widehat{G}_n^d,\widehat{\widetilde{G}}_n^d\right\}$& -
      & -
      & $\scriptstyle\left\lceil \left(1+\frac{1}{d-1}\right)\chi(G_2)c_{\epsilon,\delta}\right\rceil$ 
  \end{tabular}
  \end{ruledtabular}
  \vspace{-1pt}\footnotesize
\textbf{Notes.} $\left|\psi_1\right\rangle$, $\left|\psi_2\right\rangle$, $\scriptstyle\left|{GHZ}_n^d\right\rangle$, $\scriptstyle\left|\widetilde{GHZ}_n^d\right\rangle$, $\scriptstyle\left|{G}_n^d\right\rangle$, $\scriptstyle\left|\widetilde{G}_n^d\right\rangle$, $\scriptstyle\left|\widehat{G}_n^d\right\rangle$, $\scriptstyle\left|\widehat{\widetilde{G}}_n^d\right\rangle$, $\chi(G_1)$ and $\chi_f(G_1)$ denote, respectively,  qutrit-qubit states,  Bell-like states,  qudit GHZ states,  GHZ-like states,  graph states,  hypergraph states,  multigraph states,  multihypergraph states, chromatic number and fractional chromatic number of $G_1$. We use $c_{\epsilon,\delta}=\epsilon^{-1}\ln\delta^{-1}$, and $d$ is a prime.
  \label{tab:measurements}
\end{table}

\textit{Conclusion}.—We introduced a generalized stabilizer formalism and a corresponding verification strategy for general multi-qudit quantum states with possibly different local dimensions. By extending stabilizer testing beyond the Pauli framework, our approach verifies both prime- and composite-dimensional systems, including hybrid architectures. Using only adaptive local measurements, it certifies a broad class of states—from arbitrary two-qubit pure states and Bell-like states to GHZ, graph, hypergraph, multigraph, and multihypergraph states—with efficiencies comparable to or better than the best existing protocols. Compared with the scheme based on mutually unbiased bases, the generalized-stabilizer framework is more unified and conceptually direct, and it naturally accommodates hybrid and composite-dimensional systems. Extending this approach to additional target states, and extending  it in fully adversarial scenarios, are natural directions for future work. These results may provide a versatile and experimentally feasible path toward scalable verification of general quantum resources.


--------------------------------------------------------

\begin{acknowledgments}
This work was supported by National Natural Science Foundation of China (Grants No. 62171131), Fujian Province Natural Science Foundation (Grant No. 2022J01186 and 2023J01533), Fujian Province Young and Middle-aged Teacher Education Research Project (Grant No. JAT231018) and Open Foundation of State Key Laboratory of Networking and Switching Technology (Beijing University of Posts and Telecommunications)(SKLNST-2024-1-05).
\end{acknowledgments}


\clearpage 
\onecolumngrid
\setcounter{section}{0}
\renewcommand{\thesection}{S\arabic{section}}
\setcounter{figure}{0}
\renewcommand{\thefigure}{S\arabic{figure}}
\setcounter{table}{0}
\renewcommand{\thetable}{S\arabic{table}}
\setcounter{equation}{0}
\renewcommand{\theequation}{S\arabic{equation}}

\setcounter{page}{1}
\pagenumbering{arabic}
\begin{center}
\textbf{\large Supplemental Material for:  
Verification of General Multi-Qudit Pure States via Generalized Stabilizers}\\[4pt]
\end{center}

The Supplemental Material is organized as follows.
Appendix~A presents the proof of the proposition on the existence of generalized stabilizers.
Appendix~B provides the proof of Eq.~(6) in the main text.
Appendix~C contains the proof of the theorem stated in the main text.
Appendix~D lists the generalized stabilizers of several representative quantum states.
Appendix~E gives explicit verification examples for these states.
Finally, Appendix~F demonstrates, through a concrete example, how the verification of a composite-dimensional quantum state can be achieved by representing it as a hybrid system composed of prime-dimensional subsystems.

\TOCenable
\tableofcontents 
\section{Appendix A: Proof of the Proposition on the Existence of Generalized Stabilizers}\label{Appendix_A}

\begin{proposition}
For a multi-qudit system with prime local dimensions $d_0,\cdots,d_{n-1}$, and for any pure state $\left| \psi \right\rangle \in \mathcal{H}_{d_0}\otimes\cdots\otimes\mathcal{H}_{d_{n-1}}$, there exists a generalized stabilizer set $S_\psi$ that uniquely determines $\left| \psi \right\rangle$.
\end{proposition}

The proof follows a similar argument to that in Ref.~\cite{zhang2021efficient}.

\begin{proofS}
When $d_0, \cdots, d_{n-1}$ are prime numbers, any quantum state 
$\left| \psi \right\rangle \in \mathcal{H}_{d_0}\otimes\cdots\otimes\mathcal{H}_{d_{n-1}}$ 
can be written as 
$\left| \psi \right\rangle = U_\psi \left| 0_{d_0} \right\rangle \otimes \cdots \otimes \left| 0_{d_{n-1}} \right\rangle$, 
where $U_\psi$ is a specific unitary operator determined by $\left| \psi \right\rangle$.  
Since $\langle Z_0, \cdots ,Z_{n-1} \rangle$ is the stabilizer group of 
$\left| 0_{d_0} \right\rangle \otimes \cdots \otimes \left| 0_{d_{n-1}} \right\rangle$, 
we can derive
\begin{equation}
\mathcal{S}_\psi = \langle U_\psi Z_0^{(d_0)}U_\psi^\dagger, \cdots ,U_\psi Z_{n-1}^{(d_{n-1})}U_\psi^\dagger \rangle
\end{equation}
as the stabilizer group of $\left| \psi \right\rangle$.
Here, $Z_i$ denotes the $d_i$-dimensional Pauli-$Z$ operator acting on the $i$-th qudit,
\begin{equation}
Z^{(d_i)} = \sum_{j=0}^{d_i-1}(\omega_{d_i})^j \left| j \right\rangle \left\langle j \right|,
\end{equation}
where $\omega_{d_i}=e^{2\pi \mathbf{i}/d_i}$ and $\mathbf{i}=\sqrt{-1}$.
Identity operation $I$ is applied to all other subsystems.  
The group $\langle Z_0^{(d_0)}, \cdots ,Z_{n-1}^{(d_{n-1})} \rangle$ is generated by $\{Z_0, \cdots, Z_{n-1}\}$,  
and $\mathcal{S}_\psi$ is generated by 
$S_\psi=\{U_\psi Z_0^{(d_0)}U_\psi^\dagger, \cdots ,U_\psi Z_{n-1}^{(d_{n-1})}U_\psi^\dagger\}$. \end{proofS}

\section{Appendix B: Proof of Eq.~(5) in the main text}

In this section, we prove that the following equation holds,
\begin{equation}
    \hspace{-7mm}
    \normalsize
    \begin{aligned}
    \left| \psi \right\rangle \left\langle \psi \right|&=\frac{1}{\prod_{i=0}^{n-1}d_i}\sum_{h_0=0}^{d_0-1}\sum_{h_1=0}^{d_1-1}\cdots\sum_{h_{n-1}=0}^{d_{n-1}-1}\left(\prod_{k=0}^{n-1}g_k^{h_k}\right)=\frac{1}{\prod_{i=0}^{n-1}d_i}\prod_{i=0}^{n-1}\left(\sum_{j=0}^{d_i-1}g_i^j\right).
    \end{aligned}
    \label{eq6}
\end{equation}

\begin{proofS}
We first perform the spectral decomposition of an arbitrary element in the generalized stabilizer group, which gives
\begin{equation}
    \fontsize{8}{10}
        \begin{aligned}
           g_{(h_0,\cdots,h_{n-1})}&=\prod_{k=0}^{n-1}g_k^{h_k}=U_\psi \left(\bigotimes_{k=0}^{n-1}\left(Z_k^{(d_k)}\right)^{h_k}\right)U_\psi^\dagger=U_\psi \left[\bigotimes_{k=0}^{n-1}\left(\sum_{j_k=0}^{d_k-1}(\omega_{d_k})^{h_k\times j_k}\left| j_k \right\rangle \left\langle j_k \right|\right)\right]U_\psi^\dagger\\
           &=\sum_{j_0=0}^{d_0-1}\sum_{j_1=0}^{d_1-1}\cdots\sum_{j_{n-1}=0}^{d_{n-1}-1}\prod_{k=0}^{n-1}(\omega_{d_k})^{h_k\times j_k}U_\psi \left(\bigotimes_{k=0}^{n-1}\left| j_k \right\rangle \left\langle j_k \right|\right)U_\psi^\dagger.
        \end{aligned}
\end{equation}
The eigenvalues of $g_{(h_0,\cdots,h_{n-1})}$ can then be expressed as $\lambda_{\{j_0,\cdots,j_{n-1}\}}=\prod_{k=0}^{n-1}\omega_k^{h_k\times j_k}=e^{2\pi \mathbf{i}\sum_{k=0}^{n-1}(h_k\times j_k/d_k)}$, and the corresponding eigenspace is the subspace spanned by $V_{\lambda_{\{j_0,\cdots,j_{n-1}\}}}=\bigg\{U_\psi\bigotimes_{k=0}^{n-1}\left| i_{k} \right\rangle |i_k\in \mathbb{Z}_{d_k}\text{, for }k\in\mathbb{Z}_{n},\text{ and }\sum_{k=0}^{n-1}(h_k\times j_k/d_k)=\sum_{k=0}^{n-1}(h_k\times i_k/d_k)\bigg\}$.

Next, we have
\begin{equation}
    \fontsize{8}{10}
        \begin{aligned}
          &\frac{1}{\prod_{i=0}^{n-1}d_i}\prod_{i=0}^{n-1}\left(\sum_{j=0}^{d_i-1}g_i^j\right)=\frac{1}{\prod_{i=0}^{n-1}d_i}\sum_{h_0=0}^{d_0-1}\sum_{h_1=0}^{d_1-1}\cdots\sum_{h_{n-1}=0}^{d_{n-1}-1}\left(\prod_{k=0}^{n-1}g_k^{h_k}\right)\\
           =&\frac{1}{\prod_{i=0}^{n-1}d_i}\sum_{h_0=0}^{d_0-1}\sum_{h_1=0}^{d_1-1}\cdots\sum_{h_{n-1}=0}^{d_{n-1}-1}\left(U_\psi \left[\bigotimes_{k=0}^{n-1}\left(\sum_{j_k=0}^{d_k-1}(\omega_{d_k})^{h_k\times j_k}\left| j_k \right\rangle \left\langle j_k \right|\right)\right]U_\psi^\dagger\right)\\
           =&\frac{1}{\prod_{i=0}^{n-1}d_i}\sum_{h_0=0}^{d_0-1}\sum_{h_1=0}^{d_1-1}\cdots\sum_{h_{n-1}=0}^{d_{n-1}-1}\left(U_\psi\left(\bigotimes_{k=0}^{n-1}\left| 0_{d_k} \right\rangle \left\langle 0_{d_k}\right|\right)U_\psi^\dagger+ \sum_{(j_0,j_1,\cdots,j_{n-1})\neq (0,\cdots,0)}\left[ \prod_{k=0}^{n-1}(\omega_{d_k})^{h_k\times j_k}U_\psi\left(\bigotimes_{k=0}^{n-1}\left| j_k \right\rangle \left\langle j_k \right|\right)U_\psi^\dagger\right]\right)\\
           =&|\psi\rangle\langle\psi|+ \left[\sum_{(j_0,j_1,\cdots,j_{n-1})\neq (0,\cdots,0)}\left(\sum_{h_0=0}^{d_0-1}\sum_{h_1=0}^{d_1-1}\cdots\sum_{h_{n-1}=0}^{d_{n-1}-1}\prod_{k=0}^{n-1}(\omega_{d_k})^{h_k\times j_k}\right)U_\psi\left(\bigotimes_{k=0}^{n-1}\left| j_k \right\rangle \left\langle j_k \right|\right)U_\psi^\dagger\right]=|\psi\rangle\langle\psi|,
        \end{aligned}
\end{equation}
where
$
\sum_{h_0=0}^{d_0-1}\sum_{h_1=0}^{d_1-1}\cdots\sum_{h_{n-1}=0}^{d_{n-1}-1}\prod_{k=0}^{n-1}(\omega_{d_k})^{h_k\times j_k}=0
$ for $(j_0,j_1,\cdots,j_{n-1})\neq (0,\cdots,0)$.
Therefore, Eq.~(5) in the main text holds as stated.
\end{proofS}

\section{Appendix C: Proof of the Theorem in the Main Text}

In this section, we present the proof of the theorem given in the main text.  
Before proceeding with the proof, we first introduce the following lemma.

\begin{lemma}
\label{lemma}
For arbitrary $n$ prime numbers $d_0, \cdots, d_{n-1}$, any pure state $\left| \psi \right\rangle=U_\psi\left| 0_{d_0} \right\rangle\otimes\cdots\otimes\left| 0_{d_{n-1}} \right\rangle\in\mathcal{H}_{d_0}\otimes\cdots\otimes\mathcal{H}_{d_{n-1}}$, the $n$ generalized stabilizer generators of the state are given by $g_i=U_\psi Z_kU_\psi^\dagger$, for all $k\in\mathbb{Z}_n$.  
Let an element of the generalized stabilizer group be $g_{(h_0,\cdots,h_{n-1})}:=\prod_{k=0}^{n-1}g_k^{h_k}$, where $h_k\in\mathbb{Z}_{d_k}$ for all $k\in\mathbb{Z}_n$, and $(h_0,\cdots,h_{n-1})\neq (0,\cdots,0)$ and $\mathbf{lcm}(d_0,\cdots,d_{n-1})$ denotes the least common multiple of $d_0,\cdots,d_{n-1}$.  
Then, the quantum operator
\begin{equation}
        \begin{aligned}
           P_{(h_0,\cdots,h_{n-1})}=\frac{\sum_{l=0}^{\mathbf{lcm}(d_0,\cdots,d_{n-1})-1}\left(\prod_{k=0}^{n-1}g_k^{h_k}\right)^l}{\mathbf{lcm}(d_0,\cdots,d_{n-1})}
        \end{aligned}
        \label{eq:15-1}
\end{equation}
is the projector onto the eigenspace of $g_{(h_0,\cdots,h_{n-1})}$ corresponding to the eigenvalue $1$.
\end{lemma}

\begin{proofS}
By substituting the spectral decomposition of $\prod_{k=0}^{n-1}g_k^{h_k}$ into $P_{(h_0,\cdots,h_{n-1})}$, we obtain
\begin{equation}
        \fontsize{8}{10}
        \begin{aligned}
             P_{(h_0,\cdots,h_{n-1})}&=(\mathbf{lcm}(d_0,\cdots,d_{n-1}))^{-1}\sum_{l=0}^{\mathbf{lcm}(d_0,\cdots,d_{n-1})-1}\left[ \sum_{j_0=0}^{d_0-1}\sum_{j_1=0}^{d_1-1}\cdots\sum_{j_{n-1}=0}^{d_{n-1}-1}\left[\left(\prod_{k=0}^{n-1}(\omega_{d_k})^{h_k\times j_k\times l}\right)U_\psi\left(\bigotimes_{k=0}^{n-1}\left| j_k \right\rangle \left\langle j_k\right|\right)U_\psi^\dagger\right]\right]\\
            &=(\mathbf{lcm}(d_0,\cdots,d_{n-1}))^{-1}\left[ \sum_{j_0=0}^{d_0-1}\sum_{j_1=0}^{d_1-1}\cdots\sum_{j_{n-1}=0}^{d_{n-1}-1}\left[\left(\sum_{l=0}^{\mathbf{lcm}(d_0,\cdots,d_{n-1})-1}\left(e^{2\pi\mathbf{i}\sum_{k=0}^{n-1}\frac{h_k\times j_k}{d_k}}\right)^l\right)U_\psi\left(\bigotimes_{k=0}^{n-1}\left| j_k \right\rangle \left\langle j_k\right|\right)U_\psi^\dagger\right]\right]\\
            &=(\mathbf{lcm}(d_0,\cdots,d_{n-1}))^{-1}\left[\mathbf{lcm}(d_0,\cdots,d_{n-1})\left(\sum_{j_0,\cdots,j_{n-1}:\ \sum_{k=0}^{n-1}\frac{h_k\times j_k}{d_k}\in\mathbb{Z}}\left(U_\psi\left(\bigotimes_{k=0}^{n-1}\left| j_k \right\rangle \left\langle j_k\right|\right)U_\psi^\dagger\right)\right)\right]\\
            &=\sum_{j_0,\cdots,j_{n-1}:\ \sum_{k=0}^{n-1}\frac{h_k\times j_k}{d_k}\in\mathbb{Z}}\left(U_\psi\left(\bigotimes_{k=0}^{n-1}\left| j_k \right\rangle \left\langle j_k\right|\right)U_\psi^\dagger\right).
        \end{aligned}
        \label{eq:15-2}
\end{equation}
It can then be verified that $P_{(h_0,\cdots,h_{n-1})}$ is the projection operator onto the eigenspace of $g_{(h_0,\cdots,h_{n-1})}$ with eigenvalue $1$. Furthermore, let $\{p_0,\cdots,p_{n'-1}\}$ be the set of distinct numbers appearing in $\{d_0,\cdots,d_{n-1}\}$, where $n'\leq n$.  
Then, the condition $\sum_{k=0}^{n-1}\frac{h_k\times j_k}{d_k}\in\mathbb{Z}$, for $j_k\in d_k$, represents the complete set of solutions to the system of congruence equations 
\begin{equation}
\normalsize
\big\{\sum_{k:d_k=p_i}h_kj_k\equiv0\ (\text{mod } p_i)\,\big|\,\text{for all }i\in\mathbb{Z}_{n'}\big\},
\end{equation}
where $h_0,\cdots,h_{n-1}$ are treated as constants and $j_0,\cdots,j_{n-1}$ as unknown variables.
\end{proofS} 

We now use the conclusion of the above lemma to prove the theorem stated in the main text.
\begin{theorem}
For arbitrary $n$ prime numbers $d_0,\cdots,d_{n-1}$, consider a generalized stabilizer state $\left| \psi \right\rangle \in \mathcal{H}_{d_0}\otimes\cdots\otimes\mathcal{H}_{d_{n-1}}$.
Let
$S_{Test}:=\big\{g_{(h_0,\cdots,h_{n-1})}=\prod_{k=0}^{n-1}g_{k}^{h_{k}}\big\}$ 
denote the set of generalized stabilizers that are testable. Assume that $S_{Test}$ contains at least all generalized stabilizer generators, and that it can be partitioned into $\tau$ subsets,
$S_{Test}=\bigcup_{i=0}^{\tau-1}S_{Test_i}$,
such that each subset $S_{Test_i}$ can be tested by a single measurement. Define
$C_{Test_i}=\{(h_0,\cdots,h_{n-1})\,|\,g_{(h_0,\cdots,h_{n-1})}\in S_{Test_i}\}$.
Let $\sum_{i=0}^{\tau-1}\mu_i=1$ with $\mu_i>0$ for all $i\in\mathbb{Z}_\tau$.
If
\begin{equation}
    \normalsize
    \lambda_{(j_0,\cdots,j_{n-1})}=\sum_{i=0}^{\tau-1}\frac{\mu_i}{|S_{Test_i}|}\bigg(\sum_{\substack{h_0,\cdots,h_{n-1}:\ \sum_{k=0}^{n-1}\frac{h_k\times j_k}{d_k}\in\mathbb{Z}\\
    (h_0,\cdots, h_{n-1})\in C_{Test_i}}}1\bigg),
\end{equation}
\begin{equation}
    \normalsize
    \begin{aligned}
        &\Omega_\psi=|\psi\rangle\langle\psi|+\sum_{\substack{(j_0,\cdots,j_{n-1})\\\neq(0,\cdots,0)}}\lambda_{(j_0,\cdots,j_{n-1})}U_\psi\big(\bigotimes_{k=0}^{n-1}\left| j_k \right\rangle \left\langle j_k\right|\big)U_\psi^\dagger
    \end{aligned}
    \label{eq:18}
\end{equation}
can be used to verify the state $|\psi\rangle$. 
The number of measurements required to achieve infidelity $\epsilon$ and confidence level $1-\delta$ satisfies
\begin{equation}
    \normalsize
    \begin{aligned}
n_{opt}^\psi\leq\bigg\lceil&\epsilon^{-1}\ln\delta^{-1}\bigg(1-\min_{\sum_{i=0}^{\tau-1}\mu_i=1}\bigg\{ \max_{\substack{(j_0,\cdots,j_{n-1})\neq(0,\cdots,0),\\
j_k\in\mathbb{Z}_{d_k},\forall k\in\mathbb{Z}_{n}}}\big\{\lambda_{(j_0,\cdots,j_{n-1})}\}\bigg\}\bigg)^{-1}\bigg\rceil.
    \end{aligned}
    \label{eq:18}
\end{equation}
\end{theorem}

\begin{proofS}
From the definition of generalized stabilizers, the generalized stabilizer group of the state $\left| \psi \right\rangle$ contains $\left(\prod_{i=0}^{n-1}d_i\right)-1$ nontrivial and mutually independent operators $g_{(h_0,\cdots,h_{n-1})}$ (excluding $g_{(0,\cdots,0)}$).  
In general, among these $\left(\prod_{i=0}^{n-1}d_i\right)-1$ nontrivial elements, some may not be experimentally testable via generalized stabilizer testing.  
Therefore, the measurement operator must be constructed only from those elements that can be tested.  

Let the set of testable elements be denoted as
$
S_{Test}=\{g_{(h_{0},\cdots,h_{n-1})}\,|\,(h_0,\cdots ,h_{n-1})\in C_{Test}\},
$
where $C_{Test}$ is the set of all encoded labels corresponding to generalized stabilizers that can be tested.  
Obviously, $(0,\cdots,0)\notin C_{Test}$.  The set $S_{Test}$ must include all generalized stabilizer generators.
This requirement arises because a quantum state is fully determined by its generalized stabilizer generators.
If any generator is omitted, verification errors may occur.
This point will be illustrated later in the verification of GHZ and GHZ-like states. 
Then, we can construct the measurement operator as
\begin{equation}
    \normalsize
    \begin{aligned}
    \Omega_{\psi}=&\sum_{i=0}^{\tau-1}\frac{\mu_i}{|S_{Test_i}|}\left(\sum_{(h_0,\cdots h_{n-1})\in C_{Test_i}}P_{(h_0,\cdots,h_{n-1})}\right)\\
    =&\sum_{i=0}^{\tau-1}\frac{\mu_i}{|S_{Test_i}|}\left[\sum_{(h_0,\cdots h_{n-1})\in C_{Test_i}}\sum_{j_0,\cdots,j_{n-1}:\ \sum_{k=0}^{n-1}\frac{h_k\times j_k}{d_k}\in\mathbb{Z}}\left(U_\psi\left(\bigotimes_{k=0}^{n-1}\left| j_k \right\rangle \left\langle j_k\right|\right)U_\psi^\dagger\right)\right]\\
    =&\sum_{i=0}^{\tau-1}\frac{\mu_i}{|S_{Test_i}|}\left[\sum_{(h_0,\cdots h_{n-1})\in C_{Test_i}}\left(U_\psi\left(\bigotimes_{k=0}^{n-1}\left| 0_{d_k} \right\rangle \left\langle 0_{d_k}\right|\right)U_\psi^\dagger\right)\right.\\ &\left.+\sum_{(h_0,\cdots h_{n-1})\in C_{Test_i}}\left[\sum_{\substack{j_0,\cdots,j_{n-1}:\ \sum_{k=0}^{n-1}\frac{h_k\times j_k}{d_k}\in\mathbb{Z}\\
    (j_0,\cdots,j_{n-1})\neq(0,\cdots,0)}}\left(U_\psi\left(\bigotimes_{k=0}^{n-1}\left| j_k \right\rangle \left\langle j_k\right|\right)U_\psi^\dagger\right)\right]\right]\\
    =&\left(\sum_{i=0}^{\tau-1}\mu_i\right)|\psi\rangle\langle\psi|+\sum_{(j_0,\cdots,j_{n-1})\neq(0,\cdots,0)}\left(\sum_{i=0}^{\tau-1}\frac{\mu_i}{|S_{Test_i}|}\left(\sum_{\substack{h_0,\cdots,h_{n-1}:\ \sum_{k=0}^{n-1}\frac{h_k\times j_k}{d_k}\in\mathbb{Z}\\
    (h_0,\cdots, h_{n-1})\in C_{Test_i}}}1\right)\right)U_\psi\left(\bigotimes_{k=0}^{n-1}\left| j_k \right\rangle \left\langle j_k\right|\right)U_\psi^\dagger\\
    =&|\psi\rangle\langle\psi|+\sum_{(j_0,\cdots,j_{n-1})\neq(0,\cdots,0)}\lambda_{(j_0,\cdots,j_{n-1})}U_\psi\left(\bigotimes_{k=0}^{n-1}\left| j_k \right\rangle \left\langle j_k\right|\right)U_\psi^\dagger,
    \end{aligned}
    \label{eq:18-+}
\end{equation}
where
\begin{equation}
 \normalsize
    \lambda_{(j_0,\cdots,j_{n-1})}=\sum_{i=0}^{\tau-1}\frac{\mu_i}{|S_{Test_i}|}\left(\sum_{\substack{h_0,\cdots,h_{n-1}:\ \sum_{k=0}^{n-1}\frac{h_k\times j_k}{d_k}\in\mathbb{Z}\\
    (h_0,\cdots, h_{n-1})\in C_{Test_i}}}1\right)
\end{equation}
represents the eigenvalue corresponding to the eigenstate $U_\psi\bigotimes_{k=0}^{n-1}|j_k\rangle$.  

The smallest value of the second-largest eigenvalue of $\Omega_\psi$ is given by
\begin{equation}
    \hspace{-5mm}
    \normalsize
    \begin{aligned}
        \min_{\sum_{i=0}^{\tau-1}\mu_i=1}\{\beta(\Omega_\psi)\}&=\min_{\sum_{i=0}^{\tau-1}\mu_i=1}\left\{\max_{\substack{(j_0,\cdots,j_{n-1})\neq(0,\cdots,0)\\
        j_k\in\mathbb{Z}_{d_k},\forall k\in\mathbb{Z}_{n}}}\{\lambda_{(j_0,\cdots,j_{n-1})}\}\right\}\\
        &=\min_{\sum_{i=0}^{\tau-1}\mu_i=1}\left\{\max_{\substack{(j_0,\cdots,j_{n-1})\neq(0,\cdots,0)\\
        j_k\in\mathbb{Z}_{d_k},\forall k\in\mathbb{Z}_{n}}}\left\{\sum_{i=0}^{\tau-1}\frac{\mu_i}{|S_{Test_i}|}\left(\sum_{\substack{h_0,\cdots,h_{n-1}:\ \sum_{k=0}^{n-1}\frac{h_k\times j_k}{d_k}\in\mathbb{Z}\\
    (h_0,\cdots, h_{n-1})\in C_{Test_i}}}1\right)\right\}\right\}.
    \end{aligned}
\end{equation}
By applying the constraints $\sum_{i=0}^{\tau-1}\mu_i=1$ and $0<\mu_i<1$, one can minimize $\beta(\Omega_\psi)$ to obtain
$\min_{\mu_0,\cdots,\mu_{t-1}}[\beta(\Omega_\psi)]$, and then compute the spectral gap $\nu(\Omega_\psi)=1-\beta(\Omega_\psi)$.  

If the set $S_{Test}$ contains at least all the generalized stabilizer generators of $\left|\psi\right\rangle$,  
then according to Eq.~(2) in the main text,  
the number of measurements required for verifying the quantum state is
\begin{equation}
\normalsize
    \begin{aligned}
        n_{opt}^\psi=\left\lceil \frac{\ln\delta}{\ln[1-\nu(\Omega_\psi)\epsilon]}\right\rceil\leq\left\lceil \frac{\ln\delta^{-1}}{\nu(\Omega_\psi)\epsilon}\right\rceil=\left\lceil \epsilon^{-1}\ln\delta^{-1}\left(1-\min_{\sum_{i=0}^{\tau-1}\mu_i=1}\left\{\max_{\substack{(j_0,\cdots,j_{n-1})\neq(0,\cdots,0)\\
        j_k\in\mathbb{Z}_{d_k},\forall k\in\mathbb{Z}_{n}}}\{\lambda_{(j_0,\cdots,j_{n-1})}\}\right\}\right)^{-1}\right\rceil.
    \end{aligned}
    \label{eq:18}
\end{equation}
\end{proofS}

\section{Appendix D: Generalized Stabilizers for Several Representative Quantum States}

In this section, we present the generalized stabilizers for a series of representative quantum states,  including qutrit-qubit states, Bell-like states, qubit and qudit GHZ/GHZ-like states, graph states, hypergraph states, and the recently proposed multigraph and multihypergraph states, denoted respectively as $\left|\psi_1\right\rangle=\frac{1}{\sqrt{6}}[(|0_3\rangle+|1_3\rangle+|2_3\rangle)\otimes|0_2\rangle+(|0_3\rangle+\omega_3|1_3\rangle+\omega_3^2|2_3\rangle)\otimes|1_2\rangle]$, $\left|\psi_2\right\rangle=\sin{\theta}\left|00\right\rangle+\cos{\theta}\left|11\right\rangle$, $\scriptstyle\left|{GHZ}_n^d\right\rangle$, $\scriptstyle\left|\widetilde{GHZ}_n^d\right\rangle$, $\scriptstyle\left|{G}_n^d\right\rangle$, $\scriptstyle\left|\widetilde{G}_n^d\right\rangle$, $\scriptstyle\left|\widehat{G}_n^d\right\rangle$ and $\scriptstyle\left|\widehat{\widetilde{G}}_n^d\right\rangle$ . Here $d$ is a prime number.  
The last four graph-related quantum states require a brief introduction to graph-theoretic notation.  

An $n$-vertex graph is described as $G=(V,E)$, where $V=\mathbb{Z}_n$ denotes the vertex set (corresponding to the $n$ particles in the quantum state) and $E$ denotes the edge set.  
From Ref.~\cite{zhang2024quantum}, the edges (hyperedges) in a simple graph $G=(V,E)$ (hypergraph $\widetilde{G}=(V,\widetilde{E})$) and in a multigraph $\widehat{G}=(V,\widehat{E})$ (multihypergraph $\widehat{\widetilde{G}}=(V,\widehat{\widetilde{E}})$) are respectively represented as  
$e=\{v_0,\cdots,v_{t-1}\}$ and $\dot{e}=\left(V_{\dot{e}}\vert S_{\dot{e}}\right)=\left(v_0,\cdots,v_{t-1}\vert s_{v_0},\cdots,s_{v_{t-1}}\right)$,  
where $t=2$ for edges and $t>2$ for hyperedges.  
The set $V_{\dot{e}}=\{v_0,\cdots,v_{t-1}\}$ represents all vertices connected by $\dot{e}$, and different combinations of the parameters $s_{v_0},\cdots,s_{v_{t-1}}$ correspond to distinct $\dot{e}$ sharing the same vertex set $\{v_0,v_1,\cdots,v_{t-1}\}$.  
Here, $\dot{e}$ is not merely a set of vertices, and it contains two subsets $V_{\dot{e}}=\{v_0,\cdots,v_{t-1}\}$ and $S_{\dot{e}}=\{s_{v_0},s_{v_1},\cdots,s_{v_{t-1}}\}$, and therefore differs from the edges of simple graphs or hypergraphs.  
Thus, $e$ and $\dot{e}$ are used to distinguish the edges in qudit graph (hypergraph) states and in the multigraph (multihypergraph) states, respectively~\cite{zhang2024quantum}.  
Further details can be found in Ref.~\cite{zhang2024quantum}.  

In addition, the expressions  
`$e\in E,\, k\in e$', `$e\in \widetilde E,\, k\in e$', `$\dot{e}\in\widehat E,\, k\in V_{\dot{e}}$', and `$\dot{e}\in\widehat{\widetilde E},\,k\in V_{\dot{e}}$'  
respectively denote all edges or hyperedges connected to vertex $k$ in the graph, hypergraph, multigraph, and multihypergraph states.  
The notations $e\backslash\{k\}$ and $V_{\dot{e}}\backslash\{k\}$ represent the set of neighboring vertices of $k$ connected by an edge $e$ (or a hyperedge $\dot{e}$), where an edge connects one neighboring vertex and a hyperedge connects multiple neighboring vertices.  

We now introduce the unitary operations required for the preparation of these quantum states, which are also used to construct their generalized stabilizers.
\begin{align}
\normalsize
   U_{\psi_1}&=\mathcal{CZ}_{(0\leftarrow1)}^{(3\otimes2)}({QFT}^{(3)}\otimes H)
    ,& U_{\psi_2}&=CNOT\cdot(R_y(\pi-2\theta)\otimes I),\\
    U_{GHZ_n^2}&=\left(\prod_{j=n-1}^1 CNOT_{0,j}\right)H_0,
    & U_{\widetilde{GHZ}_n^2}&=\left(\prod_{j=n-1}^1CNOT_{0,j}\right)
    Ry(\pi-2\theta)_0,\\
    U_{GHZ_n^d}&=\left(\prod_{j=n-1}^1 CSUM_{0,j}\right)QFT_0, &
    U_{\widetilde{GHZ}_n^d}&=\left(\prod_{j=n-1}^1 CSUM_{0,j}\right)\prod_{i=0}^{d-2}R_{i,i+1}(\theta_i)_0,\\
    U_{G_n^d}&=\left(\prod_{e\in E}CZ_e^{m_e}\right)QFT^{\otimes n},& 
    U_{\widetilde{G}_n^d}&=\left(\prod_{e\in E}\widetilde{CZ}_e^{m_e}\right)QFT^{\otimes n},\\
    U_{\widehat{G}_n^d}&=\left(\prod_{e\in E}\widehat{CZ}_{\dot{e}}^{m_{\dot{e}}}\right)QFT^{\otimes n},&
    U_{\widehat{\widetilde{G}}_n^d}&=\left(\prod_{e\in E}\widehat{\widetilde{CZ}}_{\dot{e}}^{m_{\dot{e}}}\right)QFT^{\otimes n},
\end{align}
where $m_e\in\mathbb{Z}_d$ represents both the weight of an edge and the number of operations performed.  
The gates $H$, $QFT$, and $Ry(\pi-2\theta)$ denote the Hadamard gate, the quantum Fourier gate, and the $y$-axis rotation, respectively.  
The operator  
$R_{i,i+1}(\theta_i)=I_d+(\cos{\theta_i}-1)(\left|i\right\rangle \left\langle i\right|+\left|i+1\right\rangle \left\langle i+1\right|)+\sin{\theta_i}(\left|i+1\right\rangle \left\langle i\right|-\left|i\right\rangle \left\langle i+1\right|)$  
is the Givens rotation.  
It satisfies $\prod_{i=0}^{d-2}R_{i,i+1}(\theta_i)\left|0_d\right\rangle=\sum_{j=0}^{d-1}\lambda_j\left|j\right\rangle$ with $\sum_{j=0}^{d-1}\lambda_j^2=1$.  
The subscript “0” on the right indicates that the operation is performed on the first qudit.  
The gates $CNOT_{0,j}$ and $CSUM_{0,j}$ denote the qubit controlled-$X$ gate and the qudit controlled-sum gate, respectively.  

Furthermore,
\begin{align}
\normalsize
    CZ_{\{j,k\}}&= \sum_{i_0,\cdots, i_{n-1} = 0}^{d-1}\omega_d^{i_j\times i_k} \left| i_0,\cdots, i_{n-1}\right\rangle \left\langle i_0,\cdots,i_{n-1} \right|,\\
    {\widetilde{CZ}}_{\{v_0,\cdots,v_{t-1}\}}&=\sum_{i_0,\cdots,i_{n-1}=0}^{d-1} \omega_d^{\prod_{j=0}^{t-1} i_{v_j}}\left| i_0,\cdots,i_{n-1}\right\rangle \left\langle i_0,\cdots,i_{n-1}\right| ,\\
    {\widehat{CZ}}_{\left(j,k|s_j,s_k\right)}&=\sum_{i_0,\cdots,i_{n-1}=0}^{d-1}\omega_d^{{(i_j)}^{s_j}\times{(i_k)}^{s_k}} \left|i_0,\cdots,i_{n-1}\right\rangle \left\langle i_0,{\cdots,i}_{n-1}\right|,\\{\widehat{\widetilde{CZ}}}_{(v_0,\cdots,v_{t-1},|s_{v_0}\cdots,s_{v_{t-1}})}&=\sum_{i_0,\cdots,i_{n-1}=0}^{d-1} \omega_d^{\prod_{j=0}^{t-1}(i_{v_j})^{s_{v_j}}}\left|i_0,\cdots,i_{n-1}\right\rangle \left\langle i_0,\cdots,i_{n-1}\right|.
    \end{align}

When $d$ is a prime number, according to the Proposition on the existence of generalized stabilizers, the stabilizer generators of this family of quantum states can be obtained from $g_i=UZ_iU^\dagger$ as
 \begin{align}
 \normalsize
     g_k^{(\psi_1)}&=\left\{
	\begin{array}{cc}
		 X^{(3)}\otimes\left|0_2\right\rangle\left\langle 0_2\right| +\omega_3^{-1}X^{(3)}\otimes\left|1_2\right\rangle\left\langle 1_2\right|,&k=0,\\
		Z^{(3)}\otimes\left|1_2\right\rangle\left\langle 0_2\right|+\left(Z^{(3)}\right)^{-1}\otimes\left|0_2\right\rangle\left\langle 1_2\right|,&k=1,
	\end{array}
    \right.\\
    g_k^{(\psi_2)}&=\left\{
	\begin{array}{cc}
		-\cos{2\theta}(Z\otimes I)+\sin{2\theta(X\otimes X)},&k=0\\
		Z\otimes Z,&k=1
	\end{array}
    \right.\\
     g_k^{(GHZ_n^2)}&=\left\{
	\begin{array}{cc}
		X^{\otimes n},&k=0,\\
		Z_0Z_k,&k=1,\cdots,n-1,
	\end{array}
    \right.\\
    g_k^{(\widetilde{GHZ}_n^2)}&=\left\{
	\begin{array}{cc}
		-\cos{(2\theta)}Z\otimes I^{\otimes(n-1)} +\sin{(2\theta)}X^{\otimes n},&k=0,\\
		Z_0Z_k,&k=1,\cdots,n-1,
	\end{array}
    \right.\\
     g_k^{(GHZ_n^d)}&=\left\{
	\begin{array}{cc}
		X^{\otimes n},&k=0,\\
		Z_0^{-1}Z_k,&k=1,\cdots,n-1,
	\end{array}
    \right.\\
    g_k^{(\widetilde{GHZ}_n^d)}&=\left\{
	\begin{array}{cc}
		\sum_{a,b}^{d-1}B_{a,b}|a\rangle\langle b|\otimes \left(X^{a-b}\right)^{\otimes (n-1)},&k=0,\\
		Z_0^{-1}Z_k,&k=1,\cdots,n-1,
	\end{array}
    \right.\\
    g_k^{(G_n^d)}&=\left( \prod_{e\in E} {CZ}_e^{m_e}\right) X_k\left(\prod_{e' \in  E}{CZ}_{e'}^{{d-m}_{e'}}\right)= X_k\prod_{e \in E,k\in e}{CZ}_{e \backslash \left\{k\right\}}^{m_e}=X_k\prod_{e \in E,k\in e}{Z}_{e \backslash \left\{k\right\}}^{m_e},\\
    g_k^{(\widetilde{G}_n^d)}&=\left( \prod_{e\in\widetilde E} {{\widetilde{CZ}}_e}^{m_e}\right) X_k\left(\prod_{e' \in \widetilde E}{{\widetilde{CZ}}_{e'}}^{{d-m}_{e'}}\right)=X_k\prod_{e \in\widetilde E,k\in e}{{\widetilde{CZ}}_{e \backslash \left\{k\right\}}}^{m_e},\\
    g_k^{(\widehat{G}_n^d)}&=\left(\prod_{{\dot{e}}\in\widehat E} {{\widehat{CZ}}_{\dot{e}}}^{m_{\dot{e}}}\right) X_k\left(\prod_{{\dot{e}}' \in \widehat E}{{\widehat{CZ}}_{{\dot{e}}'}}^{{d-m}_{{\dot{e}}'}}\right)\\
    &=X_k\prod_{{\dot{e}} \in\widehat E,k\in V_{\dot{e}}}\bigg[\sum_{j_k=0}^{d-1}\left|j_k\right\rangle\left\langle j_k\right|\otimes {\left({^{s_{\{V_{\dot{e}}\backslash{k}\}}}Z}_{\{V_{\dot{e}}\backslash{k}\}}\right)}^{m_{\dot{e}}\times\big(\sum_{l=0}^{s_{k}-1} \binom{s_{k}}{l}\cdot{(j_k)}^l\big)}\bigg],\\
    g_k^{(\widehat{\widetilde{G}}_n^d)}&=\left(\prod_{{\dot{e}}\in \widehat{\widetilde E}} {{\widehat{\widetilde{CZ}}}_{\dot{e}}}^{m_{\dot{e}}}\right) X_k\left(\prod_{{\dot{e}}' \in \widehat{\widetilde E}}{{\widehat{\widetilde{CZ}}}_{{\dot{e}}'}}^{{d-m}_{{\dot{e}}'}}\right)\\
    &=X_k\prod_{{\dot{e}}\in \widehat{\widetilde E},k\in V_{\dot{e}}}\bigg[\sum_{j_k=0}^{d-1}\left|j_k\right\rangle\left\langle j_k\right|\otimes {{\widehat{\widetilde{CZ}}}_{(V_{\dot{e}}\backslash \{k\}\vert S_{\dot{e}}\backslash \{s_k\})}}^{m_{\dot{e}}\times\big(\sum_{l=0}^{s_{k}-1} \binom{s_{k}}{l}\cdot{(j_k)}^l\big)}\bigg],
\end{align}
where $B_{a,b}$ are the coefficients of  
$B=\left[\prod_{i=0}^{d-2}R_{i,i+1}(\theta_i)\right]Z\left[\prod_{i=0}^{d-2}R_{i,i+1}(\theta_i)\right]^\dagger=\sum_{a,b=0}^{d-1}B_{a,b}|a\rangle\langle b|$  
in the computational basis.  
Here, $\binom{S_k}{l}$ is the binomial coefficient “$S_k$ choose $l$,” and $\sum_{i=0}^{d-1}\lambda_i=1$.  
In addition,  
$^{s_j}Z:=\sum_{k=0}^{d-1}{\omega_d^{k^{s_j}}\left| k \right\rangle \left\langle k \right|}$.  

Consequently, the sets of generalized stabilizer generators for these quantum states can be expressed as  
$S^{(*)}=\{g_0^{(*)},\cdots,g_{n-1}^{(*)}\}$,  
where the superscript $*$ may represent any of the states  
$GHZ_n^d$, $\widetilde{GHZ}_n^d$, $G_n^d$, $\widetilde{G}_n^d$, $\widehat{G}_n^d$, or $\widehat{\widetilde{G}}_n^d$.  
Once the generators $g_i$ of the generalized stabilizer group are obtained, any element within the group can be expressed as  
$g_{(h_0,\cdots,h_{n-1})}=\prod_{i=0}^{n-1}(g_i)^{h_i}$ with $h_i\in \mathbb{Z}_{d_i}$.
\begin{align}
\normalsize
     g_{\{h_0,h_1\}}^{(\psi_1)}
     =&(1-h_1)\left(X^{(3)}\right)^{h_0}\otimes\left(\left|0_2\right\rangle\left\langle 0_2\right|+\omega_3^{h_0}\left|1_2\right\rangle\left\langle 1_2\right|\right)\\
     &+h_1\left[\left(X^{(3)}\right)^{h_0}\left(Z^{(3)}\right)^{-1}\otimes\left|0_2\right\rangle\left\langle 1_2\right|+\omega_3^{h_0}\left(X^{(3)}\right)^{h_0}Z^{(3)}\otimes\left|1_2\right\rangle\left\langle 0_2\right|\right],\\
    g_{\{h_0,h_1\}}^{(\psi_2)}
    =&\left\{
	\begin{array}{cc}
		-\cos{2\theta}Z^{1+h_1}\otimes Z^{h_1}+2\sin{2\theta}XZ^{h_1}\otimes XZ^{h_1},&h_0=1,\\
		Z^{h_1}\otimes Z^{h_1},&h_0=0,
	\end{array}
    \right.\\
    g_{(h_0,\cdots,h_{n-1})}^{(GHZ_n^2)}
    =&X_0^{h_0} Z_0^{\sum_{k=1}^{n-1} h_k}\prod_{j=1}^{n-1}(X_j^{h_0}Z_j^{h_j}),\\
    g_{(h_0,\cdots,h_{n-1})}^{(\widetilde{GHZ}_n^2)}
    =&\left\{
	\begin{array}{cc}
		 \sin{2\theta}X_0Z_0^{\sum_{k=1}^{n-1}h_k}\prod_{j=1}^{n-1}X_j^{h_0}Z_j^{h_j}-\cos{2\theta}Z_0^{\sum_{k=0}^{n-1}h_k}\prod_{j=1}^{n-1}Z_j^{h_j},&h_0=1,\\
		Z_0^{\sum_{k=1}^{n-1}h_k}\prod_{j=1}^{n-1}Z_j^{h_j},&h_0=0,
	\end{array}
    \right.\\
    g_{(h_0,\cdots,h_{n-1})}^{(GHZ_n^d)}
    =&X_0^{h_0}Z_0^{-\sum_{k=1}^{n-1}h_k}\prod_{j=1}^{n-1}(X_j^{h_0}Z_j^{h_j}),\\
    g_{(h_0,\cdots,h_{n-1})}^{(\widetilde{GHZ}_n^d)}
    =&\sum_{a,b}^{d-1}(B^{h_0})_{a,b}(|a\rangle\langle b|)_0Z_0^{-\sum_{k=1}^{n-1}h_k}\otimes \prod_{j=1}^{n-1}X_j^{a-b}Z_j^{h_j},\\
    g_{(h_0,\cdots,h_{n-1})}^{(G_n^d)}
    =&\left( \prod_{e\in E} {CZ}_e^{m_e}\right) \prod_{k=0}^{n-1}X_k^{h_k}\left(\prod_{e' \in  E}{CZ}_{e'}^{{d-m}_{e'}}\right)=\prod_{k=0}^{n-1}\left(X_k^{h_k}\prod_{e \in E,k\in e}{Z}_{e \backslash \left\{k\right\}}^{m_e\times h_k}\right),\\
    g_{(h_0,\cdots,h_{n-1})}^{(\widetilde{G}_n^d)}
    =&\left( \prod_{e\in\widetilde E} {{\widetilde{CZ}}_e}^{m_e}\right) \prod_{k=0}^{n-1}X_k^{h_k}\left(\prod_{e' \in \widetilde E}{{\widetilde{CZ}}_{e'}}^{{d-m}_{e'}}\right)=\prod_{k=0}^{n-1}\left(X_k^{h_k}\prod_{e \in\widetilde E,k\in e}{{\widetilde{CZ}}_{e \backslash \left\{k\right\}}}^{m_e\cdot h_k}\right),\\
    g_{(h_0,\cdots,h_{n-1})}^{(\widehat{G}_n^d)}
    =&\left(\prod_{{\dot{e}}\in\widehat E} {{\widehat{CZ}}_{\dot{e}}}^{m_{\dot{e}}}\right) \prod_{k=0}^{n-1}X_k^{h_k}\left(\prod_{{\dot{e}}' \in \widehat E}{{\widehat{CZ}}_{{\dot{e}}'}}^{{d-m}_{{\dot{e}}'}}\right),\\
    g_{(h_0,\cdots,h_{n-1})}^{(\widehat{\widetilde{G}}_n^d)}
    =&\left(\prod_{{\dot{e}}\in \widehat{\widetilde E}} {{\widehat{\widetilde{CZ}}}_{\dot{e}}}^{m_{\dot{e}}}\right) \prod_{k=0}^{n-1}X_k^{h_k}\left(\prod_{{\dot{e}}' \in \widehat{\widetilde E}}{{\widehat{\widetilde{CZ}}}_{{\dot{e}}'}}^{{d-m}_{{\dot{e}}'}}\right).
\end{align}

\section{Appendix E: Verification Examples of Quantum States}

In this section, we present the verification procedures for a series of quantum states, including qutrit-qubit states, arbitrary two-qubit states (locally unitary equivalent to $\left|\psi_2\right\rangle$, and requiring the same number of measurements~\cite{pallister2018optimal}), qubit and qudit Bell/Bell-like states, GHZ/GHZ-like states, graph states, hypergraph states, and the recently proposed multigraph and multihypergraph states. Here $d$ is a prime number for $\left|GHZ_n^d\right\rangle$, $\scriptstyle\left|\widetilde{GHZ}_n^d\right\rangle$, $\scriptstyle\left|G_n^d\right\rangle$, $\left|\widetilde{G}_n^d\right\rangle$, $\left|\widehat{G}_n^d\right\rangle$, and $\scriptstyle\left|\widehat{\widetilde{G}}_n^d\right\rangle$, and the composite-dimensional cases will be discussed separately in the next section. In our framework, the verification of qubit and qudit Bell/Bell-like states is included in the verification of qubit and qudit GHZ/GHZ-like states, and they are therefore discussed together. Before that, we first present the verification of the qubit Bell-like state $|\psi_2\rangle$, which is unitarily equivalent to arbitrary two-qubit pure state, thereby enabling the verification of arbitrary two-qubit states. The verification procedure consists of two main steps.  
First, generalized stabilizer testing is performed for these states using only single-qudit measurement settings. Second, based on the generalized stabilizer tests, the complete verification protocols are constructed, and the theoretical numbers of measurements required for verification are derived.

\subsection{E1: Verification of $\left|\psi_1\right\rangle$ and $\left|\psi_2\right\rangle$}

\subsubsection{E1.1: Verification of $\left|\psi_1\right\rangle$}

We now present the verification procedure for  
$\left|\psi_1\right\rangle=\frac{1}{\sqrt{6}}[(|0_3\rangle+|1_3\rangle+|2_3\rangle)\otimes|0_2\rangle+(|0_3\rangle+\omega_3|1_3\rangle+\omega_3^2|2_3\rangle)\otimes|1_2\rangle]$.  
For $\left|\psi_1\right\rangle$, the nontrivial generalized stabilizers can be expanded as  

\begin{align}
\normalsize
     g_{\{1,0\}}^{(\psi_1)}=&X^{(3)}\otimes\left(\left|0_2\right\rangle\left\langle 0_2\right|+\omega_3\left|1_2\right\rangle\left\langle 1_2\right|\right)=\sum_{j=0}^2\omega_3^{-j}|\kappa_j\rangle\langle \kappa_j|\otimes\left(\left|0_2\right\rangle\left\langle 0_2\right|+\omega_3\left|1_2\right\rangle\left\langle 1_2\right|\right),\\
     g_{\{2,0\}}^{(\psi_1)}=&\left(X^{(3)}\right)^{2}\otimes\left(\left|0_2\right\rangle\left\langle 0_2\right|+\omega_3^{2}\left|1_2\right\rangle\left\langle 1_2\right|\right) =\sum_{j=0}^2\omega_3^{-2j}|\kappa_j\rangle\langle \kappa_j|\otimes\left(\left|0_2\right\rangle\left\langle 0_2\right|+\omega_3^{2}\left|1_2\right\rangle\left\langle 1_2\right|\right),\\
    g_{\{0,1\}}^{(\psi_1)}=&Z^{(3)}\otimes\left|1_2\right\rangle\left\langle 0_2\right| +\left(Z^{(3)}\right)^{-1}\otimes\left|0_2\right\rangle\left\langle 1_2\right|=\left|0_3\right\rangle\left\langle 0_3\right|\otimes X^{(2)}\\
    &+\left|1_3\right\rangle\left\langle 1_3\right|\otimes\left(\omega_3\left|1_2\right\rangle\left\langle 0_2\right|+\omega_3^{-1}\left|0_2\right\rangle\left\langle 1_2\right|\right)+\left|2_3\right\rangle\left\langle 2_3\right|\otimes\left(\omega_3^{2}\left|1_2\right\rangle\left\langle 0_2\right|+\omega_3^{-2}\left|0_2\right\rangle\left\langle 1_2\right|\right)\\
    =&\left|0_3\right\rangle\left\langle 0_3\right|\otimes\left(\left| +x \right\rangle \left\langle +x \right|-\left| -x \right\rangle \left\langle -x \right|\right)\\
    &+\left|1_3\right\rangle\left\langle 1_3\right|\otimes\left(|\xi_{+}\rangle\langle\xi_{+}|-|\xi_{-}\rangle\langle\xi_{-}|\right)+\left|2_3\right\rangle\left\langle 2_3\right|\otimes\left(|\xi'_{+}\rangle\langle\xi'_{+}|-|\xi'_{-}\rangle\langle\xi'_{-}|\right),\\
     g_{\{1,1\}}^{(\psi_1)}=&X^{(3)}\left(Z^{(3)}\right)^{-1}\otimes\left|0_2\right\rangle\left\langle 1_2\right| +\omega_3X^{(3)}Z^{(3)}\otimes\left|1_2\right\rangle\left\langle 0_2\right|,\\
     g_{\{2,1\}}^{(\psi_1)}=&\left(X^{(3)}\right)^{2}\left(Z^{(3)}\right)^{-1}\otimes\left|0_2\right\rangle\left\langle 1_2\right| +\omega_3^{2}\left(X^{(3)}\right)^2Z^{(3)}\otimes\left|1_2\right\rangle\left\langle 0_2\right|,
\end{align}
where $|\kappa_j\rangle=\frac{1}{\sqrt{3}}\sum_{k=0}^2\omega_3^{jk}|k\rangle$, for $j=0,1,2$, are the eigenstates of $X^{(3)}$ with eigenvalue $\omega_3^{-j}$ and of $(X^{(3)})^{2}$ with eigenvalue $\omega_3^{-2j}$.  
The states $\left|\pm x\right\rangle=\frac{1}{\sqrt{2}}(\left|0\right\rangle\pm\left|1\right\rangle)$, $|\xi_{\pm}\rangle=\frac{1}{\sqrt{2}}\left(|0\rangle\pm\omega_3|1\rangle\right)$, and $|\xi'_{\pm}\rangle=\frac{1}{\sqrt{2}}\left(|0\rangle\pm\omega_3^2|1\rangle\right)$ are the eigenstates of $X^{(2)}$, $\omega_3\left|1_2\right\rangle\left\langle0_2\right|+\omega_3^{-1}\left|0_2\right\rangle\left\langle1_2\right|$, and $\omega_3^2\left|1_2\right\rangle\left\langle0_2\right|+\omega_3^{-2}\left|0_2\right\rangle\left\langle1_2\right|$ corresponding to eigenvalues $\pm1$, respectively. Therefore, the testing procedures for $g_{\{0,1\}}^{(\psi_1)}$ and $g_{\{0,2\}}^{(\psi_1)}$ are as follows. The qutrit is measured in the basis $\{|\kappa_0\rangle,|\kappa_1\rangle,|\kappa_2\rangle\}$,  
and the qubit is measured in the basis $\{|0\rangle,|1\rangle\}$.  The test passes when the outcomes are $|0\rangle$ and $|\kappa_0\rangle$( or $|1\rangle$ and $|\kappa_1\rangle$), respectively. For $g_{\{1,0\}}^{(\psi_1)}$, first measure the qutrit in the computational basis $\{|0\rangle,|1\rangle,|2\rangle\}$.  
If the result is $|0\rangle$ ($|1\rangle$ or $|2\rangle$), measure the qubit in the basis $\{|+\rangle,|-\rangle\}$ ($\{|\xi_+\rangle,|\xi_-\rangle\}$ or $\{|\xi'_+\rangle,|\xi'_-\rangle\}$).  
If the measurement outcome is $|+\rangle$ ($|\xi_+\rangle$ or $|\xi'_+\rangle$), the test passes. Since $\left|0_2\right\rangle\left\langle1_2\right|$ and $\left|1_2\right\rangle\left\langle0_2\right|$,  
$X^{(3)}\left(Z^{(3)}\right)^{-1}$ and $X^{(3)}Z^{(3)}$, as well as $\left(X^{(3)}\right)^{2}\left(Z^{(3)}\right)^{-1}$ and $\left(X^{(3)}\right)^{2}Z^{(3)}$, cannot be simultaneously diagonalized,  
the generalized stabilizer tests for $g_{\{1,1\}}^{(\psi_1)}$ and $g_{\{2,1\}}^{(\psi_1)}$ are not executable. Thus,  
$\normalsize S_{Test}^{\psi_1}=S_{Test_0}^{\psi_1}\bigcup S_{Test_1}^{\psi_1}\bigcup S_{Test_2}^{\psi_1}=\left\{g_{\{1,0\}}^{(\psi_1)}\right\}\bigcup\left\{g_{\{2,0\}}^{(\psi_1)}\right\}\bigcup\left\{g_{\{0,1\}}^{(\psi_1)}\right\}$,  
and  
$\normalsize C_{Test_0}^{\psi_1}=\{(1,0)\}$, $C_{Test_1}^{\psi_1}=\{(2,0)\}$, $C_{Test_2}^{\psi_1}=\{(0,1)\}$. The corresponding measurement operator for $|\psi_1\rangle$ is therefore given by  
\begin{equation}
    \normalsize
    \begin{aligned}
    \Omega_{\psi_1}&=|\psi_1\rangle\langle\psi_1|+\sum_{(j_0,j_1)\neq(0,0)}\left(\sum_{i=0}^{2}\frac{\mu_i}{|S_{Test_i}^{\psi_1}|}\left(\sum_{\substack{h_0,h_1:\ \sum_{k=0}^{1}\frac{h_k\times j_k}{d_k}\in\mathbb{Z}\\
    (h_0, h_1)\in C_{Test_i}^{\psi_1}}} 1\right)\right)U_{\psi_1}\left(\bigotimes_{k=0}^{1}\left| j_k \right\rangle \left\langle j_k\right|\right)U_{\psi_1}^\dagger\\
    &=|\psi_1\rangle\langle\psi_1|+U_{\psi_1}\big[\mu_2|10\rangle\langle10|+\mu_2|20\rangle\langle20|+(\mu_0+\mu_1)|01\rangle\langle01|\big]U_{\psi_1}^\dagger.
    \end{aligned}
    \label{eq:12}
\end{equation}
Since $\sum_{i=0}^{2}\mu_i=1$, the minimum $\min\{\beta(\Omega_{\psi_1})\}$ is achieved for $\mu_0+\mu_1=\mu_2=\frac{1}{2}$, yielding $\min\left\{\beta(\Omega_{\psi_1})\right\}=\frac{1}{2}$.  
Consequently, $\nu(\Omega_{\psi_1})=\frac{1}{2}$, and the number of measurements required for the verification of $|\psi_1\rangle$ is 
$n_{opt}^{\psi_1}\leq\lceil2\epsilon^{-1}\ln{\frac{1}{\delta}}\rceil$.

\subsubsection{E1.2: Verification of $\left|\psi_2\right\rangle$}

We now present the verification procedure for  
$\left|\psi_2\right\rangle=\sin{\theta}\left|00\right\rangle+\cos{\theta}\left|11\right\rangle$.  
Note that $|\psi_2\rangle\in\mathcal{H}_{2}\otimes\mathcal{H}_{2}$, that is, $d_0=d_1=2$.  
Expanding the nontrivial generalized stabilizers of $\left|\psi_2\right\rangle$, we obtain  
\begin{equation}
\hspace{-4mm}
    \normalsize
    \begin{aligned}
     g_{\{1,0\}}^{(\psi_2)}=&-\cos{2\theta}(Z\otimes I)+2\sin{2\theta}(X\otimes X)\\
    =&[-\cos{2\theta}Z+\sin{2\theta X}]\otimes\left| +x \right\rangle \left\langle +x \right|-[\cos{2\theta}Z+\sin{2\theta} X]\otimes\left| -x \right\rangle \left\langle -x \right|\\
   = &[\left| +\varphi_{+x} \right\rangle \left\langle +\varphi_{+x} \right|-\left| -\varphi_{+x} \right\rangle \left\langle -\varphi_{+x} \right|]\otimes\left| +x \right\rangle \left\langle +x \right|-[\left| +\varphi_{-x} \right\rangle \left\langle +\varphi_{-x} \right|-\left| -\varphi_{-x} \right\rangle \left\langle -\varphi_{-x} \right|]\otimes\left| -x \right\rangle \left\langle -x \right|,\\
    g_{\{1,1\}}^{(\psi_2)}=&-\cos{2\theta}(I\otimes Z)-2\sin{2\theta}(Y\otimes Y)\\
    =&\left| +y \right\rangle \left\langle +y \right|\otimes[-\cos{2\theta}Z-\sin{2\theta} Y]-\left| -y \right\rangle \left\langle -y \right|\otimes[\cos{2\theta}Z-\sin{2\theta} Y]\\
    =&\left| +y \right\rangle \left\langle +y \right|\otimes[\left| +\phi_{+y} \right\rangle \left\langle +\phi_{+y} \right|-\left| -\phi_{+y} \right\rangle \left\langle -\phi_{+y} \right|]-\left| -y \right\rangle \left\langle -y \right|\otimes[\left| +\phi_{-y} \right\rangle \left\langle +\phi_{-y} \right|-\left| -\phi_{-y} \right\rangle \left\langle -\phi_{-y} \right|]
    \end{aligned}
\label{eq:11}
\end{equation}
and $g_{\{0,1\}}^{(\psi_2)}=Z\otimes Z$, where $\left| \pm x \right\rangle=\frac{1}{\sqrt{2}}(\left|0\right\rangle\pm\left|1\right\rangle)$, $\left| +\varphi_{+x} \right\rangle=\sin{\theta}\left|0\right\rangle+\cos{\theta}\left|1\right\rangle$, $\left| -\varphi_{+x} \right\rangle=\cos{\theta}\left|0\right\rangle-\sin{\theta}\left|1\right\rangle$,  
$\left| +\varphi_{-x} \right\rangle=\cos{\theta}\left|0\right\rangle+\sin{\theta}\left|1\right\rangle$,  and $\left| -\varphi_{-x} \right\rangle=\sin{\theta}\left|0\right\rangle-\cos{\theta}\left|1\right\rangle$  
are the eigenstates of $X$, $-\cos{2\theta}Z+\sin{2\theta}X$, and $\cos{2\theta}Z+\sin{2\theta}X$  
with eigenvalues $\pm1$, respectively. Similarly, 
$\left| \pm y \right\rangle=\frac{1}{\sqrt{2}}(\left|0\right\rangle\pm\mathbf{i}\left|1\right\rangle)$,  
$\left| +\phi_{+y} \right\rangle=\sin{\theta}\left|0\right\rangle-\mathbf{i}\cos{\theta}\left|1\right\rangle$, $\left| -\phi_{+y} \right\rangle=\cos{\theta}\left|0\right\rangle+\mathbf{i}\sin{\theta}\left|1\right\rangle$, $\left| +\phi_{-y} \right\rangle=\cos{\theta}\left|0\right\rangle-\mathbf{i}\sin{\theta}\left|1\right\rangle$, and $\left| -\phi_{-y} \right\rangle=\sin{\theta}\left|0\right\rangle+\mathbf{i}\cos{\theta}\left|1\right\rangle$ are the eigenstates of $Y$, $-\cos{2\theta}Z-\sin{2\theta}Y$, and $\cos{2\theta}Z-\sin{2\theta}Y$ with eigenvalues $\pm1$, respectively. The testing process for $g_{\{1,0\}}^{(\psi_2)}$ is as follows.   
First, measure the second qubit (or equivalently the first one, since both qubits are symmetric in $\left|\psi_2\right\rangle$) in the basis $\{\left|+x\right\rangle,\left|-x\right\rangle\}$.  
If the measurement outcome is $\left|+x\right\rangle$ (or $\left|-x\right\rangle$),  
then measure the remaining qubit in the basis $\{\left|+\varphi_{+x}\right\rangle,\left|-\varphi_{+x}\right\rangle\}$ (or $\{\left|+\varphi_{-x}\right\rangle,\left|-\varphi_{-x}\right\rangle\}$). If the outcome is $\left|+\varphi_{+x}\right\rangle$ (or $\left|+\varphi_{-x}\right\rangle$), the test passes. The operator $g_{\{1,1\}}^{(\psi_2)}$ can be tested via adaptive single-qubit measurements  
in the bases $\{\left|+y\right\rangle,\left|-y\right\rangle\}$,  
$\{\left|+\phi_{+y}\right\rangle,\left|-\phi_{+y}\right\rangle\}$, and $\{\left|+\phi_{-y}\right\rangle,\left|-\phi_{-y}\right\rangle\}$. The operator $g_{\{0,1\}}^{(\psi_2)}$ can be tested using two nonadaptive measurements  
in the computational basis $\{\left|0\right\rangle,\left|1\right\rangle\}$, and if the outcome is $|0\rangle,|0\rangle$ or $|1\rangle,|1\rangle$, the test passes. Hence, $\normalsize S_{Test}^{\psi_2}=\left\{g_{\{0,1\}}^{(\psi_2)}\right\}\bigcup\left\{g_{\{1,0\}}^{(\psi_2)}\right\}\bigcup\left\{g_{\{1,1\}}^{(\psi_2)}\right\}$, with  
$\normalsize C_{Test_0}^{\psi_2}=\{(0,1)\}$, $C_{Test_1}^{\psi_2}=\{(1,0)\}$, and $C_{Test_2}^{\psi_2}=\{(1,1)\}$. The verification operator can thus be constructed as  
\begin{equation}
    \normalsize
    \begin{aligned}
    \Omega_{\psi_2}&=|\psi_2\rangle\langle\psi_2|+\sum_{(j_0,j_1)\neq(0,0)}\left(\sum_{i=0}^2\frac{\mu_i}{|S_{Test_i}^{\psi_2}|}\left(\sum_{\substack{h_0,h_1:\ \sum_{k=0}^{1}h_k\times j_k\text{ mod }2=0\\
    (h_0, h_1)\in C_{Test}^{\psi_2}}}\right)\right)U_{\psi_2}\left(\bigotimes_{k=0}^{1}\left| j_k \right\rangle \left\langle j_k\right|\right)U_{\psi_2}^\dagger\\
    &=|\psi_2\rangle\langle\psi_2|+U_{\psi_2}(\mu_1|01\rangle\langle01|+\mu_0|10\rangle\langle10|+\mu_2|11\rangle\langle11|)U_{\psi_2}^\dagger.
    \end{aligned}
    \label{eq:12}
\end{equation}
Since $\sum_{i=0}^2\mu_i=1$, the minimum $\min\{\beta(\Omega_{\psi_2})\}$ is obtained for  
$\mu_0=\mu_1=\mu_2=\frac{1}{3}$, yielding $\min\left\{\beta(\Omega_{\psi_2})\right\}=\frac{1}{3}$.  
Thus,  
\begin{equation}
    \begin{aligned}
        \Omega_{\psi_2}&=|\psi_2\rangle\langle\psi_2|+\frac{1}{3}U_{\psi_2}(|01\rangle\langle01|+|10\rangle\langle10|+|11\rangle\langle11|)U_{\psi_2}^\dagger\\
    &=\frac{1}{6}\left[3I\otimes I-\cos{2\theta}(Z\otimes I)+\sin{2\theta(X\otimes X)}+Z\otimes Z-\cos{2\theta}(I\otimes Z)-\sin{2\theta(Y\otimes Y)}\right],
    \end{aligned}
\end{equation}
and $\nu(\Omega_{\psi_2})=\frac{2}{3}$. The number of measurements required for verifying $|\psi_2\rangle$ is   
$n_{opt}^{\psi_2}\leq\lceil\frac{3}{2}\epsilon^{-1}\ln{\frac{1}{\delta}}\rceil$,  
which is fewer than that in Ref.~\cite{pallister2018optimal},  
where the corresponding bound is $(2+\frac{1}{2}\sin{2\theta})\epsilon^{-1}\ln{\frac{1}{\delta}}$.  
Moreover, the verification operator here is more concise and independent of the value of $\theta$. For $\theta=\frac{\pi}{4}$, $|\psi_2\rangle=\frac{1}{2}(|00\rangle+|11\rangle)$,  
which is the Bell state,  
and $\Omega_{\psi_2}(\frac{\pi}{4})=\frac{1}{2}\left(3I\otimes I+X\otimes X+Z\otimes Z+Y\otimes Y\right)$,  
exactly matching the Bell-state verification operator given in Ref.~\cite{pallister2018optimal}.  
Thus, the verification of the Bell state is a special case of the present scheme.  
Furthermore, this result shows that the number of measurements required for verifying $|\psi_2\rangle$  
is identical to that of the Bell state in Ref.~\cite{pallister2018optimal},  
regardless of the value of $\theta$.  
Since any two-qubit pure state is locally unitary equivalent to $\left|\psi_2\right\rangle$ for some $\theta$,  
the present method can verify any two-qubit pure state  
$|\psi\rangle=U_0\otimes U_1|\psi_2\rangle$  
with the same measurement cost  
$\lceil\frac{3}{2}\epsilon^{-1}\ln{\frac{1}{\delta}}\rceil$  
by using the measurement operator $(U_0\otimes U_1)\Omega_{\psi_2}(U_0\otimes U_1)^\dagger$.

\subsection{E2: Verification of qubit and qudit Bell/Bell-like states, GHZ/GHZ-like states}

As discussed in Appendix~D, it is straightforward to observe that both 
$g_{(h_0,\cdots,h_{n-1})}^{(GHZ_n^2)}$ and $g_{(h_0,\cdots,h_{n-1})}^{(GHZ_n^d)}$ 
are tensor products of Pauli operators—either the qubit Pauli operators or the qudit generalized Pauli operators.  
Therefore, generalized stabilizer testing for these two classes of states can be implemented solely through single-particle Pauli measurements. In contrast, the operators $g_{(h_0,\cdots,h_{n-1})}^{\widetilde{GHZ}_n^2}$ and 
$g_{(h_0,\cdots,h_{n-1})}^{\widetilde{GHZ}_n^d}$ 
are relatively more involved, requiring more sophisticated single-particle measurements. In what follows, we present the verification procedures for these four types of quantum states based on their respective generalized stabilizer tests.

\subsubsection{E2.1: Verification of qubit GHZ and Bell states}

The elements of the generalized stabilizer group for $\left|GHZ_n^2\right\rangle$ can be expanded as  
\begin{equation}
\normalsize
    \begin{aligned}
         g_{(0,h_1,\cdots,h_{n-1})}^{(GHZ_n^2)}
        =&Z_0^{\sum_{k=1}^{n-1}h_k}\prod_{k=1}^{n-1}Z_k^{h_k}
        =\left(|0\rangle\langle0|+(-1)^{\sum_{k=1}^{n-1}h_k}|1\rangle\langle1|\right)\otimes\bigotimes_{k=1}^{n-1}\left(|0\rangle\langle0|+(-1)^{h_k}|1\rangle\langle1|\right),\\
        g_{(1,h_1,\cdots,h_{n-1})}^{(GHZ_n^2)}
        =&X_0 Z_0^{\sum_{k=1}^{n-1} h_k}\prod_{j=1}^{n-1}(X_jZ_j^{h_j})\\
        =&(\mathbf{i})^{-2\sum_{k=1}^{n-1} h_k}\left(\mathbf{i}^{\sum_{k=1}^{n-1} h_k}X_0 Z_0^{\sum_{k=1}^{n-1} h_k}\right)\prod_{j=1}^{n-1}(\mathbf{i}^{h_j}X_jZ_j^{h_j})\\
        =&(-1)^{-\sum_{k=1}^{n-1} h_k}\left(\sum_{\pm}\pm(-1)^{\left\lfloor\frac{1}{2}\sum_{k=1}^{n-1} h_k\right\rfloor}\left|u_{\pm}^{\left(\sum_{k=1}^{n-1}h_k\text{ mod }2\right)}\right\rangle\left\langle u_{\pm}^{\left(\sum_{k=1}^{n-1}h_k\text{ mod }2\right)}\right|\right)\\
        &\otimes\bigotimes_{j=1}^{n-1}\left(\sum_{\pm}\pm(-1)^{\left\lfloor\frac{1}{2}h_j\right\rfloor}\left|u_{\pm}^{\left(h_j\right)}\right\rangle\left\langle u_{\pm}^{\left(h_j\right)}\right|\right),
    \end{aligned}
\end{equation}
where $\scriptstyle\left|u_{\pm}^{\left(0\right)}\right\rangle=|\pm x\rangle$ and $\scriptstyle\left|u_{\pm}^{\left(1\right)}\right\rangle=|\pm y\rangle$. For $g_{(0,h_1,\cdots,h_{n-1})}^{(GHZ_n^2)}$ with $h_1,\cdots,h_{n-1}\in\mathbb{Z}_2$ and $(0,h_1,\cdots,h_{n-1})\neq(0,\cdots,0)$, there are $2^{n-1}-1$ elements, each of which can be tested using a single measurement. The testing procedure is as follows.  
Measure all qubits of $\scriptstyle\left|GHZ_n^2\right\rangle$ in the computational basis $\{|0\rangle,|1\rangle\}$.  
Let the measurement outcome of the $i$-th qubit ($i\in\mathbb{Z}_n$) be denoted by $M_i$,  
where $M_i=0$ if the result is $|0\rangle$ and $M_i=1$ if the result is $|1\rangle$.  
The test passes if $
M_0{\sum_{k=1}^{n-1}h_k}+\sum_{j=1}^{n-1}{M_jh_j}=0\ \text{(mod 2)}$, for all $h_1,\cdots,h_{n-1}\in\mathbb{Z}_2$ with $(0,h_1,\cdots,h_{n-1})\neq(0,\cdots,0)$. For $g_{(1,h_1,\cdots,h_{n-1})}^{(GHZ_n^2)}$ with $h_1,\cdots,h_{n-1}\in\mathbb{Z}_2$,  
there are $2^{n-1}$ such elements.  
Since the measurement bases depend on the specific values of $h_1,\cdots,h_{n-1}$,  
each test requires an independent measurement.  
The testing procedure is as follows.  
Measure one qubit of $\left|GHZ_n^2\right\rangle$ in the basis  
$\left\{\scriptstyle\left|u_{\pm}^{\left(\sum_{k=1}^{n-1}h_k\text{ mod }2\right)}\right\rangle\right\}$,  
and measure each of the remaining $n-1$ qubits in the bases  
$\left\{\scriptstyle\left|u_{\pm}^{\left(h_j\text{ mod }2\right)}\right\rangle\right\}$ for $j\in\mathbb{Z}_n^*$.  
Let $M_i$ denote the measurement outcome of the $i$-th qubit,  
where $M_i=0$ if the result is $|+x\rangle$ or $|+y\rangle$,  
and $M_i=1$ if the result is $|-x\rangle$ or $|-y\rangle$.  
The test passes when $\sum_{i=0}^{n-1}M_i\text{ mod }2=0$. In summary, $\scriptstyle S_{Test}^{GHZ_n^2}$ can be divided into $2^{n-1}+1$ sets, where 
\begin{align}
\normalsize
     &S_{Test_0}^{{GHZ}_n^2}=\left\{ g_{(0,h_1,h_2,\cdots,h_{n-1})}^{{GHZ}_n^2}\big|h_1,h_2,\cdots,h_{n-1}\in \mathbb{Z}_2,\ (0,h_1,h_2,\cdots,h_{n-1})\neq(0,\cdots,0)\right\},\\
     &S_{Test_{1+\sum_{l=1}^{n-1}h_l\times 2^{n-1-l}}}^{{GHZ}_n^2}=\left\{ g_{(1,h_1,h_2,\cdots,h_{n-1})}^{{GHZ}_n^2}\right\},h_1,h_2,\cdots,h_{n-1}\in \mathbb{Z}_2,\\
     &C_{Test_0}^{{GHZ}_n^2}=\left\{(0,h_1,h_2,\cdots,h_{n-1})\big|h_1,h_2,\cdots,h_{n-1}\in \mathbb{Z}_2,\ (0,h_1,h_2,\cdots,h_{n-1})\neq(0,\cdots,0)\right\},\\
     &C_{Test_{1+\sum_{l=1}^{n-1}h_l\times 2^{n-1-l}}}^{{GHZ}_n^2}=\left\{(1,h_1,h_2,\cdots,h_{n-1})\right\},h_1,h_2,\cdots,h_{n-1}\in \mathbb{Z}_2.
\end{align}
The subscript $\sum_{l=1}^{n-1}h_l\times2^{n-1-l}$ corresponds to the decimal representation of the binary number $h_1h_2\cdots h_{n-1}$,  
allowing the sets to be indexed continuously as $0,1,\cdots,2^{n-1}$. Moreover,  
$\scriptstyle \left|S_{Test_0}^{GHZ_n^2}\right|=\left|C_{Test_0}^{GHZ_n^2}\right|=2^{n-1}-1$,  
and  
$\scriptstyle \left|S_{Test_j}^{GHZ_n^2}\right|=\left|C_{Test_j}^{GHZ_n^2}\right|=1$ for $j=1,2,\cdots,2^{n-1}$.  
Thus, the verification operator can be constructed as  
\begin{equation}
    \normalsize
    \begin{aligned}
   \Omega_{{GHZ}_n^2}=&\left|{GHZ}_n^2\right\rangle\left\langle {GHZ}_n^2\right|\\
    &+\sum_{\substack{j_0,\cdots,j_{n-1}=0,\\(j_0,\cdots,j_{n-1})\neq(0,\cdots,0)}}^{1}\left(\sum_{i=0}^{2^{n-1}}\frac{\mu_i}{\left|S_{Test_i}^{{GHZ}_n^2}\right|}\left(\sum_{\substack{h_0,\cdots,h_{n-1}:\ \sum_{k=0}^{n-1}h_k\times j_k\text{ mod }2=0 \\
    (h_0,\cdots, h_{n-1})\in C_{Test_i}^{{GHZ}_n^2}}}1\right)\right)U_{{GHZ}_n^2}\left(\bigotimes_{k=0}^{n-1}\left| j_k \right\rangle \left\langle j_k\right|\right)U_{{GHZ}_n^2}^\dagger\\
    =&\left|{GHZ}_n^2\right\rangle\left\langle {GHZ}_n^2\right|+\mu_0U_{{GHZ}_n^2}\left| 100\cdots0 \right\rangle \left\langle 100\cdots0\right|U_{{GHZ}_n^2}^\dagger\\
    &+\sum_{\substack{j_0,\cdots,j_{n-1}=0,\\(j_1,\cdots,j_{n-1})\neq(0,\cdots,0)}}^{1}\left[\frac{\mu_0}{2^{n-1}-1}\left(\sum_{\substack{h_1,\cdots,h_{n-1}:\ \sum_{k=1}^{n-1}h_k\times j_k\text{ mod }2=0\\
    (h_1,\cdots,h_{n-1})\neq(0,\cdots,0)}}1\right)\right.\\ &\left.+\sum_{i=1}^{2^{n-1}}\mu_i\left(\sum_{\substack{h_1,\cdots,h_{n-1}:\ j_0+\sum_{k=1}^{n-1}h_k\times j_k\text{ mod }2=0\\
    1+\sum_{l=1}^{n-1}h_l\times 2^{n-1-l}=i}}1\right)\right]U_{{GHZ}_n^2}\left(\bigotimes_{k=0}^{n-1}\left| j_k \right\rangle \left\langle j_k\right|\right)U_{{GHZ}_n^2}^\dagger,
    \end{aligned} 
\end{equation}
where $\sum_{i=0}^{2^{n-1}}\mu_i=1$. When $j_1,\cdots,j_{n-1}$ satisfy that there exists exactly one $j_{l'}=1$ ($l'\in\mathbb{Z}_n^*$) and $h_{l'}=0$, we obtain  
\begin{equation}
\normalsize
    \max_{\substack{j_0,\cdots,j_{n-1}\in\mathbb{Z}_2,\\(j_1,\cdots,j_{n-1})\neq(0,\cdots,0)}}\left\{\frac{\mu_0}{2^{n-1}-1}\left(\sum_{\substack{h_1,\cdots,h_{n-1}:\ \sum_{k=1}^{n-1}h_k\times j_k\text{ mod }2=0\\
    (h_1,\cdots,h_{n-1})\neq(0,\cdots,0)}}1\right)\right\}=\frac{(2^{n-2}-1)\mu_0}{2^{n-1}-1},
\end{equation}
and if $j_0=0$,  
\begin{equation}
\normalsize
     \sum_{i=1}^{2^{n-1}}\mu_i\left(\sum_{\substack{h_1,\cdots,h_{n-1}:\ j_0+\sum_{k=1}^{n-1}h_k\times j_k\text{ mod }2=0\\
    1+\sum_{l=1}^{n-1}h_l\times 2^{n-1-l}=i}}1\right)=\sum_{h_l=0,l\in\mathbb{Z}_{n}^*,\ l\neq l'}^{1}\mu_{1+\sum_{l=1,\ l\neq l'}^{n-1}h_l\times2^{n-1-l}}\ ,
\end{equation}
which includes the largest number of $\mu_i$, totaling $2^{n-2}$ terms for $l\in\mathbb{Z}_n^*$, $l\neq l'$, $h_l\in\mathbb{Z}_2$, and $h_0=1$.  
Since $\sum_{i=0}^{2^{n-1}}\mu_i=1$, we set $\mu_0=\frac{2^{n-1}-1}{2^n-1}$ and $\mu_i=\frac{1}{2^n-1}$ for $i\in\mathbb{Z}_{2^{n-1}}^*$.  
Then  
\begin{equation}
\normalsize
    \begin{aligned}
    \min\{\beta(\Omega_{GHZ_n^2})\}&=\mu_0=\frac{(2^{n-2}-1)\mu_0}{2^{n-1}-1}+\sum_{h_l=0,l\in\mathbb{Z}_{n}^*,\ l\neq l'}^{1}\mu_{1+\sum_{l=1,\ l\neq l'}^{n-1}h_l\times2^{n-1-l}}\\
    &=\frac{2^{n-2}-1}{2^{n-1}-1}\cdot\frac{2^{n-1}-1}{2^n-1}+\frac{2^{n-2}}{2^n-1}=\frac{2^{n-1}-1}{2^n-1}
    \end{aligned}
\end{equation}
Hence, $\nu\!\left(\Omega_{GHZ_n^2}\right)=\frac{2^{n-1}}{2^n-1}$.  
The number of measurements required for verifying $\left|GHZ_n^2\right\rangle$ is therefore  
\begin{equation}
n_{opt}^{GHZ_n^2}
=\left\lceil \frac{\ln\delta}{\ln\!\left[1-\frac{2^{n-1}}{2^n-1}\epsilon\right]}\right\rceil
\leq\left\lceil \frac{2^n-1}{2^{n-1}}\epsilon^{-1}\ln\delta^{-1}\right\rceil
=\left\lceil \!\left(2-\frac{1}{2^{n-1}}\right)\epsilon^{-1}\ln\delta^{-1}\!\right\rceil
\leq 2\epsilon^{-1}\ln\delta^{-1}.
\end{equation}
This verification efficiency is comparable to the optimal one reported in Ref.~\cite{li2020optimal},  
where $N(\Omega_I)=\frac{3}{2}\epsilon^{-1}\ln\delta^{-1}$, both achieving $O(1)$ complexity. It is straightforward to see that when $n=2$, the above verification procedure for the qubit GHZ state reduces to that for the qubit Bell state, in which case the required number of measurements satisfies $n_{opt}^{GHZ_2^2}\leq \frac{3}{2}\epsilon^{-1}\ln\delta^{-1}$.

\subsubsection{E2.2: Verification of qubit GHZ-like and Bell-like states}

We expand several elements of the generalized stabilizer group for the state $\scriptstyle\left|\widetilde{GHZ}_n^2\right\rangle$ as  
\begin{equation}
\normalsize
    \begin{aligned}
    g_{(0,h_1,\cdots,h_{n-1})}^{(\widetilde{GHZ}_n^2)}=&Z_0^{\sum_{k=1}^{n-1}h_k}\prod_{k=1}^{n-1}Z_k^{h_k}
    =\left(|0\rangle\langle0|+(-1)^{\sum_{k=1}^{n-1}h_k}|1\rangle\langle1|\right)\otimes\bigotimes_{k=1}^{n-1}\left(|0\rangle\langle0|+(-1)^{h_k}|1\rangle\langle1|\right),\\
    g_{(1,0,0,\cdots,0)}^{(\widetilde{GHZ}_n^2)}
    =&-\cos{(2\theta)}Z\otimes I^{\otimes(n-1)} +\sin{(2\theta)}X^{\otimes n}\\
    =&\frac{1}{2}\left[-\cos{(2\theta)}Z+\sin{(2\theta)}X\right]\otimes\left(I^{\otimes(n-1)}+X^{\otimes (n-1)}\right) \\
    &-\frac{1}{2}\left[\cos{(2\theta)}Z+\sin{(2\theta)}X\right]\otimes\left(I^{\otimes(n-1)}-X^{\otimes (n-1)}\right)\\
    =&\frac{1}{2}(\left| +\varphi_{+x} \right\rangle\left\langle+\varphi_{+x}\right|-\left| -\varphi_{+x} \right\rangle\left\langle-\varphi_{+x}\right|)\otimes\left(I^{\otimes(n-1)}+X^{\otimes (n-1)}\right) \\
    &-\frac{1}{2}(\left| +\varphi_{-x} \right\rangle\left\langle+\varphi_{-x}\right|-\left| -\varphi_{-x} \right\rangle\left\langle-\varphi_{-x}\right|)\otimes\left(I^{\otimes(n-1)}-X^{\otimes (n-1)}\right),
    \end{aligned}
\end{equation}
where $(0,h_1,\cdots,h_{n-1})\neq(0,\cdots,0)$, $\left| +\varphi_{+x} \right\rangle=\sin{\theta}\left| 0\right\rangle+\cos{\theta}\left| 1 \right\rangle$,  
$\left| -\varphi_{+x} \right\rangle=\cos{\theta}\left| 0\right\rangle-\sin{\theta}\left| 1 \right\rangle$,  
$\left| +\varphi_{-x} \right\rangle=\cos{\theta}\left| 0\right\rangle+\sin{\theta}\left| 1 \right\rangle$,  
and $\left| -\varphi_{-x} \right\rangle=\sin{\theta}\left| 0\right\rangle-\cos{\theta}\left| 1 \right\rangle$  
are the eigenstates of $-\cos{2\theta}Z+\sin{2\theta}X$ and $\cos{2\theta}Z+\sin{2\theta}X$  
with eigenvalues $\pm1$, respectively. For $\scriptstyle g_{(1,0,0,\cdots,0)}^{(\widetilde{GHZ}_n^2)}$, the testing procedure is as follows. First, measure any $n-1$ qubits of $\scriptstyle\left|\widetilde{GHZ}_n^2\right\rangle$  
in the basis $\{\left| \pm x \right\rangle=\frac{1}{\sqrt{2}}(\left| 0\right\rangle\pm\left| 1 \right\rangle)\}$.  
Let $M_i$ denote the measurement outcome of the $i$-th qubit,  
where $M_i=0$ for $\left| +x \right\rangle$ and $M_i=1$ for $\left| -x \right\rangle$.  
Let $i'$ denote the position of the qubit that is not measured.  
If $\sum_{i=0,i\neq i'}^{n-1}M_i=0\ \text{(mod 2)}$,  
measure the remaining qubit in the basis $\{\left| \pm\varphi_{+x} \right\rangle\}$.  
If the outcome is $\left| +\varphi_{+x} \right\rangle$ (recorded as $M_{i'}=0$), the test passes.  
If $\sum_{i=0,i\neq i'}^{n-1}M_i=1\ \text{(mod 2)}$,  
then measure the remaining qubit in the basis $\{\left| \pm\varphi_{-x} \right\rangle\}$,  
and if the result is $\left| -\varphi_{-x} \right\rangle$ (recorded as $M_{i'}=1$),  
the test passes.  
Therefore, the test passes whenever $\sum_{i=0}^{n-1}M_i=0\ \text{(mod 2)}$. For the operators  
$\normalsize g_{(0,h_1,\cdots,h_{n-1})}^{(\widetilde{GHZ}_n^2)}=\normalsize g_{(0,h_1,\cdots,h_{n-1})}^{({GHZ}_n^2)}$  
with $(0,h_1,h_2,\cdots,h_{n-1})\neq(0,\cdots,0)$,  
a total of $2^{n-1}-1$ operators can be tested simultaneously with a single measurement,  
following the same procedure as for $\normalsize g_{(0,h_1,\cdots,h_{n-1})}^{({GHZ}_n^2)}$.  
Thus, the details are omitted here. This also illustrates the necessity stated in the Theorem that the set $S_{Test}^{\psi}$  
must contain all generalized stabilizer generators of $|\psi\rangle$.  
For example, if only  
$\normalsize g_{(0,h_1,\cdots,h_{n-1})}^{(\widetilde{GHZ}_n^2)}=\normalsize g_{(0,h_1,\cdots,h_{n-1})}^{({GHZ}_n^2)}$  
were used for testing,  
then the states $\scriptstyle\left|\widetilde{GHZ}_n^2\right\rangle$  
and $\scriptstyle\left|{GHZ}_n^2\right\rangle$  
would be verified as identical quantum states. Therefore, we divide the set of testable generalized stabilizers into two subsets,  
$\scriptstyle S_{Test}^{\widetilde{GHZ}_n^2}=S_{Test_0}^{\widetilde{GHZ}_n^2}\bigcup S_{Test_1}^{\widetilde{GHZ}_n^2}$,  
where  
$\scriptstyle S_{Test_1}^{\widetilde{GHZ}_n^2}=\left\{g_{(1,0,0,\cdots,0)}^{\widetilde{GHZ}_n^2}\right\}$  
and  
$\scriptstyle C_{Test_1}^{\widetilde{GHZ}_n^2}=\{(1,0,0,\cdots,0)\}$, 
\begin{align}
        \normalsize
        S_{Test_0}^{\widetilde{GHZ}_n^2}&=\left\{ g_{(0,h_1,h_2,\cdots,h_{n-1})}^{\widetilde{GHZ}_n^2}\big|h_1,h_2,\cdots,h_{n-1}\in \mathbb{Z}_2,\ (0,h_1,h_2,\cdots,h_{n-1})\neq(0,\cdots,0)\right\},\\
        C_{Test_0}^{\widetilde{GHZ}_n^2}&=\left\{ (0,h_1,h_2,\cdots,h_{n-1})\big|h_1,h_2,\cdots,h_{n-1}\in \mathbb{Z}_2,\ (0,h_1,h_2,\cdots,h_{n-1})\neq(0,\cdots,0)\right\}.
\end{align}
Here, $\scriptstyle\left|S_{Test_1}^{\widetilde{GHZ}_n^2}\right|=\left|C_{Test_1}^{\widetilde{GHZ}_n^2}\right|=1$ and  
$\scriptstyle\left|S_{Test_0}^{\widetilde{GHZ}_n^2}\right|=\left|C_{Test_0}^{\widetilde{GHZ}_n^2}\right|=2^{n-1}-1$.  
Thus, the verification operator can be constructed as  
\begin{equation}
    \fontsize{8}{10}
    \begin{aligned}
    \Omega_{\widetilde{GHZ}_n^2}=&\left|\widetilde{GHZ}_n^2\right\rangle\left\langle \widetilde{GHZ}_n^2\right|\\
    &+\sum_{\substack{j_0,\cdots,j_{n-1}=0,\\(j_0,\cdots,j_{n-1})\neq(0,\cdots,0)}}^{1}\left(\sum_{i=0}^{1}\frac{\mu_i}{\left|S_{Test_i}^{\widetilde{GHZ}_n^2}\right|}\left(\sum_{\substack{h_0,\cdots,h_{n-1}:\ \sum_{k=0}^{n-1}h_k\times j_k\text{ mod }2=0\\
    (h_0,\cdots, h_{n-1})\in C_{Test_i}^{\widetilde{GHZ}_n^2}}}1\right)\right)U_{\widetilde{GHZ}_n^2}\left(\bigotimes_{k=0}^{n-1}\left| j_k \right\rangle \left\langle j_k\right|\right)U_{\widetilde{GHZ}_n^2}^\dagger\\
    =&\left|\widetilde{GHZ}_n^2\right\rangle\left\langle \widetilde{GHZ}_n^2\right|+\mu_0U_{\widetilde{GHZ}_n^2}\left| 100\cdots0 \right\rangle \left\langle 100\cdots0\right|U_{\widetilde{GHZ}_n^2}^\dagger\\
    &+\sum_{\substack{j_1,\cdots,j_{n-1}=0,\\(j_1,\cdots,j_{n-1})\neq(0,\cdots,0)}}^{1}\left[\frac{\mu_0}{2^{n-1}-1}\left(\sum_{\substack{h_0,\cdots,h_{n-1}:\ \sum_{k=0}^{n-1}h_k\times j_k\text{ mod }2=0\\
    (h_0,\cdots, h_{n-1})\neq(0,\cdots,0)}}1\right)+\mu_1\left(\sum_{\substack{h_0=1\\h_0\times j_0\text{ mod }2=0}}^{1}1\right)\right]U_{\widetilde{GHZ}_n^2}\left(\bigotimes_{k=0}^{n-1}\left| j_k \right\rangle \left\langle j_k\right|\right)U_{\widetilde{GHZ}_n^2}^\dagger.
    \end{aligned} 
\end{equation}
When $j_0=0$, and among $j_1,\cdots,j_{n-1}$ only one satisfies $j_{l'}=1$ ($l'\in\mathbb{Z}_n^*$) and $h_{l'}=0$, we have  
\begin{equation}
\fontsize{8}{10}
\hspace{-8mm}
    \max_{\substack{j_0,\cdots,j_{n-1}\in\mathbb{Z}_2,\\(j_1,\cdots,j_{n-1})\neq(0,\cdots,0)}}\left\{\frac{\mu_0}{2^{n-1}-1}\left(\sum_{\substack{h_1,\cdots,h_{n-1}:\ \sum_{k=1}^{n-1}h_k\times j_k\text{ mod }2=0\\
    (h_1,\cdots,h_{n-1})\neq(0,\cdots,0)}}1\right)\right\}=\frac{(2^{n-2}-1)\mu_0}{2^{n-1}-1}\text{, }\max_{\substack{j_0,\cdots,j_{n-1}\in\mathbb{Z}_2,\\(j_1,\cdots,j_{n-1})\neq(0,\cdots,0)}}{\left\{\mu_1\left(\sum_{\substack{h_0=1\\h_0\times j_0\text{ mod }2=0}}^{1}1\right)\right\}}=\mu_1.
\end{equation}
Since $\mu_0+\mu_1=1$, we take  
$\mu_0=\frac{2^{n-1}-1}{3\times2^{n-2}-1}$ and  
$\mu_1=\frac{2^{n-2}}{3\times2^{n-2}-1}$.  
Then  
\begin{equation}
\normalsize
    \begin{aligned}
    \min\left\{\beta\left(\Omega_{\scriptstyle\widetilde{GHZ}_n^2}\right)\right\}&=\mu_0=\frac{(2^{n-2}-1)\mu_0}{2^{n-1}-1}+\mu_1\\
    &=\frac{2^{n-2}-1}{2^{n-1}-1}\cdot\frac{2^{n-1}-1}{3\times2^{n-2}-1}+\frac{2^{n-2}}{3\times2^{n-1}-1}=\frac{2^{n-1}-1}{3\times2^{n-2}-1}.
    \end{aligned}
\end{equation}
Hence,  
$\nu\!\left(\Omega_{\widetilde{GHZ}_n^2}\right)=\frac{2^{n-2}}{3\times2^{n-2}-1}$.  
The number of measurements required for verifying  
$\left|\widetilde{GHZ}_n^2\right\rangle$  
is thus  
\begin{equation}
    \normalsize
n_{opt}^{\widetilde{GHZ}_n^2}
=\left\lceil \frac{\ln\delta}{\ln\!\left[1-\frac{2^{n-2}}{3\times2^{n-2}-1}\epsilon\right]}\right\rceil
\leq\left\lceil \frac{3\times2^{n-2}-1}{2^{n-2}}\epsilon^{-1}\ln\delta^{-1}\right\rceil
=\left\lceil \!\left(3-\frac{1}{2^{n-2}}\right)\epsilon^{-1}\ln\delta^{-1}\!\right\rceil
<\left\lceil 3\epsilon^{-1}\ln\delta^{-1}\right\rceil.
\end{equation}
This result exhibits an exponential improvement over Ref.~\cite{pallister2018optimal},  
where the required number of measurements is  
$[1+[(1+\tan^{2/n}{\theta})^n-\tan^{2/n}{\theta}]/(1+\tan^2{\theta})]\epsilon^{-1}\ln(\delta^{-1})$,  
and achieves a comparable optimal efficiency to that of Ref.~\cite{li2020optimal},  
where $N(\Omega_{VI})\approx2\epsilon^{-1}\ln\
\delta^{-1}$,  
both with $O(1)$ complexity. It is straightforward to see that when $n=2$, the above verification procedure for the qubit GHZ-like state reduces to that for the qubit Bell-like state, in which case the required number of measurements satisfies $n_{opt}^{\widetilde{GHZ}_2^2}\leq 2\epsilon^{-1}\ln\delta^{-1}$.

\subsubsection{E2.3: Verification of qudit GHZ and Bell states}

The elements of the generalized stabilizer group for $\left|GHZ_n^d\right\rangle$ can be expanded as  
\begin{equation}
\fontsize{8}{10}
    \begin{aligned}
        g_{(0,h_1,\cdots,h_{n-1})}^{(GHZ_n^d)}
        =&Z_0^{\sum_{k=1}^{n-1}h_k}\prod_{k=1}^{n-1}Z_k^{h_k}
        =\left(\sum_{j_0=0}^{d-1}\omega_d^{j_0\cdot\sum_{k=1}^{n-1}h_k}|j_0\rangle\langle j_0|\right)\prod_{k=1}^{n-1}\left(\sum_{j_k=0}^{d-1}\omega_d^{j_k\cdot h_k}|j_k\rangle\langle j_k|\right),\\
        g_{(h_0,h_1,\cdots,h_{n-1})}^{(GHZ_n^d)}
        =&X_0^{h_0}Z_0^{-\sum_{k=1}^{n-1}h_k}\prod_{k=1}^{n-1}(X_k^{h_0}Z_k^{h_k})\\
        =&\left(\sum_{j_0=0}^{d-1}\omega_d^{j_0}\left|u_{j_0}^{\left(\sum_{k=1}^{n-1}h_k\text{ mod }d\right)}\right\rangle\left\langle u_{j_0}^{\left(\sum_{k=1}^{n-1}h_k\text{ mod }d\right)}\right|\right)\otimes\bigotimes_{k=1}^{n-1}\left(\sum_{j_k=0}^{d-1}\omega_d^{j_k}\left|u_{j_k}^{\left(h_k\right)}\right\rangle\left\langle u_{j_k}^{\left(h_k\right)}\right|\right),
    \end{aligned}
\end{equation}
where $h_0\neq0$, and  
$\left|u_{j_0}^{(l)}\right\rangle=\frac{1}{\sqrt{d}}\sum_{i=1}^{d-1}\omega_d^{-i\cdot j_0+\frac{l\cdot h_0}{2}i(i-1)}\left|i\cdot h_0\text{ mod }d\right\rangle$,  
with $l\in\mathbb{Z}_d$. For $g_{(0,h_1,\cdots,h_{n-1})}^{({GHZ}_n^d)}$ with $(0,h_1,\cdots,h_{n-1})\neq(0,\cdots,0)$,  
a total of $d^{n-1}-1$ operators can be tested simultaneously in a single measurement.  
The testing procedure is as follows.  
Measure all qudits of $\scriptstyle\left|{GHZ}_n^d\right\rangle$ in the computational basis  
$\{|0\rangle,|1\rangle,\cdots,|d-1\rangle\}$.  
Let $M_i$ ($i\in\mathbb{Z}_n$) denotes the measurement outcome of the $i$-th qudit,  
and assign $M_i=j$ when $|j\rangle$ is observed.  
The test passes if $M_0{\sum_{k=1}^{n-1}h_k}+\sum_{j=1}^{n-1}{M_jh_j}=0\text{ (mod d)}$
holds for all $(0,h_1,h_2,\cdots,h_{n-1})\neq(0,\cdots,0)$. When $h_0\neq0$, the measurement bases for testing $g_{(h_0,h_1,\cdots,h_{n-1})}^{(GHZ_n^d)}$  
depend on the values of $h_0,h_1,\cdots,h_{n-1}$.  
The procedure is as follows.  
Measure any one qudit of $\scriptstyle\left|{GHZ}_n^d\right\rangle$ (denoted by index $i'$)  
in the basis  
$\scriptstyle\left\{\left|u_{l}^{\left(\sum_{k=1}^{n-1}h_k\text{ mod }d\right)}\right\rangle\big|l\in\mathbb{Z}_d\right\}$,  
and measure each of the remaining $n-1$ qudits  
in the bases  
$\scriptstyle\left\{\left|u_{q}^{\left(h_i\text{ mod }d\right)}\right\rangle\big|q\in\mathbb{Z}_d\right\}$  
for $i\in\mathbb{Z}_n, i\neq i'$.  
Let $M_i$ denote the measurement outcome for $\forall i\in\mathbb{Z}_n$, where $M_i=j$ when $\scriptstyle\left|u_{j}^{(l)}\right\rangle$ is observed.  
The test passes if $\sum_{i=0}^{n-1}M_i=0$. In summary, the set $\scriptstyle S_{Test}^{GHZ_n^d}$ can be divided into $(d-1)d^{n-1}+1$ subsets, where  
\begin{align}
\normalsize
     &S_{Test_0}^{{GHZ}_n^d}=\left\{ g_{(0,h_1,h_2,\cdots,h_{n-1})}^{{GHZ}_n^d}\big|h_1,h_2,\cdots,h_{n-1}\in \mathbb{Z}_d,\ (0,h_1,h_2,\cdots,h_{n-1})\neq(0,\cdots,0)\right\},\\
     &S_{Test_{1+(h_0-1)d^{n-1}+\sum_{l=1}^{n-1}h_l\times d^{n-1-l}}}^{{GHZ}_n^d}=\left\{ g_{(h_0,h_1,h_2,\cdots,h_{n-1})}^{{GHZ}_n^d}\right\},h_0,h_1,h_2,\cdots,h_{n-1}\in \mathbb{Z}_d,h_0\neq0,
\end{align}
\begin{align}
     &C_{Test_0}^{{GHZ}_n^d}=\left\{(0,h_1,h_2,\cdots,h_{n-1})\big|h_1,h_2,\cdots,h_{n-1}\in \mathbb{Z}_d,\ (0,h_1,h_2,\cdots,h_{n-1})\neq(0,\cdots,0)\right\},\\
     &C_{Test_{1+(h_0-1)d^{n-1}+\sum_{l=1}^{n-1}h_l\times d^{n-1-l}}}^{{GHZ}_n^d}=\left\{(h_0,h_1,h_2,\cdots,h_{n-1})\right\},h_0,h_1,h_2,\cdots,h_{n-1}\in \mathbb{Z}_2,h_0\neq0.
\end{align}
The subscript $(h_0-1)d^{n-1}+\sum_{l=1}^{n-1}h_l\times d^{n-1-l}$ is the decimal representation of the $d$-ary number $(h_0-1)h_1h_2\cdots h_{n-1}$, so that the sets can be labeled continuously as $0,1,\cdots,(d-1)d^{n-1}$. Moreover,  
$\scriptstyle \left|S_{Test_0}^{{GHZ}_n^d}\right|=\left|C_{Test_0}^{{GHZ}_n^d}\right|=d^{n-1}-1$,  
and  
$\scriptstyle \left|S_{Test_j}^{{GHZ}_n^d}\right|=\left|C_{Test_j}^{{GHZ}_n^d}\right|=1$  
for $j=1,2,\cdots,(d-1)d^{n-1}$.  
Thus, the verification operator can be constructed as  
\begin{equation}
    \normalsize
    \begin{aligned}
   \Omega_{{GHZ}_n^d}=&\left|{GHZ}_n^d\right\rangle\left\langle {GHZ}_n^d\right|\\
    &+\sum_{\substack{j_0,\cdots,j_{n-1}=0,\\(j_0,\cdots,j_{n-1})\neq(0,\cdots,0)}}^{d-1}\left(\sum_{i=0}^{(d-1)d^{n-1}}\frac{\mu_i}{\left|S_{Test_i}^{{GHZ}_n^d}\right|}\left(\sum_{\substack{h_0,\cdots,h_{n-1}:\ \sum_{k=0}^{n-1}h_k\times j_k\text{ mod }d=0\\
    (h_0,\cdots, h_{n-1})\in C_{Test_i}^{{GHZ}_n^d}}}1\right)\right)U_{{GHZ}_n^d}\left(\bigotimes_{k=0}^{n-1}\left| j_k \right\rangle \left\langle j_k\right|\right)U_{{GHZ}_n^d}^\dagger\\
    =&\left|{GHZ}_n^d\right\rangle\left\langle {GHZ}_n^d\right|+\mu_0\sum_{j_0=1}^{d-1}U_{{GHZ}_n^d}\left| j_000\cdots0 \right\rangle \left\langle j_000\cdots0\right|U_{{GHZ}_n^d}^\dagger\\
    &+\sum_{\substack{j_0,\cdots,j_{n-1}=0,\\(j_1,\cdots,j_{n-1})\neq(0,\cdots,0)}}^{d-1}\left[\frac{\mu_0}{d^{n-1}-1}\left(\sum_{\substack{h_1,\cdots,h_{n-1}:\ \sum_{k=1}^{n-1}h_k\times j_k\text{ mod }d=0\\
    (h_1,\cdots,h_{n-1})\neq(0,\cdots,0)}}1\right)\right.\\
    &\left.+\sum_{i=1}^{(d-1)d^{n-1}}\mu_i\left(\sum_{\substack{h_0,\cdots,h_{n-1}:\sum_{k=0}^{n-1}h_k\times j_k\text{ mod }d=0\\
    h_0\neq0,1+(h_0-1)d^{n-1}+\sum_{l=1}^{n-1}h_l\times d^{n-1-l}=i}}1\right)\right]U_{{GHZ}_n^d}\left(\bigotimes_{k=0}^{n-1}\left| j_k \right\rangle \left\langle j_k\right|\right)U_{{GHZ}_n^d}^\dagger,
    \end{aligned} 
\end{equation}
where $\sum_{i=0}^{d^{n-1}}\mu_i=1$. When $j_1,\cdots,j_{n-1}$ satisfy that exactly one $j_{l'}\neq0$ ($l'\in\mathbb{Z}_{n}^*$) and $h_{l'}=0$, we obtain  
\begin{equation}
\normalsize
    \max_{\substack{j_0,\cdots,j_{n-1}\in\mathbb{Z}_d,\\(j_1,\cdots,j_{n-1})\neq(0,\cdots,0)}}\left\{\frac{\mu_0}{d^{n-1}-1}\left(\sum_{\substack{h_1,\cdots,h_{n-1}:\ \sum_{k=1}^{n-1}h_k\times j_k\text{ mod }d=0\\
    (h_1,\cdots,h_{n-1})\neq(0,\cdots,0)}}1\right)\right\}=\frac{(d^{n-2}-1)\mu_0}{d^{n-1}-1},
\end{equation}
and if $j_0=0$,  
\begin{equation}
\normalsize
    \sum_{i=1}^{(d-1)d^{n-1}}\mu_i\left(\sum_{\substack{h_1,\cdots,h_{n-1}:\ \sum_{k=0}^{n-1}h_k\times j_k\text{ mod }d=0\\
   h_0\neq0,1+(h_0-1)d^{n-1}+\sum_{l=1}^{n-1}h_l\times d^{n-1-l}=i}}1\right)=\sum_{h_l=0,l\in\mathbb{Z}_{n}^*,\ l\neq l'}^{d-1}\left(\sum_{h_0=1}^{d-1}\mu_{1+(h_0-1)d^{n-1}+\sum_{l=1}^{n-1}h_l\times d^{n-1-l}}\right),
\end{equation}
which contains the largest number of $\mu_i$,  
a total of $(d-1)d^{n-2}$ terms for $l\in\mathbb{Z}_n^*,\ l\neq l'$, $h_l\in\mathbb{Z}_d$, and $h_0\in\mathbb{Z}_d^*$. Since $\sum_{i=0}^{(d-1)d^{n-1}}\mu_i=1$,  
we take $\mu_0=\frac{d^{n-1}-1}{d^{n}-1}$ and $\mu_i=\frac{1}{d^{n}-1}$ for $i\in\mathbb{Z}_{(d-1)d^{n-1}}^*$.  
Then  
\begin{equation}
\normalsize
    \begin{aligned}
    \min\{\beta(\Omega_{GHZ_n^d})\}&=\mu_0=\frac{(d^{n-2}-1)\mu_0}{d^{n-1}-1}+\sum_{h_l=0,l\in\mathbb{Z}_{n}^*,\ l\neq l'}^{d-1}\mu_{1+\sum_{l=1,\ l\neq l'}^{n-1}h_l\times d^{l-1}}\\
    &=\frac{d^{n-2}-1}{d^{n-1}-1}\cdot\frac{d^{n-1}-1}{d^{n}-1}+\frac{(d-1)d^{n-2}}{d^{n}-1}=\frac{d^{n-1}-1}{d^{n}-1}.
    \end{aligned}
\end{equation}
Thus,  
$\nu\!\left(\Omega_{GHZ_n^d}\right)=\frac{(d-1)d^{n-1}}{d^{n}-1}$.  
The number of measurements required for verifying $\left|GHZ_n^d\right\rangle$ is therefore  
\begin{equation}
\normalsize
n_{opt}^{GHZ_n^d}
=\left\lceil \frac{\ln\delta}{\ln\!\left[1-\frac{(d-1)d^{n-1}}{d^{n}-1}\epsilon\right]}\right\rceil
\leq\left\lceil \frac{d^{n}-1}{(d-1)d^{n-1}}\epsilon^{-1}\ln\delta^{-1}\right\rceil
=\left\lceil \!\left(1+\frac{1}{d-1}-\frac{1}{(d-1)d^{n-1}}\right)\epsilon^{-1}\ln\delta^{-1}\!\right\rceil
<\left\lceil 2\epsilon^{-1}\ln\delta^{-1}\right\rceil. 
\end{equation}
This verification efficiency is comparable to the theoretically optimal value reported in Ref.~\cite{li2020optimal},  
where $N(\Omega_{\text{III}})=\frac{d+1}{d}\epsilon^{-1}\ln\delta^{-1}$,  
both achieving $O(1)$ complexity. It is straightforward to see that when $n=2$, the above verification procedure for the qudit GHZ state reduces to that for the qudit Bell state, in which case the required number of measurements satisfies $n_{opt}^{{GHZ}_2^d}\leq \frac{d+1}{d}\epsilon^{-1}\ln\delta^{-1}$.

\subsubsection{E2.4: Verification of qudit GHZ-like and Bell-like states}

We expand several elements of the generalized stabilizer group for the state $\scriptstyle\left|\widetilde{GHZ}_n^d\right\rangle$ as  
\begin{equation}
\normalsize
    \begin{aligned}
  g_{(0,h_1,\cdots,h_{n-1})}^{(\widetilde{GHZ}_n^d)}
    =&Z_0^{\sum_{k=1}^{n-1}h_k}\prod_{k=1}^{n-1}Z_k^{h_k}
    =\left(\sum_{j_0=0}^{d-1}\omega_d^{j_0\cdot\sum_{k=1}^{n-1}h_k}|j_0\rangle\langle j_0|\right)\prod_{k=1}^{n-1}\left(\sum_{j_k=0}^{d-1}\omega_d^{j_k\cdot h_k}|j_k\rangle\langle j_k|\right),\\
   g_{(h_0,0,0,\cdots,0)}^{(\widetilde{GHZ}_n^d)}
    =&\sum_{a,b}^{d-1}(B^{h_0})_{a,b}(|a\rangle\langle b|)\otimes \prod_{j=1}^{n-1}X_j^{a-b}\\
    =&\sum_{a,b=0}^{d-1}\sum_{j_1,\cdots,j_{n-1}=0}^{d-1}(B^{h_0})_{a,b}\omega_d^{-(a-b)\sum_{k=1}^{n-1}j_k}(|a\rangle\langle b|)\otimes\bigotimes_{k=1}^{n-1}|x_{j_k}\rangle\langle x_{j_k}|\\
    =&\sum_{j_0,j_1,\cdots,j_{n-1}=0}^{d-1}\omega_d^{h_0\cdot j_0}\left|u_{j_0}^{(h_0,\sum_{k=1}^{n-1}j_k)}\right\rangle\left\langle u_{j_0}^{(h_0,\sum_{k=1}^{n-1}j_k)}\right|\otimes\bigotimes_{k=1}^{n-1}|x_{j_k}\rangle\langle x_{j_k}|,
    \end{aligned}
\end{equation}
where $(h_0,0,\cdots,0)$ and $(0,h_1,\cdots,h_{n-1})\neq(0,\cdots,0)$,  
and $|x_{j_k}\rangle=\frac{1}{\sqrt{d}}\sum_{i=0}^{d-1}\omega_d^{j_k\cdot i}|i\rangle$  
is the eigenstate of the qudit $X$ gate with eigenvalue $\omega_d^{-j_k}$. Moreover,  
\begin{equation}
\normalsize
   \left|u_{j_0}^{(h_0,\sum_{k=1}^{n-1}j_k)}\right\rangle=\sum_{a=0}^{d-1}\omega_d^{-a\cdot\sum_{k=1}^{n-1}j_k}|a\rangle\langle a|\left[\prod_{i=0}^{d-2}R_{i,i+1}(\theta_i)\right]|j_0\rangle
\end{equation}
is the eigenstate with eigenvalue $\omega_d^{h_0\cdot j_0}$ of 
\begin{equation}
\normalsize
\begin{aligned}
     &\sum_{a,b=0}^{d-1}(B^{h_0})_{a,b}\omega_d^{-(a-b)\sum_{k=1}^{n-1}j_k}(|a\rangle\langle b|)=\sum_{a=0}^{d-1}\omega_d^{-a\cdot\sum_{k=1}^{n-1}j_k}|a\rangle\langle a|\cdot B^{h_0}\cdot \left(\sum_{a=0}^{d-1}\omega_d^{-a\cdot\sum_{k=1}^{n-1}j_k}|a\rangle\langle a|\right)^\dagger\\
     =&\sum_{a=0}^{d-1}\omega_d^{-a\cdot\sum_{k=1}^{n-1}j_k}|a\rangle\langle a|\cdot \left[\prod_{i=0}^{d-2}R_{i,i+1}(\theta_i)\right]\cdot Z^{h_0}\cdot\left[\prod_{i=0}^{d-2}R_{i,i+1}(\theta_i)\right]^\dagger\cdot \left(\sum_{a=0}^{d-1}\omega_d^{-a\cdot\sum_{k=1}^{n-1}j_k}|a\rangle\langle a|\right)^\dagger\\
     =&\sum_{a=0}^{d-1}\omega_d^{-a\cdot\sum_{k=1}^{n-1}j_k}|a\rangle\langle a|\cdot \left[\prod_{i=0}^{d-2}R_{i,i+1}(\theta_i)\right]\cdot \left(\sum_{j_0=0}^{d-1}\omega_d^{h_0\cdot j_0}|j_0\rangle\langle j_0|\right)\cdot\left[\prod_{i=0}^{d-2}R_{i,i+1}(\theta_i)\right]^\dagger\cdot \left(\sum_{a=0}^{d-1}\omega_d^{-a\cdot\sum_{k=1}^{n-1}j_k}|a\rangle\langle a|\right)^\dagger,
\end{aligned}
\end{equation}
for fixed values of $j_1,\cdots,j_{n-1}$. 
For $\scriptstyle g_{(h_0,0,0,\cdots,0)}^{(\widetilde{GHZ}_n^d)}$ with $h_0=1,2,\cdots,d-1$,  
the $d-1$ elements can be tested simultaneously by a single measurement.  
The testing procedure is as follows.  
First, measure any $n-1$ qudits of $\scriptstyle\left|\widetilde{GHZ}_n^d\right\rangle$  
in the Fourier basis  
$\{|x_j\rangle=\frac{1}{\sqrt{d}}\sum_{l}\omega_d^{j\cdot l}|l\rangle,\, l\in\mathbb{Z}_d\}$.  
Let $M_i$ denotes the measurement outcome of the $i$-th qudit.  
If the outcome is $|x_j\rangle$, record $M_i=j$.  
Denote by $i'$ the position of the unmeasured qudit.  
Compute $\sum_{i=0,i\neq i'}^{n-1}M_i$,  
and measure the remaining qudit in the basis  
$\scriptstyle\left\{\left|u_{l}^{(h_0,\sum_{i=0,i\neq i'}^{n-1}M_i)}\right\rangle\big|l\in \mathbb{Z}_d\right\}$.  
If the measurement outcome is $M_{i'}=0$, the test passes. For  
$\normalsize g_{(0,h_1,\cdots,h_{n-1})}^{(\widetilde{GHZ}_n^d)}=g_{(0,h_1,\cdots,h_{n-1})}^{({GHZ}_n^d)}$  
with $(0,h_1,h_2,\cdots,h_{n-1})\neq(0,\cdots,0)$,  
a total of $d^{n-1}-1$ operators can be tested simultaneously by a single measurement,  
and the testing process is identical to that for $g_{(0,h_1,\cdots,h_{n-1})}^{({GHZ}_n^d)}$,  
which is not repeated here. Therefore, the set of testable generalized stabilizers can be divided into two subsets,  
$\scriptstyle S_{Test}^{\widetilde{GHZ}_n^d}=S_{Test_0}^{\widetilde{GHZ}_n^d}\bigcup S_{Test_1}^{\widetilde{GHZ}_n^d}$,  
where  
$\scriptstyle S_{Test_1}^{\widetilde{GHZ}_n^d}=\left\{g_{(h_0,0,0,\cdots,0)}^{\widetilde{GHZ}_n^d}\big|h_0\in\mathbb{Z}_d^*\right\}$,  
$\scriptstyle C_{Test_1}^{\widetilde{GHZ}_n^d}=\{(h_0,0,0,\cdots,0)\,|\,h_0\in\mathbb{Z}_d^*\}$, 
\begin{align}
\normalsize
       S_{Test_0}^{\widetilde{GHZ}_n^d}&=\left\{ g_{(0,h_1,h_2,\cdots,h_{n-1})}^{\widetilde{GHZ}_n^d}\big|h_1,h_2,\cdots,h_{n-1}\in \mathbb{Z}_d,\ (0,h_1,h_2,\cdots,h_{n-1})\neq(0,\cdots,0)\right\},\\
       C_{Test_0}^{\widetilde{GHZ}_n^d}&=\left\{ (0,h_1,h_2,\cdots,h_{n-1})\big|h_1,h_2,\cdots,h_{n-1}\in \mathbb{Z}_d,\ (0,h_1,h_2,\cdots,h_{n-1})\neq(0,\cdots,0)\right\}.
\end{align}
Here,  
$\scriptstyle\left|S_{Test_1}^{\widetilde{GHZ}_n^2}\right|=\left|C_{Test_1}^{\widetilde{GHZ}_n^2}\right|=d-1$ and  
$\scriptstyle\left|S_{Test_0}^{\widetilde{GHZ}_n^2}\right|=\left|C_{Test_0}^{\widetilde{GHZ}_n^2}\right|=d^{n-1}-1$.  
Hence, the verification operator can be constructed as  
\begin{equation}
    \fontsize{8}{10}
    \hspace{-5mm}
    \begin{aligned}
&\Omega_{\widetilde{GHZ}_n^d}=\left|\widetilde{GHZ}_n^d\right\rangle\left\langle \widetilde{GHZ}_n^d\right|\\
    &+\sum_{\substack{j_0,\cdots,j_{n-1}=0,\\(j_0,\cdots,j_{n-1})\neq(0,\cdots,0)}}^{d-1}\left(\sum_{i=0}^{1}\frac{\mu_i}{\left|S_{Test_i}^{\widetilde{GHZ}_n^d}\right|}\left(\sum_{\substack{h_0,\cdots,h_{n-1}:\ \sum_{k=0}^{n-1}h_k\times j_k\text{ mod }d=0\\
    (h_0,\cdots, h_{n-1})\in C_{Test_i}^{\widetilde{GHZ}_n^d}}}1\right)\right)U_{\widetilde{GHZ}_n^d}\left(\bigotimes_{k=0}^{n-1}\left| j_k \right\rangle \left\langle j_k\right|\right)U_{\widetilde{GHZ}_n^d}^\dagger\\
    &=\left|\widetilde{GHZ}_n^d\right\rangle\left\langle \widetilde{GHZ}_n^d\right|+\mu_0\sum_{j_0=1}^{d-1}U_{\widetilde{GHZ}_n^d}\left| j_000\cdots0 \right\rangle \left\langle 100\cdots0\right|U_{\widetilde{GHZ}_n^d}^\dagger\\
    &+\sum_{\substack{j_0,\cdots,j_{n-1}=0,\\(j_1,\cdots,j_{n-1})\neq(0,\cdots,0)}}^{d-1}\left[\frac{\mu_0}{2^{n-1}-1}\left(\sum_{\substack{h_1,\cdots,h_{n-1}:\ \sum_{k=1}^{n-1}h_k\times j_k\text{ mod }d=0\\
    (h_1,\cdots, h_{n-1})\neq(0,\cdots,0)}}1\right)+\frac{\mu_1}{d-1}\left(\sum_{\substack{h_0=1\\h_0\times j_0\text{ mod }d=0}}^{d-1}1\right)\right]U_{\widetilde{GHZ}_n^d}\left(\bigotimes_{k=0}^{n-1}\left| j_k \right\rangle \left\langle j_k\right|\right)U_{\widetilde{GHZ}_n^d}^\dagger.
    \end{aligned} 
\end{equation}
When $j_0=0$ and among $j_1,\cdots,j_{n-1}$ only one satisfies $j_{l'}\neq0$ ($l'\in\mathbb{Z}_{n}^*$) and $h_{l'}=0$,  
we can simultaneously obtain  
\begin{equation}
\fontsize{8}{10}
\hspace{-8mm}
\begin{aligned}
   \max_{\substack{j_0,\cdots,j_{n-1}\in\mathbb{Z}_d,\\(j_1,\cdots,j_{n-1})\neq(0,\cdots,0)}}\left\{\frac{\mu_0}{d^{n-1}-1}\left(\sum_{\substack{h_1,\cdots,h_{n-1}:\ \sum_{k=1}^{n-1}h_k\times j_k\text{ mod }d=0\\
    (h_1,\cdots,h_{n-1})\neq(0,\cdots,0)}}1\right)\right\}=\frac{(d^{n-2}-1)\mu_0}{d^{n-1}-1},\ 
    \max_{\substack{j_0,\cdots,j_{n-1}\in\mathbb{Z}_d,\\(j_1,\cdots,j_{n-1})\neq(0,\cdots,0)}}\left\{\frac{\mu_1}{d-1}\left(\sum_{\substack{h_1=1\\h_0\times j_0\text{ mod }d=0}}^{d-1}1\right)\right\}=\mu_1.
\end{aligned}
\end{equation}
Since $\mu_0+\mu_1=1$, we set  
$\mu_0=\frac{d^{n-1}-1}{2d^{n-1}-d^{n-2}-1}$ and  
$\mu_1=\frac{(d-1)d^{n-2}}{2d^{n-1}-d^{n-2}-1}$.  
Then,  
\begin{equation}
\normalsize
    \begin{aligned}
    \min\left\{\beta\left(\Omega_{\scriptstyle\widetilde{GHZ}_n^d}\right)\right\}&=\mu_0=\frac{(d^{n-2}-1)\mu_0}{d^{n-1}-1}+\mu_1\\
    &=\frac{d^{n-2}-1}{d^{n-1}-1}\cdot\frac{d^{n-1}-1}{2d^{n-1}-d^{n-2}-1}+\frac{(d-1)d^{n-2}}{2d^{n-1}-d^{n-2}-1}=\frac{d^{n-1}-1}{2d^{n-1}-d^{n-2}-1}.
    \end{aligned}
\end{equation}
Thus,  
$\nu(\Omega_{\widetilde{GHZ}_n^d})=\frac{(d-1)d^{n-2}}{2d^{n-1}-d^{n-2}-1}$.  
The number of measurements required for verifying  
$\left|\widetilde{GHZ}_n^d\right\rangle$  
is given by  
\begin{equation}
\hspace{-4mm}
\normalsize
n_{opt}^{\widetilde{GHZ}_n^d}
=\left\lceil \frac{\ln\delta}{\ln\!\left[1-\frac{(d-1)d^{n-2}}{2d^{n-1}-d^{n-2}-1}\epsilon\right]}\right\rceil
\leq\left\lceil \frac{2d^{n-1}-d^{n-2}-1}{(d-1)d^{n-2}}\epsilon^{-1}\ln\delta^{-1}\right\rceil
=\left\lceil \!\left(3-\frac{(d-2)d^{n-2}+1}{(d-1)d^{n-2}}\right)\epsilon^{-1}\ln\delta^{-1}\!\right\rceil
<\left\lceil 3\epsilon^{-1}\ln\delta^{-1}\right\rceil.
\end{equation}
For $d\geq3$,  
this result is comparable to that of Ref.~\cite{li2020optimal},  
where $N(\Omega_{\text{IV}})=2\epsilon^{-1}\ln\delta^{-1}$,  
both achieving $O(1)$ complexity. It is straightforward to see that when $n=2$, the above verification procedure for the qudit GHZ-like state reduces to that for the qudit Bell-like state, in which case the required number of measurements satisfies $n_{opt}^{\widetilde{GHZ}_2^d}\leq 2\epsilon^{-1}\ln\delta^{-1}$.

\subsection{E3: Verification of $\left|G_n^d\right\rangle$, $\left|\widetilde{G}_n^d\right\rangle$, $\left|\widehat{G}_n^d\right\rangle$, and $\scriptstyle\left|\widehat{\widetilde{G}}_n^d\right\rangle$}

A series of studies have been conducted on the verification of graph states and hypergraph states.  
Recently, we have proposed \textit{multigraph states} and \textit{multihypergraph states} \cite{zhang2024quantum} based on these two families.  
Here, by applying Theorem in main text, we present a unified verification framework for these four types of quantum states. Before proceeding, we briefly review the notions of \textbf{independent cover} and \textbf{graph coloring} in graph theory. In any undirected graph $G=(V,E)$, which can be a simple graph, multigraph, hypergraph, or multihypergraph, an independent set $A_i=\{v_{(i,0)},v_{(i,1)},\cdots,v_{(i,p_i-1)}\}$ (where $p_i$ is the number of vertices in $A_i$)  
is a subset of the vertex set $V=\mathbb{Z}_n$ such that no two vertices in $A_i$ are connected by an edge.  
If $\bigcup_{i=0}^{k-1}A_i=V$, then $\{A_0,\cdots,A_{k-1}\}$ is called an \textbf{independent cover} of $G$.  \textbf{Graph coloring} of a graph assigns a color to each vertex such that adjacent vertices have distinct colors.  
The smallest number of colors required is the \textit{chromatic number} $\chi(G)$.  
In terms of independent sets, this can be described as  
a collection of pairwise disjoint independent sets $\{A_0,\cdots,A_{k-1}\}$  
forming a cover of $V$, where each independent set is assigned a distinct color.  
The minimum number of such sets gives  
$\chi(G)=\min\bigg\{k:\bigcup_{i=0}^{k-1}A_i=V,\ \forall i\neq j,\ A_i\cap A_j=\varnothing\bigg\}$.

Since qubit graph and hypergraph states can be obtained by setting $d=2$  
in the qudit definitions of graph and hypergraph states,  
their verification procedures are included within our qudit verification framework (also with $d=2$).  
On the other hand, multigraph and multihypergraph states are defined only for qudit systems~\cite{zhang2024quantum}.  
Therefore, in the following, we focus exclusively on the verification of qudit states. By expanding the generators of the generalized stabilizer group  
for $\left|G_n^d\right\rangle$, $\left|\widetilde{G}_n^d\right\rangle$,  
$\left|\widehat{G}_n^d\right\rangle$, and $\scriptstyle\left|\widehat{\widetilde{G}}_n^d\right\rangle$, we obtain  
\begin{equation}
\hspace{-5mm}
\fontsize{8}{10}
    \begin{aligned}
    \left(g_k^{(G_n^d)}\right)^{h_k}&=\left( \prod_{e\in E} {CZ}_e^{m_e}\right) X_k^{h_k}\left(\prod_{e' \in  E}{CZ}_{e'}^{{d-m}_{e'}}\right)= X_k^{h_k}\prod_{e \in E,k\in e}{CZ}_{e \backslash \left\{k\right\}}^{m_e\cdot h_k}=X_k^{h_k}\prod_{e \in E,k\in e}{Z}_{e \backslash \left\{k\right\}}^{m_e\cdot h_k}\\[-1mm]
    &=\sum_{j_k=0}^{d-1}\omega_d^{j_k\cdot h_k}|x_{j_k}\rangle\langle x_{j_k}|\otimes\prod_{e \in E,k\in e}\left(\sum_{j_{e\backslash\{k\}}=0}^{d-1}\omega_d^{m_e\cdot h_k\cdot j_{e\backslash\{k\}}}\left|j_{e\backslash\{k\}}\right\rangle\left\langle j_{e\backslash\{k\}}\right|\right),\\[-1mm]
    \left(g_k^{(\widetilde{G}_n^d)}\right)^{h_k}&=\left( \prod_{e\in\widetilde E} {{\widetilde{CZ}}_e}^{m_e}\right) X_k^{h_k}\left(\prod_{e' \in \widetilde E}{{\widetilde{CZ}}_{e'}}^{{d-m}_{e'}}\right)=X_k^{h_k}\prod_{e \in\widetilde E,k\in e}{{\widetilde{CZ}}_{e \backslash \left\{k\right\}}}^{m_e\cdot h_k}\\[-1mm]
    &=\sum_{j_k=0}^{d-1}\omega_d^{j_k\cdot h_k}|x_{j_k}\rangle\langle x_{j_k}|\otimes\prod_{e \in E,k\in e}\left(\sum_{t\in e\backslash\{k\} ,j_{t}=0}^{d-1}\omega_d^{m_e\cdot h_k\cdot\left(\sum_{t\in e\backslash\{k\}}j_{t}\right)}\bigotimes_{t\in e\backslash\{k\}}\left|j_{t}\right\rangle\left\langle j_{t}\right|\right),\\[-1mm]
    \left(g_k^{(\widehat{G}_n^d)}\right)^{h_k}&=\left(\prod_{{\dot{e}}\in\widehat E} {{\widehat{CZ}}_{\dot{e}}}^{m_{\dot{e}}}\right) X_k^{h_k}\left(\prod_{{\dot{e}}' \in \widehat E}{{\widehat{CZ}}_{{\dot{e}}'}}^{{d-m}_{{\dot{e}}'}}\right)=X_k^{h_k}\prod_{{\dot{e}} \in\widehat E,k\in V_{\dot{e}}}\left[\sum_{j_k=0}^{d-1}\left|j_k\right\rangle\left\langle j_k\right|\otimes {\left({^{s_{\{V_{\dot{e}}\backslash{k}\}}}Z}_{\{V_{\dot{e}}\backslash{k}\}}\right)}^{m_{\dot{e}}\cdot\big(\sum_{l=0}^{s_{k}-1} \binom{s_{k}}{l}\cdot{(j_k)}^l\cdot{(h_{k})}^{s_k-l}\big)}\right]\\[-1mm]
    &=X_k^{h_k}\sum_{j_k=0}^{d-1}\left(\left|j_k\right\rangle\left\langle j_k\right|\otimes\prod_{\dot{e} \in\widehat E,k\in V_{\dot{e}}}\sum_{\substack{j_{V_{\dot{e}}\backslash\{k\}}=0}}^{d-1}\omega_d^{m_{\dot{e}}\cdot\big(\sum_{l=0}^{s_{k}-1} \binom{s_{k}}{l}\cdot{(j_k)}^l\cdot{(h_{k})}^{s_k-l}\big)\cdot \left(j_{V_{\dot{e}}\backslash\{k\}}\right)^{s_{V_{\dot{e}}\backslash\{k\}}}}\left|j_{V_{\dot{e}}\backslash\{k\}}\right\rangle\left\langle j_{V_{\dot{e}}\backslash\{k\}}\right|\right)\\[-1mm]
    &=\sum_{\substack{j_k,j_{V_{\dot{e}}\backslash\{k\}}=0\\ \dot{e} \in\widehat E,k\in V_{\dot{e}}}}^{d-1}\left(\omega_d^{j_k}\left|\kappa_{j_k}^{\left(m_{\dot{e}}\cdot\sum_{\dot{e} \in\widehat E,k\in V_{\dot{e}}}\left(j_{V_{\dot{e}}\backslash\{k\}}\right)^{s_{V_{\dot{e}}\backslash\{k\}}}\right)}\right\rangle\left\langle \kappa_{j_k}^{\left(m_{\dot{e}}\cdot\sum_{\dot{e} \in\widehat E,k\in V_{\dot{e}}}\left(j_{V_{\dot{e}}\backslash\{k\}}\right)^{s_{V_{\dot{e}}\backslash\{k\}}}\right)}\right|\otimes\prod_{\dot{e} \in\widehat E,k\in V_{\dot{e}}}\left|j_{V_{\dot{e}}\backslash\{k\}}\right\rangle\left\langle j_{V_{\dot{e}}\backslash\{k\}}\right|\right),\\[-1mm]
    \left(g_k^{(\widehat{\widetilde{G}}_n^d)}\right)^{h_k}&=\left(\prod_{{\dot{e}}\in \widehat{\widetilde E}} {{\widehat{\widetilde{CZ}}}_{\dot{e}}}^{m_{\dot{e}}}\right) X_k^{h_k}\left(\prod_{{\dot{e}}' \in \widehat{\widetilde E}}{{\widehat{\widetilde{CZ}}}_{{\dot{e}}'}}^{{d-m}_{{\dot{e}}'}}\right)=X_k^{h_k}\prod_{{\dot{e}}\in \widehat{\widetilde E},k\in V_{\dot{e}}}\left(\sum_{j_k=0}^{d-1}\left|j_k\right\rangle\left\langle j_k\right|\otimes {{\widehat{\widetilde{CZ}}}_{(V_{\dot{e}}\backslash \{k\}\vert S_{\dot{e}}\backslash \{s_k\})}}^{m_{\dot{e}}\cdot\big(\sum_{l=0}^{s_{k}-1} \binom{s_{k}}{l}\cdot{(j_k)}^l\cdot{(h_{k})}^{s_k-l}\big)}\right)\\[-1mm]
    &=X_k^{h_k}\prod_{{\dot{e}}\in \widehat{\widetilde E},k\in V_{\dot{e}}}\left(\sum_{j_k=0}^{d-1}\left|j_k\right\rangle\left\langle j_k\right|\otimes\sum_{j_{V_{\dot{e}}\backslash\{k\}}=0}^{d-1}\omega_d^{m_{\dot{e}}\cdot\big(\sum_{l=0}^{s_{k}-1} \binom{s_{k}}{l}\cdot{(j_k)}^l\cdot{(h_{k})}^{s_k-l}\big)\cdot\left(\sum_{t\in V_{\dot{e}}\backslash\{k\}}(j_{t})^{s_{t}}\right)}\bigotimes_{t\in V_{\dot{e}}\backslash\{k\}}\left|j_{t}\right\rangle\left\langle j_{t}\right|\right)\\[-1mm]
    &=X_k^{h_k}\sum_{j_k=0}^{d-1}\left(\left|j_k\right\rangle\left\langle j_k\right|\otimes\sum_{\substack{j_{V_{\dot{e}}\backslash\{k\}}=0\\ \dot{e} \in\widehat{\widetilde E},k\in V_{\dot{e}}}}^{d-1}\omega_d^{m_{\dot{e}}\cdot\big(\sum_{l=0}^{s_{k}-1} \binom{s_{k}}{l}\cdot{(j_k)}^l\cdot{(h_{k})}^{s_k-l}\big)\cdot \sum_{\dot{e} \in\widehat{\widetilde E},k\in V_{\dot{e}}}\left(\sum_{t\in V_{\dot{e}}\backslash\{k\}}(j_{t})^{s_{t}}\right)}\prod_{\dot{e} \in\widehat{\widetilde E},k\in V_{\dot{e}}}\bigotimes_{t\in V_{\dot{e}}\backslash\{k\}}\left|j_{t}\right\rangle\left\langle j_{t}\right|\right)\\[-1mm]
     &=\sum_{\substack{j_t=0,t\in V_{\dot{e}}=0\\ \dot{e} \in\widehat{\widetilde E},k\in V_{\dot{e}}}}^{d-1}\left(\omega_d^{j_k}\left|\kappa_{j_k}^{\left(m_{\dot{e}}\cdot\sum_{\dot{e} \in\widehat{\widetilde E},k\in V_{\dot{e}}}\left(\sum_{t\in {\dot{e}}\backslash\{k\}}(j_{t})^{s_{t}}\right)\right)}\right\rangle\left\langle \kappa_{j_k}^{\left(m_{\dot{e}}\cdot\sum_{\dot{e} \in\widehat{\widetilde E},k\in V_{\dot{e}}}\left(\sum_{t\in V_{\dot{e}}\backslash\{k\}}(j_t)^{s_{t}}\right)\right)}\right|\otimes\prod_{\dot{e} \in\widehat{\widetilde E},k\in V_{\dot{e}}}\bigotimes_{t\in V_{\dot{e}}\backslash\{k\}}\left|j_{t}\right\rangle\left\langle j_{t}\right|\right),
    \end{aligned}
\end{equation}
where $h_k\neq0$, and  
$\left|\kappa_{j_k}^{\left(q\right)}\right\rangle
=\frac{1}{\sqrt{d}}\sum_{i=0}^{d-1}
\omega_d^{-j_k\cdot i+q\cdot (h_k)^{s_k}\cdot (i)^{s_k}}
|h_k\cdot i\rangle$
is the eigenstate of   
$T_q=X^{h_k}\sum_{i=0}^{d-1}\omega_d^{q\cdot \sum_{l=0}^{s_k-1}\binom{s_{k}}{l}i^l h_k^{s_k-l}}|i\rangle\langle i|$
corresponding to the eigenvalue $\omega_k^{j_k}$, i.e.,
\begin{equation}
    \begin{aligned}
    T_q\left|\kappa_{j_k}^{\left(q\right)}\right\rangle=&X^{h_k}\sum_{i=0}^{d-1}\omega_d^{q\cdot \sum_{l=0}^{s_k-1}\binom{s_{k}}{l}i^l h_k^{s_k-l}}|i\rangle\langle i| \frac{1}{\sqrt{d}}\sum_{i'=0}^{d-1}\omega_d^{-j_k\cdot i'+q\cdot (h_k)^{s_k}\cdot (i')^{s_k}}|h_k\cdot i'\rangle\\
       =&\frac{1}{\sqrt{d}}\sum_{i=0}^{d-1}\omega_d^{-j_k\cdot i+q(h_k\cdot i)^{s_k}+q[(h_k(i+1))^{s_k}-(h_k\cdot i)^{s_k}]}|h_k\cdot(i+1)\rangle\\
       =&\frac{1}{\sqrt{d}}\sum_{i=0}^{d-1}\omega_d^{-j_k\cdot (i+1-1)+q(h_k\cdot(i+1))^{s_k}}|h_k\cdot (i+1)\rangle\\
       =&\omega_d^{j_k}\frac{1}{\sqrt{d}}\sum_{i+1=0}^{d-1}\omega_d^{-j_k\cdot (i+1)+q(h_k\cdot (i+1))^{s_k}}|h_k\cdot (i+1)\rangle=\omega_d^{j_k}\left|\kappa_{j_k}^{\left(q\right)}\right\rangle.
    \end{aligned}
\end{equation}

The generalized stabilizers $\scriptstyle\left(g_k^{(G_n^d)}\right)^{h_k}$, $\scriptstyle\left(g_k^{(\widetilde{G}_n^d)}\right)^{h_k}$, $\scriptstyle\left(g_k^{(\widehat{G}_n^d)}\right)^{h_k}$, and $\scriptstyle\left(g_k^{(\widehat{\widetilde{G}}_n^d)}\right)^{h_k}$ share similar structures, and their testing procedures are correspondingly similar. The common part of the procedure is to measure all neighboring vertices (particles) of vertex $k$ in the computational basis $\{|0\rangle,|1\rangle,\cdots,|d-1\rangle\}$. The measurement outcome of the $i$-th vertex is denoted as $M_i$. The differences among these procedures are as follows.  

- For $\scriptstyle\left(g_k^{(G_n^d)}\right)^{h_k}$, vertex $k$ of $\left|G_n^d\right\rangle$ is measured in the Fourier basis  
  $\{|x_j\rangle=\frac{1}{\sqrt{d}}\sum_{l}\omega_d^{j\cdot l}|l\rangle,\,l\in\mathbb{Z}_d\}$,  
  with the measurement result recorded as $M_k$.  
  The test passes if  
  \begin{equation}
     h_k\cdot M_k+h_k\sum_{e\in E, k\in e}m_eM_{e\backslash\{k\}}=0.
  \end{equation}

- For $\scriptstyle\left(g_k^{(\widetilde{G}_n^d)}\right)^{h_k}$, vertex $k$ is also measured in the Fourier basis,  
  and the test passes if  
  \begin{equation}
      h_k\cdot M_k+h_k\sum_{e\in E, k\in e}\left(m_e\sum_{t\in e\backslash\{k\}}M_{e\backslash\{k\}}\right)=0.
  \end{equation}

- For $\scriptstyle\left(g_k^{(\widehat{G}_n^d)}\right)^{h_k}$,  
  after obtaining the measurement outcomes $M_{\dot{e}\backslash{k}}$ of all neighboring vertices  
  ($\dot{e}\in\widehat E,\,k\in V_{\dot{e}}$),  
  vertex $k$ is measured in the basis  
  $\scriptstyle\left\{\left|\kappa_{j_k}^{\left(q\right)}\right\rangle \big|q=m_{\dot{e}}\cdot\sum_{\dot{e} \in\widehat E,k\in V_{\dot{e}}}\left(M_{V_{\dot{e}}\backslash\{k\}}\right)^{s_{V_{\dot{e}}\backslash\{k\}}}, j_k\in\mathbb{Z}_d\right\}$.  
  The test passes if the measurement result is $\scriptstyle\left|\kappa_{0}^{\left(q\right)}\right\rangle$.

- For $\scriptstyle\left(g_k^{(\widehat{\widetilde{G}}_n^d)}\right)^{h_k}$,  
  the measurement outcomes $M_t$ of neighboring vertices $t\in\dot{e}\backslash\{k\}$  
  ($\dot{e} \in\widehat{\widetilde E},\,k\in V_{\dot{e}}$) are first obtained.  
  Then vertex $k$ is measured in the basis  
  $\scriptstyle\left\{\left|\kappa_{j_k}^{\left(q'\right)}\right\rangle \big|q'=\left(m_{\dot{e}}\cdot\sum_{\dot{e} \in\widehat{\widetilde E},k\in V_{\dot{e}}}\left(\sum_{t\in {\dot{e}}\backslash\{k\}}(M_{t})^{s_{t}}\right)\right), j_k\in\mathbb{Z}_d\right\}$,  
  and the test passes if the result is $\scriptstyle\left|\kappa_{0}^{\left(q'\right)}\right\rangle$. 
  
Overall, for all four cases—  
$\scriptstyle\left(g_k^{(G_n^d)}\right)^{h_k}$,  
$\scriptstyle\left(g_k^{(\widetilde{G}_n^d)}\right)^{h_k}$,  
$\scriptstyle\left(g_k^{(\widehat{G}_n^d)}\right)^{h_k}$, and  
$\scriptstyle\left(g_k^{(\widehat{\widetilde{G}}_n^d)}\right)^{h_k}$—  
the testing process consists of measuring the neighboring particles of vertex $k$ in the computational basis  
and vertex $k$ itself in a specific measurement basis. Hence, for two non-adjacent vertices $k$ and $k'$,  
if they share a common neighbor, the common neighbor needs to be measured only once in the computational basis,  
and the measurement result can be reused.  
If they have no common neighbor, their testing operations are independent within a single global measurement.  
Therefore, $\scriptstyle\left(g_k^{(G_0)}\right)^{h_k}$ and $\scriptstyle\left(g_{k'}^{(G_0)}\right)^{h_{k'}}$  
can be tested simultaneously, where  
$G_0\in\scriptstyle\left\{G_n^d,\widetilde{G}_n^d,\widehat{G}_n^d,\widehat{\widetilde{G}}_n^d\right\}$. That is, for all $k\in A_i^{G_0}$,  
the stabilizers $\scriptstyle\left(g_k^{(G_0)}\right)^{h_k}$ can be jointly tested through a single measurement on $|G_0\rangle$.  
Furthermore, since the testing bases of  
$\scriptstyle\left(g_k^{(G_n^d)}\right)^{h_k}$ and $\scriptstyle\left(g_k^{(\widetilde{G}_n^d)}\right)^{h_k}$  
do not depend on $h_k$,  
while the bases for $\scriptstyle\left(g_k^{(\widehat{G}_n^d)}\right)^{h_k}$ and $\scriptstyle\left(g_k^{(\widehat{\widetilde{G}}_n^d)}\right)^{h_k}$  
do, we have  
$G_1\in\scriptstyle\left\{G_n^d,\widetilde{G}_n^d\right\}$ and  
$G_2\in\scriptstyle\left\{\widehat{G}_n^d,\widehat{\widetilde{G}}_n^d\right\}$, giving
\begin{equation}
\hspace{-5mm}
\fontsize{8}{10}
    \begin{aligned}
    S_{Test_i}^{G_1}&=\left\{g_{(h_0,h_1,h_2,\cdots,h_{n-1})}^{G_1}\big|h_{v_{(i,t_i)}}\in \mathbb{Z}_d,t_i\in\mathbb{Z}_{p_i},(h_{v_{(i,0)}},h_{v_{(i,2)}},\cdots,h_{v_{(i,p_i-1)}})\neq(0,\cdots,0),h_{v'}=0,v'\in V\backslash A_i^{G_1}\right\},
    \\S_{Test_{(\hat{i},l_{\hat{i}})}}^{G_2}&=\left\{g_{(h_0,h_1,h_2,\cdots,h_{n-1})}^{G_2}\big|h_{v_{(\hat{i},t_{\hat{i}})}}=0\text{ or }h_{v_{(\hat{i},t_{\hat{i}})}}=k_{v_{(\hat{i},t_{\hat{i}})}},t_{\hat{i}}\in\mathbb{Z}_{p_{\hat{i}}},(h_{v_{(\hat{i},0)}},h_{v_{(\hat{i},1)}},\cdots,h_{v_{(\hat{i},p_{\hat{i}}-1)}})\neq(0,\cdots,0),h_{v'}=0,v'\in V \backslash A_{\hat{i}}^{G_2}\right\},
    \end{aligned}
\end{equation}
where $i\in\mathbb{Z}_{\chi(G_1)}$, ${\hat{i}}\in\mathbb{Z}_{\chi(G_2)}$,  
$t_{\hat{i}}\in\mathbb{Z}_{p_{\hat{i}}}$, $k_{v_{(\hat{i},t_{\hat{i}})}}\in \mathbb{Z}_d^*$,  
and $l_{\hat{i}}=\sum_{t_{\hat{i}}=0}^{p_{\hat{i}}-1}(k_{v_{(\hat{i},t_{\hat{i}})}}-1)(d-1)^{t_{\hat{i}}}$.  
Moreover,  
$\scriptstyle C_{Test_i}^{G_1}=\left\{(h_0,h_1,h_2,\cdots,h_{n-1})\big|g_{(h_0,\cdots,h_{n-1})}^{G_1}\in S_{Test_i}^{G_1}\right\}$ and  
$\scriptstyle C_{Test_{({\hat{i}},l_{\hat{i}})}}^{G_2}=\left\{(h_0,h_1,\cdots,h_{n-1})\big|g_{(h_0,\cdots,h_{n-1})}^{G_2}\in S_{Test_{({\hat{i}},l_{\hat{i}})}}^{G_2}\right\}$,  
with $\left|S_{Test_i}^{G_1}\right|=\left|C_{Test_i}^{G_1}\right|=d^{p_i}-1$ and  
$\left|S_{Test_{{\hat{i}},l_{\hat{i}}}}^{G_2}\right|=\left|C_{Test_{{\hat{i}},l_{\hat{i}}}}^{G_2}\right|=2^{p_{\hat{i}}}-1$. Here, $S_{Test_i}^{G_1}$ includes $\chi(G_1)$ sets for $i\in\mathbb{Z}_{\chi(G_1)}$,  
and $S_{Test_{({\hat{i}},l_{\hat{i}})}}^{G_2}$ includes  
$\sum_{{\hat{i}}=0}^{\chi(G_2)-1}(d-1)^{p_{\hat{i}}}$ sets  
for $\hat{i}\in\mathbb{Z}_{\chi(G_2)}$ and $l_{\hat{i}}\in\mathbb{Z}_{(d-1)^{p_{\hat{i}}}}$.  
Each set of stabilizers can be tested simultaneously by a single measurement.  
Therefore, according to the Theorem, the measurement operators for verifying $|G_1\rangle$ and $|G_2\rangle$ can be written as  
\begin{equation}
\normalsize
    \begin{aligned}
    \Omega_{G_1}=&|{G_1}\rangle\langle{G_1}|+\sum_{(j_0,\cdots,j_{n-1})\neq(0,\cdots,0)}\left(\sum_{i=0}^{\chi(G_1)-1}\frac{\mu_i}{|S_{Test_i}^{G_1}|}\left(\sum_{\substack{h_0,\cdots,h_{n-1}:\ \sum_{k=0}^{n-1}\frac{h_k\times j_k}{d_k}\in\mathbb{Z}\\
    (h_0,\cdots, h_{n-1})\in C_{Test_i}^{G_1}}}1\right)\right)U_\psi\left(\bigotimes_{k=0}^{n-1}\left| j_k \right\rangle \left\langle j_k\right|\right)U_\psi^\dagger\\
    =&|{G_1}\rangle\langle{G_1}|+\sum_{(j_0,\cdots,j_{n-1})\neq(0,\cdots,0)}\left(\sum_{i=0}^{\chi(G_1)-1}\frac{\mu_i}{d^{p_i}-1}\left(\sum_{\substack{h_0,\cdots,h_{n-1}:\ \sum_{k=0}^{n-1}\frac{h_k\times j_k}{d_k}\in\mathbb{Z}\\
    (h_0,\cdots, h_{n-1})\in C_{Test_i}^{G_1}}}1\right)\right)U_\psi\left(\bigotimes_{k=0}^{n-1}\left| j_k \right\rangle \left\langle j_k\right|\right)U_\psi^\dagger,\\
    \Omega_{G_2}=&|{G_2}\rangle\langle{G_2}|+\sum_{(j_0,\cdots,j_{n-1})\neq(0,\cdots,0)}\left(\sum_{i=0}^{\chi(G_2)-1}\left(\sum_{l=0}^{(d-1)^{p_{\hat{i}}}-1}\frac{\mu_{({\hat{i}},l_{\hat{i}})}}{|S_{Test_{({\hat{i}},l_{\hat{i}})}}^{G_2}|}\sum_{\substack{h_0,\cdots,h_{n-1}:\ \sum_{k=0}^{n-1}\frac{h_k\times j_k}{d_k}\in\mathbb{Z}\\
    (h_0,\cdots, h_{n-1})\in C_{Test_{({\hat{i}},l_{\hat{i}})}}^{G_2}}}1\right)\right)U_\psi\left(\bigotimes_{k=0}^{n-1}\left| j_k \right\rangle \left\langle j_k\right|\right)U_\psi^\dagger\\
    =&|{G_2}\rangle\langle{G_2}|+\sum_{(j_0,\cdots,j_{n-1})\neq(0,\cdots,0)}\left(\sum_{{\hat{i}}=0}^{\chi(G_2)-1}\left(\sum_{l_{\hat{i}}=0}^{(d-1)^{p_{\hat{i}}}-1}\frac{\mu_{({\hat{i}},l_{\hat{i}})}}{2^{p_{\hat{i}}}-1}\sum_{\substack{h_0,\cdots,h_{n-1}:\ \sum_{k=0}^{n-1}\frac{h_k\times j_k}{d_k}\in\mathbb{Z}\\
    (h_0,\cdots, h_{n-1})\in C_{Test_{({\hat{i}},l_{\hat{i}})}}^{G_2}}}1\right)\right)U_\psi\left(\bigotimes_{k=0}^{n-1}\left| j_k \right\rangle \left\langle j_k\right|\right)U_\psi^\dagger,
    \end{aligned}
\end{equation}
where $\sum_{i=0}^{\chi(G_1)-1}\mu_i=1$ and  
$\sum_{i=0}^{\chi(G_2)-1}\sum_{l_{\hat{i}}=0}^{(d-1)^{p_{\hat{i}}}-1}\mu_{({\hat{i}},l_{\hat{i}})}=1$. When $j_0,\cdots,j_{n-1}$ satisfies that only $j_{k'}\neq0$ ($k'\in\mathbb{Z}_n^*$) and $h_{k'}=0$,  
and vertex $k'$ corresponds to $v_{(i',t_{i'}')}\in A_{i'}^{G_1}$ and $v_{(\hat{i}',t_{\hat{i}'}')}\in A_{{\hat{i}}'}^{G_2}$, we obtain  
\begin{equation}
\normalsize
    \begin{aligned}
    &\sum_{i=0}^{\chi(G_1)-1}\frac{\mu_i}{d^{p_i}-1}\sum_{\substack{h_0,\cdots,h_{n-1}:\ \sum_{k=0}^{n-1}\frac{h_k\times j_k}{d_k}\in\mathbb{Z}\\
    (h_0,\cdots, h_{n-1})\in C_{Test_i}^{G_1}}}1
    =\sum_{i=0,i\neq i'}^{\chi(G_1)-1}\frac{\mu_i}{d^{p_i}-1}\sum_{\substack{h_{v_{(i,t_i)}}=0,t_i\in\mathbb{Z}_{p_i}\\
    \left(h_{v_{(i,0)}},h_{v_{(i,1)}},\cdots,h_{v_{(i,p_i-1)}}\right)\neq(0,\cdots,0)}}^{d-1}1\\
    &+\frac{\mu_{i'}}{d^{p_{i'}}-1}\sum_{\substack{h_{v_{(i',t_{i'})}}=0,t_{i'}\in\mathbb{Z}_{p_{i'}}\backslash\{t_{i'}'\}\\
    \left(h_{v_{(i',0)}},h_{v_{(i',1)}},\cdots,h_{v_{(i',p_{i'}-1)}}\right)\neq(0,\cdots,0)}}^{d-1}1=\sum_{i=0,i\neq i'}^{\chi(G_1)-1}\mu_i+\frac{(d^{p_{i'}-1}-1)\mu_{i'}}{d^{p_{i'}}-1}=1-\frac{(d-1)d^{p_{i'}-1}\mu_{i'}}{d^{p_{i'}}-1},
    \end{aligned}
\end{equation}
\begin{equation}
\normalsize
    \begin{aligned}    &\sum_{i=0}^{\chi(G_2)-1}\sum_{l_{\hat{i}}=0}^{(d-1)^{p_{\hat{i}}}-1}\frac{\mu_{({\hat{i}},l_{\hat{i}})}}{2^{p_{\hat{i}}}-1}\sum_{\substack{h_0,\cdots,h_{n-1}:\ \sum_{k=0}^{n-1}\frac{h_k\times j_k}{d_k}\in\mathbb{Z},\\
    (h_0,\cdots, h_{n-1})\in C_{Test_{({\hat{i}},l_{\hat{i}})}}^{G_2}}}1
    =\sum_{{\hat{i}}=0,{\hat{i}}\neq {\hat{i}}'}^{\chi(G_2)-1}\sum_{l_{\hat{i}}=0}^{(d-1)^{p_{\hat{i}}}-1}\frac{\mu_{({\hat{i}},l_{\hat{i}})}}{2^{p_{\hat{i}}}-1}\sum_{\substack{h_{v_{(\hat{i},t_{\hat{i}})}}=0\text{ or }h_{v_{(\hat{i},t_{\hat{i}})}}=k_{v_{(\hat{i},t_{\hat{i}})}},t_{\hat{i}}\in\mathbb{Z}_{p_{\hat{i}}},\\
    l_{\hat{i}}=\sum_{t_{\hat{i}}=0}^{p_{\hat{i}}-1}(k _{v_{t_{\hat{i}}}}-1)(d-1)^{t_{\hat{i}}},\\
    \left(h_{v_{(\hat{i},0)}},h_{v_{(\hat{i},1)}},\cdots,h_{v_{(\hat{i},p_{\hat{i}}-1)}}\right)\neq(0,\cdots,0)}}1\\[-4mm] &+\sum_{\substack{k_{v_{(\hat{i}',t_{\hat{i}'})}}=1,\\t_{\hat{i}'}\in\mathbb{Z}_{p_{\hat{i}'}}\backslash\{t_{\hat{i}'}'\}}}^{d-1}\frac{\mu_{\left(\hat{i}',\sum_{t_{\hat{i}'}=0}^{p_{\hat{i}}-1}(k _{v_{t_{\hat{i}'}}}-1)(d-1)^{t_{\hat{i}'}}\right)}}{2^{p_{\hat{i}'}}-1}\sum_{\substack{h_{v_{(\hat{i}',t_{\hat{i}})}}=0\text{ or }h_{v_{(\hat{i}',t_{\hat{i}'})}}=k_{v_{(\hat{i}',t_{\hat{i}'})}},\\
    \left(h_{v_{(\hat{i}',0)}},h_{v_{(\hat{i}',1)}},\cdots,h_{v_{(\hat{i}',p_{\hat{i}'}-1)}}\right)\neq(0,\cdots,0)}}1\\[1mm]
    &=1-\sum_{l_{\hat{i}'}=0}^{(d-1)^{p_{\hat{i}'}}-1}\mu_{({\hat{i}'},l_{\hat{i}'})}+\sum_{\substack{k_{v_{(\hat{i}',t_{\hat{i}'})}}=1,\\t_{\hat{i}'}\in\mathbb{Z}_{p_{\hat{i}'}}\backslash\{t_{\hat{i}'}'\}}}^{d-1}\frac{(2^{p_{\hat{i}'}-1}-1)\cdot\mu_{\left(\hat{i}',\sum_{t_{\hat{i}'}=0}^{p_{\hat{i}}-1}(k _{v_{t_{\hat{i}'}}}-1)(d-1)^{t_{\hat{i}'}}\right)}}{2^{p_{\hat{i}'}}-1},
    \end{aligned}
\end{equation}
which contain the largest number of $\mu_i$ and $\mu_{(\hat{i},l)}$ values, respectively.  
Hence,  
\begin{equation}
   \normalsize
   \mu_i=\frac{d-d^{1-p_i}}{d\cdot\chi(G_1)-\sum_{j=0}^{\chi(G_1)-1}d^{1-p_j}} 
\end{equation} for $i\in\mathbb{Z}_{\chi(G_1)}$,   
\begin{equation}
    \normalsize
    \mu_{(\hat{i},l_{\hat{i}})}=\frac{(2^{p_{\hat{i}}}-1)}{(d-1)^{p_{\hat{i}}}\left[(2^{p_{\hat{i}}}-1)(d-1)-(2^{p_{\hat{i}}-1}-1)\right]}\bigg/\sum_{j=0}^{\chi(G_2)-1}\frac{(2^{p_j}-1)}{(2^{p_j}-1)(d-1)-(2^{p_j-1}-1)}
\end{equation} for $\hat{i}\in\mathbb{Z}_{\chi(G_2)}$, $l_{\hat{i}}\in\mathbb{Z}_{(d-1)^{p_{\hat{i}}}}$, and 
\begin{equation}
\fontsize{8}{10}
    \begin{aligned}
    \min\{\beta\left(\Omega_{G_1}\right)\}&=\min_{\sum_{i=0}^{\chi(G_1)-1}\mu_i=1,\mu_i\geq0}\left\{\max_{i\in\mathbb{Z}_{\chi(G_1)}}\left\{1-\frac{(d-1)d^{p_{i}-1}\mu_{i}}{d^{p_{i}}-1}\right\}\right\}=1-\frac{d-1}{d\cdot\chi(G_1)-\sum_{j=0}^{\chi(G_1)-1}d^{1-p_j}},\\
        \min\{\beta\left(\Omega_{G_2}\right)\}&=\min_{\sum_{\hat{i}=0}^{\chi(G_2)-1}\sum_{l_{\hat{i}}=0}^{(d-1)^{p_{\hat{i}}}-1}\mu_{({\hat{i}},l_{\hat{i}})}=1,\mu_{({\hat{i}},l_{\hat{i}})}\geq0}\left\{\max_{\hat{i}\in\mathbb{Z}_{\chi(G_2)},t_{\hat{i}}'\in\mathbb{Z}_{p_{\hat{i}}}}\left\{1-\sum_{l_{\hat{i}}=0}^{(d-1)^{p_{\hat{i}}}-1}\mu_{({\hat{i}},l_{\hat{i}})}\right.\right.\\& \left.\left.+\sum_{\substack{k_{v_{(\hat{i},t_{\hat{i}})}}=1,\\t_{\hat{i}}\in\mathbb{Z}_{p_{\hat{i}}}\backslash\{t_{\hat{i}}'\}}}^{d-1}\frac{(2^{p_{\hat{i}}-1}-1)\cdot\mu_{\left(\hat{i},\sum_{t_{\hat{i}}=0}^{p_{\hat{i}}-1}(k _{v_{t_{\hat{i}}}}-1)(d-1)^{t_{\hat{i}}}\right)}}{2^{p_{\hat{i}}}-1}\right\}\right\}=1-\frac{1}{\sum_{\hat{i}=0}^{\chi{G_2}-1}\frac{(2^{p_{\hat{i}}}-1)(d-1)}{(2^{p_{\hat{i}}}-1)(d-1)-(2^{p_{\hat{i}}-1}-1)}}.
    \end{aligned}
\end{equation}
Then  
\begin{equation}
    \nu(\Omega_{G_1})=\frac{d-1}{d\cdot\chi(G_1)-\sum_{j=0}^{\chi(G_1)-1}d^{1-p_j}}
\end{equation} and  
\begin{equation}
   \nu(\Omega_{G_2})=\left[\sum_{\hat{i}=0}^{\chi(G_2)-1}\frac{(2^{p_{\hat{i}}}-1)(d-1)}{(2^{p_{\hat{i}}}-1)(d-1)-(2^{p_{\hat{i}}-1}-1)}\right]^{-1}.
\end{equation}  
Therefore,
\begin{equation}
\fontsize{8}{10}
    \begin{aligned}
    n_{opt}^{G_1}&=\left\lceil \ln\delta\cdot\left[\ln\left(1-\frac{d-1}{d\cdot\chi(G_1)-\sum_{j=0}^{\chi(G_1)-1}d^{1-p_j}}\epsilon\right)\right]^{-1}\right\rceil\\
        &\leq\left\lceil \frac{d\cdot\chi(G_1)-\sum_{j=0}^{\chi(G_1)-1}d^{1-p_j}}{d-1}\epsilon^{-1}\ln\delta^{-1}\right\rceil=\left\lceil \left(\chi(G_1)+\frac{\chi(G_1)}{d-1}-\frac{\sum_{j=0}^{\chi(G_1)-1}d^{1-p_j}}{d-1}\right)\epsilon^{-1}\ln\delta^{-1}\right\rceil&<\left\lceil \left(\chi(G_1)+\frac{\chi(G_1)}{d-1}\right)\epsilon^{-1}\ln\delta^{-1}\right\rceil,
        \end{aligned}
\end{equation}
\begin{equation}
\fontsize{8}{10}
    \begin{aligned}
        n_{opt}^{G_2}&=\left\lceil \ln\delta\cdot\left[\ln\left(1-\left(1\bigg/\sum_{\hat{i}=0}^{\chi{G_2}-1}\frac{(2^{p_{\hat{i}}}-1)(d-1)}{(2^{p_{\hat{i}}}-1)(d-1)-(2^{p_{\hat{i}}-1}-1)}\right)\epsilon\right)\right]^{-1}\right\rceil\\
        &\leq\left\lceil \left(\sum_{\hat{i}=0}^{\chi{G_2}-1}\frac{(2^{p_{\hat{i}}}-1)(d-1)}{(2^{p_{\hat{i}}}-1)(d-1)-(2^{p_{\hat{i}}-1}-1)}\right)\epsilon^{-1}\ln\delta^{-1}\right\rceil=\left\lceil \left(\chi(G_2)+\sum_{\hat{i}=0}^{\chi{G_2}-1}\frac{(2^{p_{\hat{i}}-1}-1)}{(2^{p_{\hat{i}}}-1)(d-1)-(2^{p_{\hat{i}}-1}-1)}\right)\epsilon^{-1}\ln\delta^{-1}\right\rceil\\
        &<\left\lceil \left(\chi(G_2)+\frac{\chi(G_2)}{d-1}\right)\epsilon^{-1}\ln\delta^{-1}\right\rceil.
    \end{aligned}
\end{equation}
In summary, for  
$G_1\in\scriptstyle\left\{G_n^d,\widetilde{G}_n^d\right\}$ and  
$G_2\in\scriptstyle\left\{\widehat{G}_n^d,\widehat{\widetilde{G}}_n^d\right\}$,  
when $G_0\in\scriptstyle\left\{G_n^d,\widetilde{G}_n^d,\widehat{G}_n^d,\widehat{\widetilde{G}}_n^d\right\}$,  
the number of measurements required for verifying $|G_0\rangle$ satisfies  
\begin{equation}
    n_{opt}^{G_0}<\left\lceil \left(\chi(G_0)+\frac{\chi(G_0)}{d-1}\right)\epsilon^{-1}\ln\delta^{-1}\right\rceil.
\end{equation}
Hence, qubit and qudit  graph states, hypergraph states,  
and the recently proposed multigraph and multihypergraph states  
can all be verified with a number of measurements scaling approximately with  
the chromatic number of the corresponding graph.  
This is comparable to the fractional-coloring–based efficient verification scheme for hypergraph states proposed in Ref.~\cite{zhu2019efficient}, since for any graph, the difference between its chromatic number and fractional chromatic number is only a constant factor. Nevertheless, our scheme is universal.

\section{Appendix F: Verification of Composite-Dimensional Quantum States}

Given a qudit generalized Bell-like state with $d=6$, $|\psi_3\rangle=\frac{1}{\sqrt{6}}\sum_{j=0}^{5}\omega_3^j|jj\rangle$, since $6=2\times3$, it can be equivalently represented as
\begin{equation}
    |\psi_3\rangle=\frac{1}{\sqrt{6}}\sum_{j_2=j_0=0}^1\sum_{j_3=j_1=0}^2\omega_3^{j_0\cdot j_1}|j_0j_1j_2j_3\rangle.
\end{equation}
Note that this does not imply the actual preparation of particles in the Hilbert space 
$\mathcal{H}_2\otimes\mathcal{H}_3\otimes\mathcal{H}_2\otimes\mathcal{H}_3$, 
but rather an equivalent mathematical representation. 
In what follows, we use this representation to construct quantum measurement operators that still act within $\mathcal{H}_6\otimes\mathcal{H}_6$. 
The feasibility lies in the fact that the measurement operators in $\mathcal{H}_6\otimes\mathcal{H}_6$ and in 
$\mathcal{H}_2\otimes\mathcal{H}_3\otimes\mathcal{H}_2\otimes\mathcal{H}_3$ 
are mathematically isomorphic. 
Therefore, we can define measurement operators in 
$\mathcal{H}_2\otimes\mathcal{H}_3\otimes\mathcal{H}_2\otimes\mathcal{H}_3$ 
and use them to measure quantum states in $\mathcal{H}_6\otimes\mathcal{H}_6$.  

First, we give the mathematical construction of $|\psi_3\rangle$. 
Its preparation operation (for mathematical representation only) in the space 
$\mathcal{H}_2\otimes\mathcal{H}_3\otimes\mathcal{H}_2\otimes\mathcal{H}_3$ 
is
\begin{equation}
    \begin{aligned}
        U_{\psi_3}=CSUM_{1,3}^{3\otimes3}\cdot CNOT_{0,2}^{2\otimes2}\cdot CZ_{0\rightarrow{1}}^{2\otimes3}\cdot(I_{0}\otimes QFT_{1})\cdot\left(H_{0}\otimes I_{1}\right).
    \end{aligned}
\end{equation}
According to definition and proposition of generalized stabilizers in the main text, 
the generalized stabilizer group of $|\psi_3\rangle$ has four generators
\begin{equation}
\normalsize
    \begin{aligned}
     g_0^{(\psi_3)}&=U_{\psi_3}Z_0^{(2)}U_{\psi_3}^\dagger=|0\rangle\langle1|_0\otimes \left(Z_1^{(3)}\right)^\dagger\otimes\left(X_2^{(2)}\right)^\dagger\otimes I_3^{(3)}+|1\rangle\langle0|_0\otimes Z_1^{(3)}\otimes X_2^{(2)}\otimes I_3^{(3)}\\[-1mm]
        &=\left(|0\rangle\langle1|_0\otimes \left(Z_1^{(3)}\right)^\dagger+|1\rangle\langle0|_0\otimes Z_1^{(3)}\right)\otimes X_2^{(2)}\otimes I_3^{(3)}\\[-1mm]
        &=\left(\sum_{i=0}^2\sum_{\pm}\pm|u_0(i,\pm)\rangle\langle u_0(i,\pm)|\right)\otimes\left(\sum_{j=0}^2\sum_{\pm}\pm|u_0'(j,\pm)\rangle\langle u_0'(j,\pm)|\right),\\[-1mm]
        g_1^{(\psi_3)}&=U_{\psi_3}Z_1^{(3)}U_{\psi_3}^\dagger=\left(|0\rangle\langle0|+\omega_3^{-1}|1\rangle\langle1|\right)_0\otimes \left(X_1^{(3)}\right)^\dagger\otimes I_2^{(2)}\otimes \left(X_3^{(3)}\right)^\dagger\\[-1mm]
        &=\left(\sum_{j_0=0}^1\sum_{j_1=0}^2\omega_3^{j_1-j_0}|u_1(j_0,j_1)\rangle\langle u_1(j_0,j_1)|\right)\otimes\left(\sum_{j_0=0}^1\sum_{j_0}^2\omega_3^{j_3}|u_1(j_2,j_3)\rangle\langle u_1(j_2,j_3)|\right),
    \end{aligned}
\end{equation}
\begin{equation}
\normalsize
    \begin{aligned}g_2^{(\psi_3)}&=U_{\psi_3}Z_2^{(2)}U_{\psi_3}^\dagger=Z_0^{(2)}\otimes I_1^{(3)}\otimes Z_2^{(2)}\otimes I_3^{(3)}\\[-1mm]
        &=\left(\sum_{j_0=0}^1\sum_{j_1=0}^2(-1)^{j_0}|j_0,j_1\rangle\langle j_0,j_1|\right)\otimes\left(\sum_{j_0=0}^1\sum_{j_0}^2(-1)^{j_2}|j_2,j_3\rangle\langle j_2,j_3|\right)=\left(\sum_{j=0}^5(-1)^{\lfloor j/3\rfloor}|j\rangle\langle j|\right)^{\otimes2},\\[-1mm]
        g_3^{(\psi_3)}&=U_{\psi_3}Z_3^{(3)}U_{\psi_3}^\dagger=I_0^{(2)}\otimes \left(Z_{1}^{(3)}\right)^\dagger\otimes I_2^{(2)}\otimes Z_3^{(3)}\\[-1mm]
        &=\left(\sum_{j_0=0}^1\sum_{j_1=0}^2\omega_3^{-j_1}|j_0,j_1\rangle\langle j_0,j_1|\right)\otimes\left(\sum_{j_0=0}^1\sum_{j_0}^2\omega_3^{j_3}|j_2,j_3\rangle\langle j_2,j_3|\right)\\[-1mm]
        &=\left(\sum_{k_0=0}^5\omega_3^{-(k_0\text{ mod }3)}|k_0\rangle\langle k_0|\right)\otimes\left(\sum_{k_1=0}^5\omega_3^{k_1\text{ mod }3}|k_0\rangle\langle k_0|\right),
    \end{aligned}
\end{equation}
where the superscripts $(2)$ and $(3)$ denote the dimension of the Hilbert space for each operation, 
and the subscripts $0,1,2,3$ specify the position of the particles (consistent with the convention used in the main text). 
Note that these are only mathematical representations rather than physical particles. 
In addition, we have
\begin{align}
   |u_0(i,\pm)\rangle&=\frac{1}{\sqrt{2}}(|0,i\rangle\pm\omega_3^i|1,i\rangle)=\frac{1}{\sqrt{2}}(|i\rangle\pm\omega_3^i|i+3\text{ mod }6\rangle),\ i\in \mathbb{Z}_3,\\[-1mm]
        |u_0'(j,\pm)\rangle&=\frac{1}{\sqrt{2}}(|0,j\rangle\pm|1,j\rangle)=\frac{1}{\sqrt{2}}(|j\rangle\pm|j+3\text{ mod }6\rangle),\ j\in \mathbb{Z}_3,\\[-2mm]
        |u_1(a,b)\rangle&=|a\rangle\otimes\frac{1}{\sqrt{3}}\sum_{l=0}^2\omega_3^{l\cdot b}|l\rangle=\frac{1}{\sqrt{3}}\sum_{l=0}^2\omega_3^{l\cdot b}|3\cdot a+l\text{ mod }6\rangle,a\in\mathbb{Z}_2,b\in\mathbb{Z}_3.
\end{align}

In summary, the testing procedure for $g_0^{(\psi_3)}$ is as follows. 
Measure the first particle of $|\psi_3\rangle$ in the basis $\{|u_0(i,\pm)\rangle\}$, 
and the second particle in the basis $\{|u_0'(i,\pm)\rangle\}$.  
If the outcomes are $\{|u_0(i,+)\rangle,|u_0'(i,+)\rangle\}$ 
or $\{|u_0(i,-)\rangle,|u_0'(i,-)\rangle\}$, 
the test is passed. For $g_1^{(\psi_3)}$, 
measure both particles in the basis $\{|u_1(a,b)\rangle\}$, 
and if the measurement outcomes $\{|u_1(j_2,j_3)\rangle,|u_1(a,b)\rangle\}$ satisfy 
$j_1-j_0+j_3\text{ mod }3=0$, 
the test is passed. The testing procedures for $g_2^{(\psi_3)}$ and $g_3^{(\psi_3)}$ 
both use the computational basis $\{|0\rangle,|1\rangle,\cdots,|5\rangle\}$ 
to measure the two particles of $|\psi_3\rangle$, 
and both can be tested simultaneously in one measurement.  
Denote the measurement results as $M_0, M_1\in\mathbb{Z}_6$.  
If $\lfloor M_0/3\rfloor+\lfloor M_1/3\rfloor\text{ mod }2=0$, 
then the $g_2^{(\psi_3)}$ test passes.  
If $M_1-M_0\text{ mod }3=0$, 
then the $g_3^{(\psi_3)}$ test passes. Hence, we can set 
$S_{Test}^{\psi_3}=\big\{g_0^{(\psi_3)},g_1^{(\psi_3)},g_2^{(\psi_3)},g_3^{(\psi_3)}\big\}
$. Careful readers may notice that $S_{Test}^{\psi_3}$ can be further enlarged, 
but here we only choose the simplest configuration to demonstrate the verification of composite-dimensional quantum states.  
Based on the above analysis, we can further express
$S_{Test}^{\psi_3}=\bigcup_{i=0}^2S_{Test_i}^{\psi_3}
=\big\{g_0^{(\psi_3)}\big\}\bigcup\{g_1^{(\psi_3)}\big\}\bigcup\{g_2^{(\psi_3)},g_3^{(\psi_3)}\big\}$, with $C_{Test_0}^{\psi_3}=\{(1,0,0,0)\},\quad
C_{Test_1}^{\psi_3}=\{(0,1,0,0)\},\quad
C_{Test_2}^{\psi_3}=\{(0,0,1,0),(0,0,0,1)\}$. Therefore, according to the Theorem in main text, 
the measurement operator for verifying $|\psi_3\rangle$ can be constructed as
\begin{equation}
\hspace{-10mm}
\normalsize
    \begin{aligned}
    &\Omega_{\psi_3}\\[-4mm]
    =&|\psi_3\rangle\langle\psi_3|+\sum_{\substack{j_0,j_2\in\mathbb{Z}_2;j_1,j_3\in\mathbb{Z}_3\\(j_0,j_1,j_2,j_3)\neq(0,0,0,0)}}\left(\sum_{i=0}^{2}\frac{\mu_i}{|S_{Test_i}^{\psi_3}|}\left(\sum_{\substack{h_0,h_1:\ \sum_{k=0}^{1}\frac{h_k\times j_k}{d_k}\in\mathbb{Z}\\
    (h_0,h_1,h_3,h_4)\in C_{Test_i}^{\psi_3}}} 1\right)\right)U_{\psi_3}\left(\bigotimes_{k=0}^{3}\left| j_k \right\rangle \left\langle j_k\right|\right)U_{\psi_3}^\dagger\\[-1mm]
    =&|\psi_3\rangle\langle\psi_3|+U_{\psi_3}\bigg[\mu_0\sum_{\substack{j_2\in\mathbb{Z}_2,j_1,j_3\in\mathbb{Z}_3\\
    (j_1,j_2,j_3)\neq(0,0,0)}}|j_1,3 j_2+j_3\text{ mod }6\rangle\langle j_1,3 j_2+j_3\text{ mod }6|\\[-1mm]
    & +\mu_1\sum_{\substack{j_0,j_2\in\mathbb{Z}_2;j_3\in\mathbb{Z}_3\\
    (j_0,j_2,j_3)\neq(0,0,0)}}|3 j_0,3 j_2+j_3\text{ mod }6\rangle\langle3 j_0,3 j_2+j_3\text{ mod }6|\\[-1mm]
    &+\frac{\mu_2}{2}\bigg(\sum_{\substack{j_0\in\mathbb{Z}_2;j_1,j_3\in\mathbb{Z}_3\\
    (j_0,j_1,j_3)\neq(0,0,0)}}|3 j_0+j_1\text{ mod }6,j_3\rangle\langle3 j_0+j_1\text{ mod }6,j_3|+\sum_{\substack{j_0,j_2\in\mathbb{Z}_2;j_1\in\mathbb{Z}_3\\
    (j_0,j_1,j_2)\neq(0,0,0)}}|3 j_0+j_1\text{ mod }6,3j_2\rangle\langle3 j_0+j_1\text{ mod }6,3j_2|\bigg)\bigg]U_{\psi_3}^\dagger,
    \end{aligned}
\end{equation}
where $\sum_{i=0}^2\mu_i=1$ and $\mu_0,\mu_1,\mu_2>0$.  
When $2\mu_0=2\mu_1=\mu_2=\frac{1}{2}$,  
we have $\min\{\beta(\Omega_{\psi_1})\}=\mu_0=\mu_1=\frac{1}{2}\mu_2=\frac{1}{4}$.  
In this case, $\nu(\Omega_{\psi_3})=\frac{3}{4}$,  
and the number of measurements required for verifying $|\psi_3\rangle$ satisfies
$n_{opt}^{\psi_3}\leq\left\lceil\frac{4}{3}\epsilon^{-1}\ln{\frac{1}{\delta}}\right\rceil$.

\end{document}